\documentclass[12pt]{article}
\usepackage{graphicx,amsmath,amssymb}
\usepackage{bm}
\usepackage{color}

\usepackage{capt-of,subcaption}
\usepackage{array,multirow}

\usepackage[T1]{fontenc}

\newcounter{mysubsubsection}

\newcommand{\spc}{{\ }}
\newcommand{\pr}[1]{{\sc{\lowercase{#1}}}}

\newcommand{\gras}[1]{\boldsymbol{#1}}

\newcommand{\execversion}{273y}
\newcommand{\codeversion}{2.73y}
\newcommand{\fissionversion}{7b}
\newcommand{\fitsversion}{15}
\newcommand{\functionalversion}{3}
\newcommand{\hfbthoversion}{200j}
\newcommand{\interfaceversion}{4}
\newcommand{\lipkinversion}{7c}
\newcommand{\modulesversion}{17b}
\newcommand{\mpiioversion}{5b}
\newcommand{\mpimanagerversion}{4b}
\newcommand{\pnpversion}{6}
\newcommand{\shellversion}{4b}
\newcommand{\sizeversion}{2}
\newcommand{\scalapackversion}{3}

\newcounter{leteq}

\newenvironment{eqnalpha}{\setcounter{leteq}{1}

\begin{eqnarray}}{\end{eqnarray}%
}

\newenvironment{eqnalphalabel}[1]{\setcounter{leteq}{1}
\raisebox{0cm}[0cm][0cm]{\begin{minipage}{1cm}%
\begin{eqnarray}\label{#1}&&\nonumber\end{eqnarray}\end{minipage}}

\begin{eqnarray}}{\end{eqnarray}%
}

\newcommand{\bnl}{\begin{eqnalpha}}
\newcommand{\enl}{\end{eqnalpha}}
\newcommand{\bnll}[1]{\begin{eqnalphalabel}{#1}}
\newcommand{\enll}{\end{eqnalphalabel}}

\newcommand{\keyw}{{\bf Keyword:}}
\newcommand{\key}[1]{\vspace{1ex}\noindent\keyw{\spc}{\tk{#1}}
                         \newline\phantom{\keyw{\spc}{\tk{#1}}}{\spc}}
\newcommand{\keyp}[1]{\newline\phantom{\keyw{\spc}{\tk{#1}}}{\spc}}
\newcommand{\keywo}[1]{{keyword {\tk{#1}}}}
\newcommand{\keyspace}{\vspace{2ex}}

\newcommand{\be}{\begin{equation}}
\newcommand{\ee}{\end{equation}}
\newcommand{\ba}{\begin{array}}
\newcommand{\ea}{\end{array}}
\newcommand{\bn}{\begin{eqnarray}}
\newcommand{\en}{\end{eqnarray}}
\newcommand{\bc}{\begin{center}}
\newcommand{\ec}{\end{center}}
\newcommand{\bi}{\begin{itemize}}
\newcommand{\ei}{\end{itemize}}

\newcommand{\thalf}{{\textstyle{\frac{1}{2}}}}

\newcommand{\tquar}{{\textstyle{\frac{1}{4}}}}

\newcommand{\ti}[1]{#1\index{#1}}
\newcommand{\tii}[2]{#1\index{#1 #2}}
\newcommand{\tv}[1]{{\tt{#1}}\index{#1}}
\newcommand{\tk}[1]{{\tt{#1}}\index{#1}}
\newcommand{\ts}[1]{{\pr{#1}}\index{#1}}
\newcommand{\tf}[1]{{\tt{#1}}\index{#1}}
\renewcommand{\ti}[1]{#1}
\renewcommand{\tii}[2]{#1}
\renewcommand{\tv}[1]{\textcolor{red}    {{\tt{#1}}}}
\renewcommand{\tk}[1]{\textcolor{blue}   {{\tt{#1}}}}
\renewcommand{\ts}[1]{\textcolor{green}  {{\pr{#1}}}}
\renewcommand{\tf}[1]{\textcolor{magenta}{{\tt{#1}}}}
\renewcommand{\tv}[1]{{\tt{#1}}}
\renewcommand{\tk}[1]{{\tt{#1}}}
\renewcommand{\ts}[1]{{\pr{#1}}}
\renewcommand{\tf}[1]{{\tt{#1}}}

\begin{document}

\vspace{0.5cm}
\begin{center}
        {\bf\Large
                     Solution of the Skyrme-Hartree-Fock-Bogolyubov equations in
                     the Cartesian deformed harmonic-oscillator basis. \\[1ex]
                     (VIII) \pr{hfodd} {(v\codeversion)}: a new version of the
                     program.
        }

\vspace{5mm}
        {\large
                       N. Schunck,$^{a}$\footnote
                       {E-mail: schunck1@llnl.gov}
                       J.~Dobaczewski,$^{b,c,d,e}$
                       W.~Satu{\l}a,$^{d,e}$
                       P.~B\k{a}czyk,$^{d}$
                       J.~Dudek,$^{f,g}$
                       Y.~Gao,$^{c}$
                       M.~Konieczka,$^{d}$
                       K.~Sato,$^{h}$
                       Y.~Shi,$^{c,i,j}$
                     X.B.~Wang,$^{c,k}$
                 and T.R.~Werner$^{d}$
        }

\vspace{3mm}
        {\it
          $^a$Nuclear and Chemical Sciences Division, Lawrence Livermore National Laboratory Livermore, CA 94551, USA \\
          $^b$Department of Physics, University of York, Heslington, York YO10 5DD, United Kingdom \\
          $^c$Department of Physics, P.O. Box 35 (YFL), FI-40014 University of Jyv\"askyl\"a, Finland \\
          $^d$Institute of Theoretical Physics, Faculty of Physics, University of Warsaw, \\
              ul. Pasteura 5, PL-02093 Warsaw, Poland \\
          $^e$Helsinki Institute of Physics, P.O. Box 64, FI-00014 University of Helsinki, Finland \\
          $^f$Universit\'e de Strasbourg, CNRS, IPHC UMR 7178, F-67000 Strasbourg, France \\
          $^g$Institute of Physics, Marie Curie-Sk{\l}odowska University, PL-20031 Lublin, Poland \\
          $^h$Department of Physics, Osaka City University, Osaka, 558-8585, Japan \\
          $^i$National Superconducting Cyclotron Laboratory,
              Michigan State University, East Lansing, Michigan, 48824-1321, USA\\
          $^j$Department of Physics, Harbin Institute of Technology, Harbin 150001, China \\
          $^k$School of Science, Huzhou University, Huzhou, 313000, P.R. China \\
        }
\end{center}

\vspace{5mm}
\hrule

\vspace{2mm}
\noindent{\bf Abstract}

We describe the new version {(v\codeversion)} of the code \pr{hfodd} which solves
the nuclear Skyrme Hartree-Fock or Skyrme Hartree-Fock-Bogolyubov
problem by using the Cartesian deformed harmonic-oscillator basis. In the new
version, we have implemented the following new features:
(i) full proton-neutron mixing in the particle-hole channel for Skyrme functionals,
(ii) the Gogny force in both particle-hole and particle-particle channels,
(iii) linear multi-constraint method at finite temperature,
(iv) fission toolkit including the constraint on the number of particles in the neck
between two fragments, calculation of the interaction energy between
fragments, and calculation of the nuclear and Coulomb energy of each
fragment,
(v) the new version 200d of the code \pr{hfbtho}, together with an
enhanced interface between \pr{hfbtho} and \pr{hfodd},
(vi) parallel capabilities, significantly extended by adding several
restart options for large-scale jobs,
(vii) the Lipkin translational energy correction method with pairing,
(viii) higher-order Lipkin particle-number corrections,
(ix) interface to a program plotting single-particle energies or Routhians,
(x) strong-force isospin-symmetry-breaking terms,
and
(xi) the Augmented Lagrangian Method for calculations with 3D constraints
on angular momentum and isospin.
Finally, an important bug related to the calculation of
the entropy at finite temperature and several other little significant errors
of the previous published version were corrected.
\vspace{2mm}
\hrule

\vspace{2mm}

\newpage
\vspace{5mm}
\noindent{\bf\large NEW VERSION PROGRAM SUMMARY}

\bigskip\noindent{\it Title of the program:} \pr{hfodd}
                      {(v\codeversion)}

\bigskip\noindent{\it Catalogue number:}
                   ....

\bigskip\noindent{\it Program obtainable from:}
                      CPC Program Library, \
                      Queen's University of Belfast, N. Ireland
                      (see application form in this issue)

\bigskip\noindent{\it Reference in CPC for earlier version of program:}
                      N. Schunck, J. Dobaczewski, J. McDonnell, W. Satu{\l}a,
                      J. Sheikh, A. Staszczak, M. Stoitsov, and P. Toivanen,
                      Comput.\ Phys.\ Commun.\ {\bf 183} (2012) 166-192.

\bigskip\noindent{\it Catalogue number of previous version:}
                      ADFL\_v2\_1

\bigskip\noindent{\it Licensing provisions:} GPL v3

\bigskip\noindent{\it Does the new version supersede the previous one:} yes

\bigskip\noindent{\it Computers on which the program has been tested:}
                      Intel Pentium-III, Intel Xeon, AMD-Athlon, AMD-Opteron,
                      Cray XT4, Cray XT5

\bigskip\noindent{\it Operating systems:} UNIX, LINUX, Windows$^{\text{xp}}$

\bigskip\noindent{\it Programming language used:} FORTRAN-90

\bigskip\noindent{\it Memory required to execute with typical data:} 10 Mwords

\bigskip\noindent{\it No. of bits in a word:}
                      The code is written in single-precision for the use on
                      a 64-bit processor. The compiler option {\tt{}-r8} or
                      {\tt{}+autodblpad} (or equivalent) must be used to
                      promote all real and complex single-precision
                      floating-point items to double precision when the code
                      is used on a 32-bit machine.

\bigskip\noindent{\it Has the code been vectorised?:} Yes

\bigskip\noindent{\it Has the code been parallelized?:} Yes

\bigskip\noindent{\it No.{\spc}of lines in distributed program:}
                      159 633 (of which 62 714 are comments and separators)

\bigskip\noindent{\it Keywords:}
                      Hartree-Fock; Hartree-Fock-Bogolyubov; Skyrme interaction;
                      Self-consistent mean field;
                      Nuclear many-body problem; Superdeformation;
                      Quadrupole deformation; Octupole deformation; Pairing;
                      Nuclear radii; Single-particle spectra;
                      Nuclear rotation; High-spin states;
                      Moments of inertia; Level crossings; Harmonic oscillator;
                      Coulomb field; Pairing; Point symmetries;
                      Yukawa interaction; Angular-momentum projection;
                      Generator Coordinate Method; Schiff moments;
                      Isospin mixing; Isospin projection, Finite temperature;
                      Shell correction; Lipkin method; Multi-threading; Hybrid
                      programming model; High-performance computing.

\bigskip\noindent{\it Nature of physical problem}

\noindent
The nuclear mean field and an analysis of its symmetries in realistic cases are
the main ingredients of a description of nuclear states. Within the Local
Density Approximation, or for a zero-range velocity-dependent Skyrme
interaction, the nuclear mean field is local and velocity dependent. The
locality allows for an effective and fast solution of the self-consistent
Hartree-Fock equations, even for heavy nuclei, and for various nucleonic
($n$-particle $n$-hole) configurations, deformations, excitation energies, or
angular momenta. Similarly, Local Density Approximation in the particle-particle
channel, which is equivalent to using a zero-range interaction, allows for a
simple implementation of pairing effects within the Hartree-Fock-Bogolyubov
method. For finite-range interactions, like Coulomb, Yukawa, or Gogny
interaction, the nuclear mean field becomes nonlocal, but using the
spatial separability of the deformed harmonic-oscillator basis in
three Cartesian directions, the self-consistent calculations can be
efficiently performed.

\bigskip\noindent{\it Method of solution}

\noindent
The program uses the Cartesian harmonic oscillator basis to expand
single-particle or single-quasiparticle wave functions of neutrons
and protons interacting by means of the Skyrme or Gogny effective
interactions and zero-range or finite-range pairing interactions. The
expansion coefficients are determined by the iterative
diagonalization of the mean-field Hamiltonians or Routhians which
depend non-linearly on the local or nonlocal neutron, proton, or
mixed proton-neutron densities. Suitable constraints are used to
obtain states corresponding to a given configuration, deformation or
angular momentum. The method of solution has been presented in: J.
Dobaczewski and J. Dudek, Comput.\ Phys.\ Commun.\ {\bf 102} (1997)
166.

\bigskip\noindent{\it Summary of revisions}

\noindent
\begin{enumerate}
\setlength{\itemsep}{-1ex}
\item Full proton-neutron mixing in the particle-hole channel for Skyrme functionals was implemented.
\item The Gogny force was implemented in both particle-hole and
particle-particle channels.
\item Linear multi-constraint method based on the cranking approximation of
the RPA matrix was extended at finite temperature.
\item Fission toolkit including the constraint on the number of particles in
the neck between two fragments, calculation of the interaction energy
between fragments, and calculation of the nuclear and Coulomb energy of
each fragment.
\item The \pr{hfbtho} module was updated to version 200d, and an enhanced
interface between \pr{hfbtho} and \pr{hfodd} was implemented.
\item Parallel capabilities were significantly extended by adding several
restart options for large-scale jobs.
\item The Lipkin translational energy correction method with pairing
was implemented.
\item Higher-order Lipkin particle-number corrections were implemented.
\item Interface to a program plotting single-particle energies or Routhians
were added.
\item Strong-force isospin-symmetry-breaking terms were implemented.
\item The Augmented Lagrangian Method for calculations with 3D constraints
on angular momentum and isospin was implemented.
\item An important bug related to the calculation of
the entropy at finite temperature and several other little significant errors
of the previous published version were corrected.
\end{enumerate}

\bigskip\noindent{\it Restrictions on the complexity of the problem}

\noindent

\bigskip\noindent{\it Typical running time}

\noindent

\bigskip\noindent{\it Unusual features of the program}

\noindent
The user must have access to (i) the LAPACK subroutines \pr{ZHPEV},
\pr{ZHPEVX}, \pr{ZHEEVR}, or \pr{ZHEEVD}, which diagonalize complex hermitian
matrices, (ii) the LAPACK subroutines \pr{DGETRI} and \pr{DGETRF} which invert
arbitrary real matrices, (iii) the LAPACK subroutines \pr{DSYEVD}, \pr{DSYTRF}
and \pr{DSYTRI} which compute eigenvalues and eigenfunctions of real symmetric
matrices and (iv) the LINPACK subroutines \pr{ZGEDI} and \pr{ZGECO}, which
invert arbitrary complex matrices and calculate determinants, (v) the BLAS
routines \pr{DCOPY}, \pr{DSCAL}, \pr{DGEEM} and \pr{DGEMV} for double-precision
linear algebra and \pr{ZCOPY}, \pr{ZDSCAL}, \pr{ZGEEM} and \pr{ZGEMV} for
complex linear algebra, or provide another set of subroutines that can perform
such tasks. The BLAS and LAPACK subroutines can be obtained from the Netlib
Repository at the University of Tennessee, Knoxville:
\verb+http://netlib2.cs.utk.edu/+.

\bigskip

\bigskip

{\bf\large LONG WRITE-UP}

\bigskip

%%%%%%%%%%%%%%%%%%%%%%%%%%%%%%%%%%%%%%%%%%%%%%%%%%%%%%%%%%%%%%%%%%%%%%%%%%%%%%%%
%%%%%%%%%%%%%%%%%%%%%%%%%%%%%%%%%%%%%%%%%%%%%%%%%%%%%%%%%%%%%%%%%%%%%%%%%%%%%%%%
%%%%%%%%%%%%%%%%%%%%%%%%%%%%%%%%%%%%%%%%%%%%%%%%%%%%%%%%%%%%%%%%%%%%%%%%%%%%%%%%
%%%%%%%%%%%%%%%%%%%%%%%%%%%%%%%%%%%%%%%%%%%%%%%%%%%%%%%%%%%%%%%%%%%%%%%%%%%%%%%%

\section{Introduction}
\label{sec:introduction}

The method of solving the Hartree-Fock (HF) equations in the Cartesian harmonic
oscillator (HO) basis was described in the publication, Ref.~\cite{[Dob97c]}.
Six versions of the code \pr{hfodd} were previously published in six independent
publications:
(v1.60r)~\cite{[Dob97d]},(v1.75r)~\cite{[Dob00c]}, (v2.08i)~\cite{[Dob04]},
(v2.08k)~\cite{[Dob05]}, (v2.40h)~\cite{[Dob09d]}, and (v2.49t)~\cite{[Sch12]}.
Version (v2.08i)~\cite{[Dob04]} introduced solutions of the Hartree-Fock-Bogolyubov (HFB) equations.
Below we refer to these publications by using roman capitals II--VII.
The User's Guide for version (v2.40v) is available in Ref.~\cite{[Dob09f]} and
the code home page is at
\verb+http://www.fuw.edu.pl/~dobaczew/hfodd/hfodd.html+.
The present paper is a long write-up of the new version {(v\codeversion)} of the
code \pr{hfodd}. This extended version features the full proton-neutron
mixing in the particle-hole channel for Skyrme functionals; full Gogny force in both
the particle-hole an particle-particle channels; linear
multi-constraint method at finite temperature; fission toolkit
including the constraint on the number of particles in the neck
between two fragments, calculation of the interaction energy between
fragments, and calculation of the nuclear and Coulomb energy of each
fragment; enhanced interface to the new version 200d of the code
\pr{hfbtho}; enhanced hybrid MPI/OpenMP parallel programming model
with several restart options for large-scale calculations on
massively parallel computers; the Lipkin translational energy
correction method with pairing; higher-order Lipkin particle-number
corrections; interface to a program plotting single-particle energies
or Routhians; strong-force isospin-symmetry-breaking (ISB) terms;
and the Augmented Lagrangian Method (ALM) for calculations with 3D constraints
on angular momentum and isospin.
In serial mode, it
remains fully compatible with all previous versions. Information
provided in previous publications \cite{[Dob97d]}-\cite{[Sch12]} thus
remains valid, unless explicitly mentioned in the present long
write-up.

In Section \ref{sec:modifications} we review the modifications
introduced in version {(v\codeversion)} of the code \pr{hfodd}. Section
\ref{sec:input_file} lists all additional new input keywords and data values,
introduced in version {(v\codeversion)}. In serial mode, the structure of the
input data file remains the same as in the previous versions, see
Section I-3. In parallel mode, two input files, with strictly
enforced names, must be used: \tf{hfodd.d} has the same keyword
structure as all previous \pr{hfodd} input files, with the
restriction that not all keywords can be activated (see updated list
in Section~\ref{subsubsec:hfodd.d}); \tf{hfodd\_mpiio.d} contains
processor-dependent data, see Section~\ref{subsubsec:hfodd_mpiio.d}.

%%%%%%%%%%%%%%%%%%%%%%%%%%%%%%%%%%%%%%%%%%%%%%%%%%%%%%%%%%%%%%%%%%%%%%%%%%%%%%%%
%%%%%%%%%%%%%%%%%%%%%%%%%%%%%%%%%%%%%%%%%%%%%%%%%%%%%%%%%%%%%%%%%%%%%%%%%%%%%%%%
%%%%%%%%%%%%%%%%%%%%%%%%%%%%%%%%%%%%%%%%%%%%%%%%%%%%%%%%%%%%%%%%%%%%%%%%%%%%%%%%
%%%%%%%%%%%%%%%%%%%%%%%%%%%%%%%%%%%%%%%%%%%%%%%%%%%%%%%%%%%%%%%%%%%%%%%%%%%%%%%%

\section{Modifications introduced in version {(v\codeversion)}}
\label{sec:modifications}

%%%%%%%%%%%%%%%%%%%%%%%%%%%%%%%%%%%%%%%%%%%%%%%%%%%%%%%%%%%%%%%%%%%%%%%%%%%%%%%%
%%%%%%%%%%%%%%%%%%%%%%%%%%%%%%%%%%%%%%%%%%%%%%%%%%%%%%%%%%%%%%%%%%%%%%%%%%%%%%%%

\subsection{Proton-neutron mixing Hartree-Fock theory}
\label{subsec:pnmixing}

The code has been extended to treat Skyrme energy density functionals
(EDFs) that include proton-neutron mixing (p-n) in the particle-hole
channel. Such a generalization leads to single-particle states that
are no longer pure proton or neutron states but mixtures thereof. In
turn, these give rise to isovector densities where all three
components are possibly non zero, in contrast to the standard p-n
separable EDFs that depend only on one isovector density, which is
the difference between neutron and proton densities.

Generalized functionals are built according to the general rules
defined by Perli\'nska {\it et al.\/}~\cite{[Per04]} and include all
terms up to the next-to-leading order that are allowed by symmetries
or, equivalently, up to second-order in derivatives of densities. In
the limit of no Coulomb and no strong-force ISB terms, the theory
becomes invariant under the rotation in the isospin space, which
constitutes an invaluable test of numerical implementations. The most
general isoscalar-scalar EDFs are of the following form (see
Eqs.~(39) and (40) in Ref.~\cite{[Per04]}):
\begin{equation}
\mathcal{H}=\frac{\hbar^2}{2m}\tau_0({\bm r})+\sum_{t=0,1}\chi_t({\bm r}),
\end{equation}
where the so-called isoscalar and isovector parts, or more precisely, the parts
of EDF depending, respectively, on the isoscalar and isovector densities, are
\begin{align}
\chi_0(r) &=
 C_0^{\rho } \rho_0^2
+C_0^{\Delta \rho} \rho_0 \Delta \rho_0
+C_0^{\tau} \rho_0 \tau_0
+C_0^{J_0} J_0^2
+C_0^{J_1} {\bm J}_0^2
+C_0^{J_2} \underline{\sf J}_0^2
+C_0^{\nabla J} \rho_0 \nabla \cdot {\bm J}_0
+C_0^{s} {\bm s}_0^2 \notag \\
&+C_0^{\Delta s}{\bm s}_0 \cdot \Delta {\bm s}_0
+C_0^{T} {\bm s}_0 \cdot {\bm T}_0
+C_0^{j} {\bm j}_0^2
+C_0^{\nabla j} {\bm s}_0 \cdot {\left(\nabla \times \bm j_0 \right)}
+C_0^{\nabla s} \left({\nabla \cdot \bm s}_0\right)^2, \label{eq:chi_0}
%+C_0^{F} {\bm s}_0 \cdot {\bm F}_0
\end{align}
and
\begin{align}
\chi_1(r) &=
 C_1^{\rho } \vec \rho^2
+C_1^{\Delta \rho} \vec \rho \circ \Delta \vec \rho
+C_1^{\tau} \vec \rho \circ \vec \tau
+C_1^{J_0} \vec J^2
+C_1^{J_1} \vec {\bm J}^2
+C_1^{J_2} \vec {\underline{ \sf J}}^2
+C_1^{\nabla J} \vec \rho \circ \nabla \cdot \vec{ \bm J}
+C_1^{s} \vec {\bm s}^2 \notag \\
&+C_1^{\Delta s}\vec {\bm s}  \cdot \circ \Delta \vec{\bm s}
+C_1^{T} \vec{\bm s} \cdot \circ \vec{\bm T}
+C_1^{j} \vec{\bm j}^2
+C_1^{\nabla j} \vec{\bm s} \cdot \circ {\left( \nabla \times \vec {\bm j} \right)}
+C_1^{\nabla s} \left({\nabla \cdot \vec{\bm s}}\right)^2. \label{eq:chi_1}
%+C_1^{F} \vec {\bm s} \cdot \vec{\bm F}
\end{align}
The particle, kinetic, spin, spin-kinetic, current, and spin-current
densities are denoted as $\rho, \tau, {\bm s}, {\bm T}, {\bm j},$ and
${\underline{ \sf J}}$, respectively. In version (v\codeversion), the tensor-kinetic density ${\bm F}$,
which appears in Eqs.~(39) and (40) of Ref.~\cite{[Per04]} is not yet implemented.
Boldfaced and underlined symbols refer, respectively, to vector
and tensor densities in space. Isoscalar densities are labeled
with the subscripts $0$, whereas isovector densities are marked with arrows.
Scalar products in space are denoted by a dot; in isospace by a
circle. Coupling constants $C$ of the EDF can be either expressed in
terms of the original Skyrme force parameters, as given in Eq.~(62) and Table I
of Ref.~\cite{[Per04]}, or read (modified) from the input data file.

In the p-n-mixing calculations, we employ the three-dimensional
cranking method in isospace (isocranking~\cite{[Sat01]}) to enforce
the total isospin of the system. The technique is analogous to the
well known cranking method in real space which is successfully used
in high-spin physics. It is realized by adding the isocranking term
to the mean-field Hamiltonian $\hat {h}$,
\begin{equation}
\hat{h}'
=\hat {h} -\vec{\lambda} \circ \vec{t}
= \hat {h}  - \lambda_1\hat{t}_1-\lambda_2\hat{t}_2-\lambda_3\hat{t}_3,
\label{eq:isocranking}
\end{equation}
where the single-particle isospin operators, $\vec{t}=\thalf\vec{\tau}$, are
expressed by means of the Pauli matrices $\hat{\tau}_k~(k=1,2,3 )$ in isospace.
By adjusting the isocranking frequencies $\vec{\lambda}$, one can control both
the length and direction of the isospin vector. In the code, the isovector
frequency, $\vec{\lambda}$, is parameterized as follows
\begin{align}
\vec \lambda  =(\lambda_1, \lambda_2, \lambda_3)
              =(\lambda^{\prime}\sin \theta^\prime \cos \phi,
              \lambda^{\prime}\sin \theta^\prime \sin \phi,
              \lambda^{\prime}\cos \theta^\prime +\lambda_{\rm off}). \label{eq:isofrequency}
\end{align}
For $\lambda_{\rm off}=0$, it corresponds to the standard spherical
coordinate-type parametrization. Offset frequency $\lambda_{\rm off}$ is introduced to facilitate
calculations with the Coulomb interaction. Its proper choice allows us to
compensate for the effective contribution of the $T_3$-dependent electrostatic
interaction to the third component of the isocranking term. In this way, it helps
to avoid crossings of the single-particle levels in function of the tilting angles
and, consequently, to keep the total isospin fixed~\cite{[Sat13c]}.
This trick is invaluable when performing self-consistent calculations for
different members of an isobaric multiplet as demonstrated in
Ref.~\cite{[Sat13c]}, where the strategies of choosing the value of
$\lambda_{\rm off}$ are discussed in detail. Table \ref{tab:A014_with_Coulomb}
shows an example of results calculated for the $T\simeq 1$ states in $A=14$ isobars with the
Coulomb interaction included. We used $\lambda_{\rm off}=-1.45$ and
$\lambda^\prime=7.20$\,MeV. The $T_3\simeq 0$ state consists of the
single-particle states in which proton and neutron components are almost
equally mixed.

\begin{table}[!ht]
\caption{
Total energies calculated for the $T \simeq 1$ triplet states in $A=14$ isobars,
with $(\lambda_{\rm off},\lambda^\prime)= (-1.45, 7.20)$\,MeV.
The Coulomb interaction is treated exactly both in the direct and
exchange channels~\protect\cite{[Dob09d]}. Angle $\theta$ is the polar angle of
$\vec \lambda$ and angle $\theta'$ is defined in Eq.~(\protect\ref{eq:isofrequency}).
Expectation values of the $T_1$ and $T_3$ components of the total
isospin and its polar angles $\theta_T$ are also shown.
}
\label{tab:A014_with_Coulomb}
\begin{center}
\begin{tabular}{rrrrrrrrrr}
\hline
 $\theta^\prime$  &   $\theta_T$     &$\theta$      & $\langle \hat T_1\rangle$ & $\langle \hat T_3 \rangle$  & $E_{\rm tot}$ [MeV]  \\ \hline
 $0^\circ$        &   $0^\circ$      &$0^\circ$     & 0.00000  & 1.00000
                 & -108.491518    \\
 $90^\circ$       &   $89.97^\circ$  &$101.4^\circ$ & 1.00015  & 0.00047 & -105.685256      \\
 $180^\circ$      &   $180^\circ$    &$180^\circ$   & 0.00000  &-1.00000 & -102.680239    \\
\hline
\end{tabular}
\end{center}
\end{table}

Finally, let us recall that for the phase convention used in
\pr{hfodd}~\cite{[Dob97c]}, the time-reversal operator reads $\hat
T=-i \hat \sigma_y \hat K$ and depends on the $y$-component of the
spin Pauli matrix $\hat \sigma_y$ and complex conjugation operator
$\hat K$. Hence, any calculation involving the $2$-component of the
isocranking term, which is purely imaginary, should be performed in
time-reversal-symmetry-breaking mode. For $\lambda_2 = 0$, since the
other two components $-\lambda_1 \hat t_1$ and $-\lambda_3 \hat t_3$
are real, the time-reversal symmetry can be conserved. Note also that
the Coulomb interaction is axially symmetric in isospace. It implies
that the total EDF including the Coulomb term is always invariant
under the rotation about the $3$-isoaxis, which allows us to set the
azimuthal angle $\phi=0$ without any loss of generality.

%%%%%%%%%%%%%%%%%%%%%%%%%%%%%%%%%%%%%%%%%%%%%%%%%%%%%%%%%%%%%%%%%%%%%%%%%%%%%%%%

\subsection{The Gogny force}
\label{subsec:gogny}

In version {(v\codeversion)}, in the case without proton-neutron
mixing, the local finite-range Gogny force was implemented in both
the particle-hole and particle-particle channel. We recall that the
Gogny force reads \cite{[Dec80]},
\begin{equation}
\hat{V} =
\sum_{i=1,2}
e^{-\frac{(\gras{r}_{1} - \gras{r}_{2})^{2}}{\mu_{i}^{2}}}
\left( W_{i}\hat{1}_{\sigma}\hat{1}_{\tau} + B_{i}\hat{P}_{\sigma}\hat{1}_{\tau}
     - H_{i}\hat{1}_{\sigma}\hat{P}_{\tau} - M_{i}\hat{P}_{\sigma}\hat{P}_{\tau} \right),
\end{equation}
where $\hat{1}_{\sigma}$ and $\hat{1}_{\tau}$ are the spin and
isospin unity operators, and $\hat{P}_{\sigma}$ and $\hat{P}_{\tau}$ are the standard spin and
isospin exchange operators. In \pr{hfodd}, the spin-isospin particle-hole
expansion of the Gogny force is used, that is, the antisymmetrized potential is
written as
\begin{equation}
\label{Vfield}
\hat{\bar{V}}
 =
\hat{V}(1 - \hat{P}_{\sigma}\hat{P}_{\tau}\hat{P}_{M})
 =
\sum_{i=1,2}
e^{-\frac{(\gras{r}_{1} - \gras{r}_{2})^{2}}{\mu_{i}^{2}}}
\sum_{\mu k} \left( V_{ST}^{(iD)}+V_{ST}^{(iE)}\hat{P}_{M} \right)
\hat{\sigma}_{\mu}^{(1)}\hat{\sigma}_{\mu}^{(2)}\;
\hat{\tau}_{k}^{(1)}    \hat{\tau}_{k}^{(2)}
=\hat{{V}}_{\text{dir}}+\hat{{V}}_{\text{exc}},
\end{equation}
where $\hat{P}_{M}$ is the standard space exchange operator, $\hat{\sigma}_{\mu}$
($\mu=0,x,y,z$) and $\hat{\tau}_{k}$ ($k=0,1,2,3$) are the spin and isospin
identity and Pauli matrices,
\begin{equation}
\label{eq:sigma_std}
\hat{{\sigma}}_{0} = \left(\ba{rr}   1  &   0 \\
                                     0  &   1  \ea\right),\quad
\hat{{\sigma}}_{x} = \left(\ba{rr}   0  &   1 \\
                                     1  &   0  \ea\right),\quad
\hat{{\sigma}}_{y} = \left(\ba{rr}   0  &  -i \\
                                     i  &   0  \ea\right),\quad
\hat{{\sigma}}_{z} = \left(\ba{rr}   1  &   0 \\
                                           0  &  -1  \ea\right),
\end{equation}
and the direct, $V_{ST}^{(iD)}$,
and exchange, $V_{ST}^{(iE)}$, strength parameters of the
direct, $\hat{{V}}_{\text{dir}}=\hat{V}$,
and exchange, $\hat{{V}}_{\text{exc}}=-\hat{V}\hat{P}_{\sigma}\hat{P}_{\tau}\hat{P}_{M}$, interaction can be
expressed in terms of parameters $W_{i}$, $B_{i}$, $H_{i}$, and $M_{i}$
of the Gogny force as
\begin{eqnarray}
V^{(iD)}_{00} &=& \phantom{-\thalf}W_{i} + \thalf B_{i} - \thalf H_{i} - \tquar M_{i}, \\
V^{(iD)}_{01} &=&                                       - \thalf H_{i} - \tquar M_{i}, \\
V^{(iD)}_{10} &=& \phantom{-}              \thalf B_{i}                - \tquar M_{i}, \\
V^{(iD)}_{11} &=&                                                      - \tquar M_{i},
\end{eqnarray}
and
\begin{eqnarray}
V^{(iE)}_{00} &=& \phantom{-\thalf}M_{i} + \thalf H_{i} - \thalf B_{i} - \tquar W_{i}, \\
V^{(iE)}_{01} &=&                                       - \thalf B_{i} - \tquar W_{i}, \\
V^{(iE)}_{10} &=& \phantom{-}              \thalf H_{i}                - \tquar W_{i}, \\
V^{(iE)}_{11} &=&                                                      - \tquar W_{i},
\end{eqnarray}
with values of the total spin $S$ and isospin $T$ given by $S,T=0$ for $\mu,k=0$ and $S,T=1$ for $\mu=x,y,z$ and $k=1,2,3$.

Similarly, the spin-isospin particle-particle
expansion of the Gogny force, suitable for calculations in the pairing channel,
is written as
\begin{equation}
\label{Vpair}
\hat{V}
 =
\sum_{i=1,2}
e^{-\frac{(\gras{r}_{1} - \gras{r}_{2})^{2}}{\mu_{i}^{2}}}
\sum_{\mu k} V_{ST}^{(iP)}
\hat{\tilde{\sigma}}_{\mu}^{(L)*}\hat{\tilde{\sigma}}_{\mu}^{(R)}\;
\hat{\tilde{\tau}}_{k}^{(L)*}    \hat{\tilde{\tau}}_{k}^{(R)},
\end{equation}
where the spin and isospin identity and Pauli matrices in the particle-particle channel
are defined as
\begin{eqnarray}
\left(\hat{\tilde{\sigma}}_{\mu}\right)_{\sigma_1,\sigma_2} &=&  -\sigma_1\left(\hat{\sigma}_{\mu}\right)_{-\sigma_1,\sigma_2}, \\
\left(\hat{\tilde{\tau  }}_{k  }\right)_{\tau  _1,\tau  _2} &=&  -\tau  _1\left(\hat{\tau  }_{k  }\right)_{-\tau  _1,\tau  _2},
\end{eqnarray}
that is,
\begin{equation}
\label{eq:sigma_tilde}
\hat{\tilde{\sigma}}_{0} = \left(\ba{rr}   0  &  -1 \\
                                           1  &   0  \ea\right),\quad
\hat{\tilde{\sigma}}_{x} = \left(\ba{rr}  -1  &   0 \\
                                           0  &   1  \ea\right),\quad
\hat{\tilde{\sigma}}_{y} = \left(\ba{rr}  -i  &   0 \\
                                           0  &  -i  \ea\right),\quad
\hat{\tilde{\sigma}}_{z} = \left(\ba{rr}   0  &   1 \\
                                           1  &   0  \ea\right),
\end{equation}
and matrices denoted by $(L)$ and $(R)$ correspond to the bra and ket indices of the interaction, respectively.
Since the spin and isospin identity and exchange operators recoupled to the particle-particle
representation read (see Eqs.~(65) in Ref.~\cite{[Per04]}),
\begin{eqnarray}
\hat{1}_{\sigma} &=&\phantom{-} \thalf \hat{\tilde{\sigma}}_{0  }^{(L)*}\hat{\tilde{\sigma}}_{0  }^{(R)}
              +\thalf \sum_{\mu=x,y,z} \hat{\tilde{\sigma}}_{\mu}^{(L)*}\hat{\tilde{\sigma}}_{\mu}^{(R)}, \\
\hat{P}_{\sigma} &=&           -\thalf \hat{\tilde{\sigma}}_{0  }^{(L)*}\hat{\tilde{\sigma}}_{0  }^{(R)}
              +\thalf \sum_{\mu=x,y,z} \hat{\tilde{\sigma}}_{\mu}^{(L)*}\hat{\tilde{\sigma}}_{\mu}^{(R)}, \\
\hat{1}_{\tau  } &=&\phantom{-} \thalf \hat{\tilde{\tau  }}_{0  }^{(L)*}\hat{\tilde{\tau  }}_{0  }^{(R)}
              +\thalf \sum_{k  =1,2,3} \hat{\tilde{\tau  }}_{k  }^{(L)*}\hat{\tilde{\tau  }}_{k  }^{(R)}, \\
\hat{P}_{\tau  } &=&           -\thalf \hat{\tilde{\tau  }}_{0  }^{(L)*}\hat{\tilde{\tau  }}_{0  }^{(R)}
              +\thalf \sum_{k  =1,2,3} \hat{\tilde{\tau  }}_{k  }^{(L)*}\hat{\tilde{\tau  }}_{k  }^{(R)},
\end{eqnarray}
we obtain the pairing strength parameters, $V_{ST}^{(iP)}$,
expressed in terms of parameters $W_{i}$, $B_{i}$, $H_{i}$, and $M_{i}$
of the Gogny force as
\begin{eqnarray}
V^{(iP)}_{00} &=& \tquar\left(W_{i} -  B_{i} + H_{i} - M_{i}\right) , \\
V^{(iP)}_{01} &=& \tquar\left(W_{i} -  B_{i} - H_{i} + M_{i}\right) , \\
V^{(iP)}_{10} &=& \tquar\left(W_{i} +  B_{i} + H_{i} + M_{i}\right) , \\
V^{(iP)}_{11} &=& \tquar\left(W_{i} +  B_{i} - H_{i} - M_{i}\right) .
\end{eqnarray}

In \pr{hfodd}, the matrix elements of the particle-hole (mean-field) potential $\Gamma$ and particle-particle (pairing) potential
$\Delta$ \cite{[Rin80]} are computed directly in the configuration space,
\begin{eqnarray}
\label{eq:gogny_aux1}
\Gamma_{ac}^{\text{dir}} & = &
\sum_{bd}
\langle ab | \hat{{V}}_{\text{dir}} | cd\rangle\rho_{db}, \\
\label{eq:gogny_aux2}
\Gamma_{ac}^{\text{exc}} & = &
\sum_{bd}
\langle ab | \hat{{V}}_{\text{exc}} | cd\rangle\rho_{db}, \\
\label{eq:gogny_aux3}
\Delta_{ab} & = &
\sum_{cd}
\langle ab | \hat{V}| cd\rangle \kappa_{cd},
\end{eqnarray}
where $\Gamma_{ac}=\Gamma_{ac}^{\text{dir}}+\Gamma_{ac}^{\text{exc}}$.
In these expressions, $\rho_{db}$ and $\kappa_{cd}$ are the one-body
density matrix and pairing tensor in the configuration space. Note that to
calculate the particle-hole and particle-particle potentials we use
the antisymmetrized (\ref{Vfield}) and nonantisymmetrized (\ref{Vpair}) interactions,
respectively, and the former ones are split into the direct and exchange
contributions. Similarly, the total potential energy is split
into the direct, exchange, and pairing contributions,
\begin{equation}
E_{\text{pot}}=E_{\text{dir}}+E_{\text{exc}}+E_{\text{pair}},
\end{equation}
where
\begin{eqnarray}
\label{eq:gogny_dir}
E_{\text{dir}} & = &
\thalf\sum_{ac}\rho_{ca}\Gamma_{ac}^{\text{dir}} =
\thalf\sum_{abcd}
\rho_{ca}\langle ab | \hat{{V}}_{\text{dir}} | cd\rangle\rho_{db}, \\
\label{eq:gogny_exc}
E_{\text{exc}} & = &
\thalf\sum_{ac}\rho_{ca}\Gamma_{ac}^{\text{exc}} =
\thalf\sum_{abcd}
\rho_{ca}\langle ab | \hat{{V}}_{\text{exc}} | cd\rangle\rho_{db}, \\
\label{eq:gogny_pair}
E_{\text{pair}} & = &
\thalf\sum_{ab}\kappa^*_{ab}\Delta_{ab} =
\thalf\sum_{abcd}
\kappa^*_{ab}\langle ab | \hat{V}| cd\rangle \kappa_{cd}.
\end{eqnarray}

Basis states of the configuration space used in
\pr{hfodd}~\cite{[Dob97c]} are generically denoted by, e.g., $|d\rangle
\equiv |\gras{n}\rangle \otimes |s\rangle \otimes | \tau\rangle$,
where $\gras{n}=(n_x,n_y,n_z)$ are the HO quantum numbers, $s$ stands
for the $y$-simplex of the basis state and $\tau$ for its isospin
projection. In this basis, matrix elements of the density matrix
and pairing tensor read $\rho_{\gras{n}s\tau,\gras{m}u\chi}$ and
$\kappa_{\gras{n}'s'\tau',\gras{n}s\tau}$, respectively, and
Eqs.~(\ref{eq:gogny_dir})--(\ref{eq:gogny_pair}) can be written as
\begin{eqnarray}
\label{eq:gogny2_dir}
E_{\text{dir}} & = &
\thalf\sum_i\sum_{\mu k} V_{ST}^{(iD)}
\sum_{\gras{m}'\gras{m}\gras{n}'\gras{n}}
\rho^{\mu k}_{\gras{n}',\gras{m}'}G^{i}_{\gras{m}'\gras{m}\gras{n}'\gras{n}} \rho^{\mu k}_{\gras{n},\gras{m}}, \\
\label{eq:gogny2_exc}
E_{\text{exc}} & = &
\thalf\sum_i \sum_{\mu k} V_{ST}^{(iE)}
\sum_{\gras{m}'\gras{m}\gras{n}'\gras{n}}
\rho^{\mu k}_{\gras{n}',\gras{m}'}G^{i}_{\gras{m}'\gras{m}\gras{n}\gras{n}'} \rho^{\mu k}_{\gras{n},\gras{m}}, \\
\label{eq:gogny2_pair}
E_{\text{pair}} & = &
\thalf\sum_i \sum_{\mu k} V_{ST}^{(iP)}
\sum_{\gras{m}'\gras{m}\gras{n}'\gras{n}}
\kappa^{\mu k*}_{\gras{m}',\gras{m}}G^{i}_{\gras{m}'\gras{m}\gras{n}'\gras{n}} \kappa^{\mu k}_{\gras{n}',\gras{n}},
\end{eqnarray}
where the spin-isospin components of the density matrix
and pairing tensor are defined as
\begin{eqnarray}
\label{eq:denmat_spiso}
\rho^{\mu k}_{\gras{n},\gras{m}} & = & \sum_{s\tau,u\chi}
\rho_{\gras{n}s\tau,\gras{m}u\chi}
\hat{\sigma}^{\mu}_{us}\;\hat{\tau}^{k}_{\chi\tau} , \\
\label{eq:pairte_spiso}
\kappa^{\mu k}_{\gras{n}',\gras{n}} & = & \sum_{s'\tau',s\tau}
\kappa_{\gras{n}'s'\tau',\gras{n}s\tau}
\hat{\tilde{\sigma}}^{\mu}_{s's}\;\hat{\tilde{\tau}}^{k}_{\tau'\tau} .
\end{eqnarray}
Matrix elements of the Gaussian potential can be calculated as outlined
in Ref.~\cite{[Dob09d]}, that is,
\begin{equation}
G^{i}_{\gras{m}'\gras{m}\gras{n}'\gras{n}}=
\langle\gras{m}'\gras{m}|e^{-\frac{(\gras{r}_{1} - \gras{r}_{2})^{2}}{\mu_{i}^{2}}}|\gras{n}'\gras{n}\rangle=
G^{i}_{{m}'_x{m}_x{n}'_x{n}_x}G^{i}_{{m}'_y{m}_y{n}'_y{n}_y}G^{i}_{{m}'_z{m}_z{n}'_z{n}_z},
\end{equation}
and the time-consuming 12-dimensional summations over $\gras{m}'\gras{m}\gras{n}'\gras{n}$
can be effectively performed \cite{[Dob09d]}. We also note that the three
Eqs.~(\ref{eq:gogny2_dir})--(\ref{eq:gogny2_pair}) can be rewritten in the identical form,
\begin{eqnarray}
\label{eq:gogny3_dir}
E_{\text{dir}} & = &
\thalf\sum_i\sum_{\mu k} V_{ST}^{(iD)}
\sum_{\gras{m}'\gras{m}\gras{n}'\gras{n}}
\rho^{\mu k}_{\gras{n}',\gras{m}'}G^{i,\text{dir}}_{\gras{m}'\gras{m}\gras{n}'\gras{n}} \rho^{\mu k}_{\gras{n},\gras{m}}, \\
\label{eq:gogny3_exc}
E_{\text{exc}} & = &
\thalf\sum_i \sum_{\mu k} V_{ST}^{(iE)}
\sum_{\gras{m}'\gras{m}\gras{n}'\gras{n}}
\rho^{\mu k}_{\gras{n}',\gras{m}'}G^{i,\text{exc}}_{\gras{m}'\gras{m}\gras{n}'\gras{n}} \rho^{\mu k}_{\gras{n},\gras{m}}, \\
\label{eq:gogny3_pair}
E_{\text{pair}} & = &
\thalf\sum_i \sum_{\mu k} V_{ST}^{(iP)}
\sum_{\gras{m}'\gras{m}\gras{n}'\gras{n}}
\kappa^{\mu k*}_{\gras{n}',\gras{m}'}G^{i,\text{pair}}_{\gras{m}'\gras{m}\gras{n}'\gras{n}} \kappa^{\mu k}_{\gras{n},\gras{m}},
\end{eqnarray}
with
\begin{equation}
\label{eq:Gmatri}
G^{i,\text{dir }}_{\gras{m}'\gras{m}\gras{n}'\gras{n}}=G^{i}_{\gras{m}'\gras{m}\gras{n}'\gras{n}},\quad
G^{i,\text{exc }}_{\gras{m}'\gras{m}\gras{n}'\gras{n}}=G^{i}_{\gras{m}'\gras{m}\gras{n}\gras{n}'},\quad
G^{i,\text{pair}}_{\gras{m}'\gras{m}\gras{n}'\gras{n}}=G^{i}_{\gras{n}'\gras{m}'\gras{n}\gras{m}}.
\end{equation}
Therefore, all the three potential energies (\ref{eq:gogny3_dir})--(\ref{eq:gogny3_pair})
can be calculated by the same routine, provided it is fed with matrix elements (\ref{eq:Gmatri}) with properly
exchanged indices.

Finally we note that the above derivations are valid for an
arbitrary proton-neutron mixing~\cite{[Sat13c],[She14]}.
Without the proton-neutron mixing, which is the option implemented in \pr{hfodd} {(v\codeversion)}
for finite-range interactions,
the basis states are
either pure neutron, $\tau=n$, or pure proton states, $\tau=p$, that is,
\begin{eqnarray}
\label{eq:denmat_nomix}
\rho_{\gras{n}s\tau,\gras{m}u\chi} & = & \rho^\tau_{\gras{n}s,\gras{m}u}\delta_{\tau,\chi}, \\
\label{eq:pairte_nomix}
\kappa_{\gras{n}'s'\tau',\gras{n}s\tau} & = & \kappa^\tau_{\gras{n}'s',\gras{n}s}\delta_{\tau',\tau}.
\end{eqnarray}
In this case, the spin-isospin components of the density matrix (\ref{eq:denmat_spiso}) only
involve the $k=0$ and $k=z$ terms, see Eq.~(\ref{eq:sigma_std}), and those of the pairing tensor (\ref{eq:pairte_spiso}) only
involve the $k=x$ and $k=y$ terms, see Eq.~(\ref{eq:sigma_tilde}), and can be expressed in terms
of the neutron and proton densities as,
\begin{eqnarray}
\label{eq:denmat_pn}
  \rho^{\mu 0}_{\gras{n},\gras{m}} & = & \phantom{-}
  \rho^{\mu n}_{\gras{n},\gras{m}}
 +\rho^{\mu p}_{\gras{n},\gras{m}}, \quad
  \rho^{\mu z}_{\gras{n},\gras{m}}  =  \phantom{-i}
  \rho^{\mu n}_{\gras{n},\gras{m}}
 -\rho^{\mu p}_{\gras{n},\gras{m}}, \\
\label{eq:pairte_pn}
  \kappa^{\mu x}_{\gras{n}',\gras{n}} & = &
 -\kappa^{\mu n}_{\gras{n}',\gras{n}}
 +\kappa^{\mu p}_{\gras{n}',\gras{n}}, \quad
  \kappa^{\mu y}_{\gras{n}',\gras{n}} =
-i\kappa^{\mu n}_{\gras{n}',\gras{n}}
-i\kappa^{\mu p}_{\gras{n}',\gras{n}}.
\end{eqnarray}
We see that both isoscalar ($T=0$) and isovector ($T=1$) coupling
constants define the potential energies in the particle-hole
channel, and only the isovector ones define those in the particle-particle
channel. Therefore, Eqs.~(\ref{eq:gogny3_dir})--(\ref{eq:gogny3_pair})
can now be written as
\begin{eqnarray}
\label{eq:gogny4_dir}
E_{\text{dir}} & = &
\thalf\sum_i\sum_{\mu \tau\tau'} V_{S\tau\tau'}^{(iD)}\sum_{\gras{m}'\gras{m}\gras{n}'\gras{n}}
\rho^{\mu \tau}_{\gras{n}',\gras{m}'}G^{i,\text{dir}}_{\gras{m}'\gras{m}\gras{n}'\gras{n}} \rho^{\mu \tau'}_{\gras{n},\gras{m}}, \\
\label{eq:gogny4_exc}
E_{\text{exc}} & = &
\thalf\sum_i\sum_{\mu \tau\tau'} V_{S\tau\tau'}^{(iE)}\sum_{\gras{m}'\gras{m}\gras{n}'\gras{n}}
\rho^{\mu \tau}_{\gras{n}',\gras{m}'}G^{i,\text{exc}}_{\gras{m}'\gras{m}\gras{n}'\gras{n}} \rho^{\mu \tau'}_{\gras{n},\gras{m}}, \\
\label{eq:gogny4_pair}
E_{\text{pair}} & = &
\phantom{\thalf}\sum_i \sum_{\mu \tau} V_{S1}^{(iP)}
\sum_{\gras{m}'\gras{m}\gras{n}'\gras{n}}
\kappa^{\mu \tau*}_{\gras{n}',\gras{m}'}G^{i,\text{pair}}_{\gras{m}'\gras{m}\gras{n}'\gras{n}} \kappa^{\mu \tau}_{\gras{n},\gras{m}},
\end{eqnarray}
where the particle-hole coupling constants, $V_{S\tau\tau'}^{(iD)}$ and $V_{S\tau\tau'}^{(iE)}$,
are defined as
\begin{eqnarray}
V_{Snn}^{(iD)}=V_{Spp}^{(iD)}=V_{S0}^{(iD)}+ V_{S1}^{(iD)},&\quad&V_{Snp}^{(iD)}=V_{Spn}^{(iD)}=V_{S0}^{(iD)}- V_{S1}^{(iD)}, \\
V_{Snn}^{(iE)}=V_{Spp}^{(iE)}=V_{S0}^{(iE)}+ V_{S1}^{(iE)},&\quad&V_{Snp}^{(iE)}=V_{Spn}^{(iE)}=V_{S0}^{(iE)}- V_{S1}^{(iE)},
\end{eqnarray}
and explicitly read
\begin{eqnarray}
V^{(iD)}_{0nn} = V^{(iD)}_{0pp} &=& \phantom{\thalf}W_{i} + \thalf B_{i} -        H_{i} - \thalf M_{i}, \\
V^{(iD)}_{0np} = V^{(iD)}_{0pn} &=& \phantom{\thalf}W_{i} + \thalf B_{i}                              , \\
V^{(iD)}_{1nn} = V^{(iD)}_{1pp} &=&                         \thalf B_{i}                - \thalf M_{i}, \\
V^{(iD)}_{1np} = V^{(iD)}_{1pn} &=&                         \thalf B_{i}                              ,
\end{eqnarray}
and
\begin{eqnarray}
V^{(iE)}_{0nn} = V^{(iE)}_{0pp} &=& \phantom{\thalf}M_{i} + \thalf H_{i} -        B_{i} - \thalf W_{i}, \\
V^{(iE)}_{0np} = V^{(iE)}_{0pn} &=& \phantom{\thalf}M_{i} + \thalf H_{i}                              , \\
V^{(iE)}_{1nn} = V^{(iE)}_{1pp} &=&                         \thalf H_{i}                - \thalf W_{i}, \\
V^{(iE)}_{1np} = V^{(iE)}_{1pn} &=&                         \thalf H_{i}                              .
\end{eqnarray}

\begin{table}[!ht]
\begin{center}
\caption{Benchmark of HFB calculations with the Gogny force in $^{120}$Sn for
the D1S parametrization; see text for numerical details.
}
\begin{tabular}{ccccc}
\hline
                                        &          \pr{hfodd}           &           Spherical Code      \\
\hline
$E_{\text{tot}}$ [MeV]                  &-1369.5013{\bf\color{blue} 26} &-1369.5013{\bf\color{red} 30}  \\
$E_{\text{kin}}$ [MeV]                  & 2304.73357{\bf\color{blue} 3} & 2304.73357{\bf\color{red} 6}  \\
$E^{\text{(dir)}}_{\text{Gogny}}$ [MeV] &-7239.2658{\bf\color{blue} 96} &-7239.2658{\bf\color{red} 06}  \\
$E^{\text{(exc)}}_{\text{Gogny}}$ [MeV] & -327.9558{\bf\color{blue} 07} & -327.9558{\bf\color{red} 12}  \\
$E_{\text{SO}}$ [MeV]                   &  -69.51{\bf\color{blue} 7989} &  -69.51{\bf\color{red} 8001}  \\
$r^{\text{(n)}}_{\text{rms}}$ [fm]      &    4.55458                    &    4.55458                    \\
$r^{\text{(p)}}_{\text{rms}}$ [fm]      &    4.44765                    &    4.44765                    \\
$E_{\text{pair}}$ [MeV]                 &  -18.748{\bf\color{blue} 583} &  -18.748{\bf\color{red} 603}  \\
%$\Delta^{\text{(n)}}$ [MeV]             &    2.139{\bf\color{blue} 750} &    2.093{\bf\color{red} 547}  \\
$\lambda^{\text{(n)}}$ [MeV]            &   -7.18972                    &   -7.18972                    \\
\hline
\end{tabular}
\label{table:gogny}
\end{center}
\end{table}

In Table~\ref{table:gogny}, we show a benchmark comparison of \pr{hfodd}
against a spherical-harmonic-oscillator Gogny code developed in Madrid
and used in studies of neutron-rich nuclei~\cite{[Sch08a],[Sch08b]}.
Calculations were performed in $^{120}$Sn with the D1S interaction in a
spherical basis with $N_{\mathrm{shell}}=10$ full shells and an oscillator
length $b = 2.0390475$ fm. The direct and exchange part of the Coulomb force
were switched off for the
test. For the kinetic energy term, we use the default value hard-coded in
\pr{hfodd} for the D1S parametrization, $\hbar^{2}/2m = 20.736676229$
MeV.fm$^{2}$; see the test input and output files provided with the source
code. We emphasize that a possible source of numerical differences is in
the density-dependent term of the interaction. Implementations of this term
depend on the Gauss quadrature integration schemes, which are different:
Gauss-Laguerre in the spherical code and Gauss-Hermite in \pr{hfodd}.
Nevertheless, the largest {\em relative} difference observed in Table
\ref{table:gogny} is for the pairing energy, where the 20 eV difference
between the two codes represent about 0.0001 \% relative error.

%%%%%%%%%%%%%%%%%%%%%%%%%%%%%%%%%%%%%%%%%%%%%%%%%%%%%%%%%%%%%%%%%%%%%%%%%%%%%%%%
%%%%%%%%%%%%%%%%%%%%%%%%%%%%%%%%%%%%%%%%%%%%%%%%%%%%%%%%%%%%%%%%%%%%%%%%%%%%%%%%

\subsection{Linear multi-constraint method at finite temperature}
\label{subsec:rpa}

Version {(v\codeversion)} of the code \pr{hfodd} features multiple linear
constraints for the multipole moments and Gaussian-neck operators both at zero
and finite temperature. At each iteration, the Lagrange multipliers
are readjusted based on the cranking approximation of the QRPA matrix.
This method was very briefly sketched in Ref.~\cite{[Dec80]} and only
quickly summarized in Refs.~\cite{[You09]} and \cite{[Sch12]}. Since
the extension of this technique to finite temperature has not been
presented so far, we take this opportunity to provide a complete
derivation of this very powerful method at $T\geq 0$.

%%%%%%%%%%%%%%%%%%%%%%%%%%%%%%%%%%%%%%%%%%%%%%%%%%%%%%%%%%%%%%%%%%%%%%%%%%%%%%%%

\subsubsection{HFB equations with multiple constraints}

We introduce the one-body
constraint operators $\hat{F}_{a}$,
\begin{equation}
\hat{F}_{a} = \sum_{ij} F_{a;ij} c_{i}^{\dagger}c_{j}
=
\sum_{ij} \langle i | \hat{F}_{a} | j\rangle c_{i}^{\dagger}c_{j}.
\end{equation}
Solving the finite-temperature HFB equations with constraints on the expectation values
$\bar{F}_{a}$ of $\hat{F}_{a}$ is achieved by minimizing the Routhian:
\begin{equation}
\mathcal{E} = E - \text{Tr}\left[ \Lambda( \mathcal{R}^{2} - \mathcal{R} ) \right]  -
\sum_{a} \lambda_{a} \left[ \text{Tr}(\hat{F}_{a}\rho) - \bar{F}_{a} \right].
\end{equation}
Here, $E = \langle\Psi| \hat{H}_{0} |\Psi\rangle/\langle\Psi|\Psi\rangle$ with
$|\Psi\rangle$ the quasi-particle vacuum, $\hat{H}_{0}$ is the original
effective two-body Hamiltonian, $\mathcal{R}$ is the generalized density
matrix and $\rho$ the one-body density matrix, and
$\gras{\lambda} \equiv (\lambda_{1}, \dots, \lambda_{N})$ is the set of
Lagrange multipliers. Using the representation of the operator $\hat{F}_{a}$ in
the doubled single-particle basis,
\begin{equation}
\mathcal{F}_{a} = \left(\begin{array}{cc}
F_{a} & 0 \\
0 & -F_{a}^{*}
\end{array}\right),
\end{equation}
one can show that $\mathcal{E}$ becomes
\begin{equation}
\mathcal{E}
= E
- \text{Tr}\left[ \Lambda( \mathcal{R}^{2} - \mathcal{R} ) \right]
- \sum_{a} \lambda_{a}
\left[ \frac{1}{2}\text{Tr}(F_{a})
+ \frac{1}{2}\text{Tr}(\mathcal{F}_{a}\mathcal{R}) - \bar{F}_{a} \right].
\end{equation}
We impose that the variations $\delta\mathcal{E}$ vanish under variations
$\delta\mathcal{R}$ of the generalized density matrix, which yields the
familiar Bogoliubov equations
\begin{equation}
\left[ \mathcal{R}, \mathcal{H} - \frac{1}{2}\sum_{a} \lambda_{a}\mathcal{F}_{a} \right]
= 0.
\label{eq:RHcommutator}
\end{equation}
with $\mathcal{H}$ the HFB matrix.

%%%%%%%%%%%%%%%%%%%%%%%%%%%%%%%%%%%%%%%%%%%%%%%%%%%%%%%%%%%%%%%%%%%%%%%%%%%%%%%%

\subsubsection{Variations around the HFB minimum}

In the next step, we denote by
$\mathcal{R}^{(0)}$ a particular solution to the HFB equations, i.e., the
generalized density matrix that diagonalizes the HFB matrix $\mathcal{H}^{(0)}$
under the set of constraints $\hat{F}_{a}$. Under small variations
$\delta \mathcal{R}$ of the density matrix, we can write formally
\begin{equation}
\mathcal{R} = \mathcal{R}^{(0)} + \delta \mathcal{R}, \ \ \ \
\mathcal{H} = \mathcal{H}^{(0)} + \delta \mathcal{H}, \ \ \ \
\lambda_{a} = \lambda_{a}^{(0)} + \delta\lambda_{a},\ \ \ \forall a.
\label{eq:rpa_perturbation}
\end{equation}
Substituting Eqs.~(\ref{eq:rpa_perturbation}) into Eq.~(\ref{eq:RHcommutator}),
and keeping only the terms up to first order in $\delta\mathcal{R}$, we obtain
\begin{equation}
\left[
\delta\mathcal{R}, \mathcal{H}^{(0)} - \frac{1}{2}\sum_{a} \lambda_{a}\mathcal{F}_{a}
\right]
+
[ \mathcal{R}^{(0)}, \delta\mathcal{H} ]
-
\frac{1}{2} \sum_{a} \delta\lambda_{a} [ \mathcal{R}^{(0)}, \mathcal{F}_{a} ] = 0.
\label{eq:masterEq}
\end{equation}
In the cranking approximation of the QRPA, we neglect the variation of the HFB matrix
under a change of the generalized density (second term). Equation
(\ref{eq:masterEq}) is the central part of the method: it relates the
variations of the Lagrange parameters to the variations of the generalized
density.

%%%%%%%%%%%%%%%%%%%%%%%%%%%%%%%%%%%%%%%%%%%%%%%%%%%%%%%%%%%%%%%%%%%%%%%%%%%%%%%%

\subsubsection{Extension to finite temperature}

At finite temperature, the Fermi-Dirac statistical occupation factors of quasi-particles
depend on the q.p.\ energies, which themselves are implicit functions
of the generalized density matrix. In other words, variations of the
generalized density matrix induce changes in the q.p.\ energies,
which, in turn, affect the occupation factors, and hence the
generalized density. This implies that the variation of $\mathcal{R}$
should now be written as
\begin{equation}
\label{eq:thermal}
\delta\mathcal{R} \rightarrow
\delta\mathcal{R}|_{f^{(0)}}
+
\left. \frac{\partial\mathcal{R}}{\partial f}\right|_{f^{(0)}}\!\!\! \delta f,
\end{equation}
where $f^{(0)}$ is the set of occupation factors corresponding to the
unperturbed generalized density $R^{(0)}$. In Eq.~(\ref{eq:thermal}),
$\delta f$ is a matrix of the same size as $\mathcal{R}$. Equation
(\ref{eq:masterEq}) then becomes
\begin{equation}
\left[ \delta \mathcal{R}|_{f^{(0)}}, \mathcal{H}^{(0)}
- \frac{1}{2}\sum_{a} \lambda_{a}\mathcal{F}_{a}  \right]
+
\left[ \left.\frac{\partial\mathcal{R}}{\partial f}\right|_{f^{(0)}}
\!\!\!\delta f, \mathcal{H}^{(0)} - \frac{1}{2}\sum_{a} \lambda_{a}\mathcal{F}_{a} \right]
-
\frac{1}{2} \sum_{a} \delta\lambda_{a} [ \mathcal{R}^{(0)}, \mathcal{F}_{a} ] = 0.
\label{eq:rpa_tosolve}
\end{equation}
Hereafter, we will refer to equation (\ref{eq:rpa_tosolve}) as the master equation.

%%%%%%%%%%%%%%%%%%%%%%%%%%%%%%%%%%%%%%%%%%%%%%%%%%%%%%%%%%%%%%%%%%%%%%%%%%%%%%%%

\subsubsection{Solutions to the master equation}

Equation~(\ref{eq:rpa_tosolve})
is best solved in the q.p.\ basis, where several of the matrices involved
take special forms. In particular, we have
\begin{equation}
\delta\mathcal{R}|_{f^{(0)}} =
\left(\begin{array}{cc}
R^{11} & R^{12} \\
R^{21} & R^{22}
\end{array}\right),\ \ \
\left.\frac{\partial\mathcal{R}}{\partial f}\right|_{f^{(0)}}
\!\!\!\delta f =
\left(\begin{array}{cc}
\delta f & 0 \\
0 & -\delta f
\end{array}\right),\ \ \
\mathcal{F}_{a} =
\left(\begin{array}{cc}
F_{a}^{11} & F_{a}^{12} \\
F_{a}^{21} & F_{a}^{22}
\end{array}\right),\
\end{equation}
and
\begin{equation}
\mathcal{H}^{(0)}  - \frac{1}{2}\sum_{a} \lambda_{a}\mathcal{F}_{a}=
\left(\begin{array}{cc}
E & 0 \\
0 & -E
\end{array}\right).
\end{equation}
The non-trivial structure of the matrix of the generalized density comes from
the fact that it is not a projector at $T>0$. After some trivial algebra, we
find the following relations
\begin{eqnarray}
R^{11}_{\mu\nu} & = & +\frac{1}{E_{\nu} - E_{\mu}}
\left[
\frac{1}{2}\sum_{a}\delta\lambda_{a}(f_{\mu} - f_{\nu} ) F_{a;\mu\nu}^{11}
\right]
- \delta f_{\mu\nu},\ \ \mu\neq\nu, \label{eq:RtoF1}
\medskip\\
R^{12}_{\mu\nu} & = & - \frac{1}{E_{\nu} + E_{\mu}}
\left[
\frac{1}{2}\sum_{a}\delta\lambda_{a} (1 + f_{\mu} + f_{\nu})
F_{a;\mu\nu}^{12}
\right].
\label{eq:RtoF2}
\end{eqnarray}
These equations contain the variations of the statistical occupation factors
$\delta f_{\nu\mu}$, which read
\begin{equation}
\delta f_{\nu\mu}
= \delta_{\mu\nu}\delta f_{\nu}
= \frac{\partial f_{\nu}}{\partial E_{\nu}} \delta E_{\nu}
= -\frac{\beta e^{\beta E_{\nu}}}{\left( 1 + e^{\beta E_{\nu}} \right)^{2}}\delta E_{\nu}
= -\beta f_{\nu}(f_{\nu} - 1)\delta E_{\nu}.
\end{equation}
Using similar arguments, we can show that variations of the q.p.\ energies
are related to the term $F_{a;\mu\nu}^{11}$ as
\begin{equation}
\delta E_{\nu} = -\frac{1}{2}\sum_{b} \delta\lambda_{b} F_{b;\nu\nu}^{11}.
\end{equation}

%%%%%%%%%%%%%%%%%%%%%%%%%%%%%%%%%%%%%%%%%%%%%%%%%%%%%%%%%%%%%%%%%%%%%%%%%%%%%%%%

\subsubsection{Readjustment of the Lagrange parameters}

We now assume that at the
iteration $n$ of the self-consistent loop, the deviation between the actual and
requested value of the constraint operators are
\begin{equation}
\delta F_{a} = \bar{F}_{a} - \langle\hat{F}_{a}\rangle^{(n)}, \ \ \forall a.
\end{equation}
Using the vector of constraint operators $\gras{\hat{F}}$ and the related
vector $\gras{\lambda}$ of values of the linear constraints, we define the vector of
perturbations $\delta\gras{\lambda}$ such that
\begin{equation}
\langle \gras{\hat{F}}(\gras{\lambda}+\delta\gras{\lambda}) \rangle =
\langle \Psi(\gras{\lambda}+\delta\gras{\lambda}) |
\gras{\hat{F}}| \Psi(\gras{\lambda}+\delta\gras{\lambda}) \rangle.
\end{equation}
This yields
\begin{equation}
\delta\gras{F}  = \frac{1}{2}\text{Tr}(\gras{F}\delta \mathcal{R}) .
\end{equation}
In the q.p.\ basis, the trace can be expressed as
\begin{equation}
\delta F_{a}
=
\mathfrak{Re}
\left( F_{a}^{11}R^{11} - F_{a}^{12}R^{12\;*}  \right),
\label{eq:trace_aux}
\end{equation}
with $\mathfrak{Re}$ denoting the real part. Inserting relations
(\ref{eq:RtoF1})-(\ref{eq:RtoF2}) into (\ref{eq:trace_aux}), we can introduce the matrix $M \equiv M_{ab}$
that has the following elements,
\begin{equation}
M_{ab} =
-\frac{1}{2} \sum_{\mu\nu}
F_{a;\mu\nu}^{11}\left(
\frac{f_{\nu} - f_{\mu} }{E_{\nu} - E_{\mu}} + \delta_{\nu\mu}\beta f_{\nu}(f_{\nu} - 1)
\right)
F_{b;\nu\mu}^{11}
-\frac{1}{2} \sum_{\mu\nu}
F_{a;\mu\nu}^{12}\frac{1 + f_{\nu} + f_{\mu}}{E_{\mu} + E_{\nu}} F_{b;\nu\mu}^{12\;*}.
\label{eq:matrixConsT}
\end{equation}
This is an $N\times N$ matrix, where $N$ is the number of constraints, and it
verifies
\begin{equation}
\delta\gras{F} = M\delta\gras{\lambda}.
\end{equation}
At every iteration $n$, the variation $\delta\gras{F}$ represents the
deviations between the requested values of the constraints and their actual
values. Matrix $M$ can be easily computed and inverted; hence the unknown
quantity $\delta\gras{\lambda}$ can be obtained and used to iterate the
Lagrange parameters. At the limit $T\rightarrow 0$, i.e.,
$f_{\mu} \rightarrow 0$, the term proportional to $F_{a}^{11}$
disappears.

%%%%%%%%%%%%%%%%%%%%%%%%%%%%%%%%%%%%%%%%%%%%%%%%%%%%%%%%%%%%%%%%%%%%%%%%%%%%%%%%

\subsubsection{Implementation in \pr{hfodd}}

In version {(v\codeversion)}, the method described above has been implemented
in the case where simplex symmetry is conserved. In the simplex-conserving basis
used in the code, the matrices thus take the following block structure
\begin{equation}
V =
\left( \begin{array}{cc}
V_{+} & 0 \\
0 & V_{-}
\end{array}\right), \ \ \
U =
\left( \begin{array}{cc}
0 & U_{+} \\
U_{-} & 0
\end{array}\right), \ \ \
F_{a} =
\left( \begin{array}{cc}
F_{+}^{(a)} & 0 \\
0 & F_{-}^{(a)}
\end{array}\right).
\end{equation}
It follows that, in the q.p.\ basis, we have the following structure for the
matrix of the constrained operator,
\begin{equation}
\begin{array}{l}
F_{a}^{11} =
\left(\begin{array}{cc}
U_{-}^{\dagger}F_{-}^{(a)}U_{-} - V_{+}^{\dagger}F_{+}^{(a)\;*}V_{+} & 0 \\
0 & U_{+}^{\dagger}F_{+}^{(a)}U_{+} - V_{-}^{\dagger}F_{-}^{(a)\;*}V_{-}
\end{array}\right)
=
\left(\begin{array}{cc}
F_{+}^{11} & 0 \\
0 & F_{-}^{11}
\end{array}\right),
\medskip\\
F_{a}^{12} =
\left( \begin{array}{cc}
0 & U_{-}^{\dagger}F_{-}^{(a)}V_{-}^{*}  - V_{+}^{\dagger}F_{+}^{(a)\;*}U_{+}^{*}\\
U_{+}^{\dagger}F_{+}^{(a)}V_{+}^{*} - V_{-}^{\dagger}F_{-}^{(a)\;*}U_{-}^{*} & 0
\end{array}\right)
=
\left(\begin{array}{cc}
0 & F_{+}^{12}  \\
-F_{+}^{12\;T}
\end{array}\right).
\end{array}
\end{equation}
Forming the auxiliary matrices
\begin{equation}
\begin{array}{l}
G^{11}_{\mu\nu;\pm} = \displaystyle
-\left[
\frac{f_{\mu} - f_{\nu} }{E_{\mu} - E_{\nu}} + \delta_{\mu\nu}\beta f_{\mu}(f_{\mu} - 1)
\right]
F_{\mu\nu;\pm}^{11}
\medskip\\
G^{12}_{\mu\nu;\pm} = \displaystyle
\frac{1 + f_{\mu} + f_{\nu}}{E_{\mu} + E_{\nu}} F_{\mu\nu;\pm}^{12},
\end{array}
\end{equation}
we can show that the matrix of the constraints becomes
\begin{equation}
M_{ab} =
\frac{1}{2}\text{Tr}( F_{+}^{11}G_{+}^{11})
+
\frac{1}{2}\text{Tr}( F_{-}^{11}G_{-}^{11})
+
\text{Tr}( F_{+}^{12\;T}G_{+}^{12\;*} ).
\end{equation}

\begin{figure}[!ht]
\center
\includegraphics[width=0.8\linewidth]{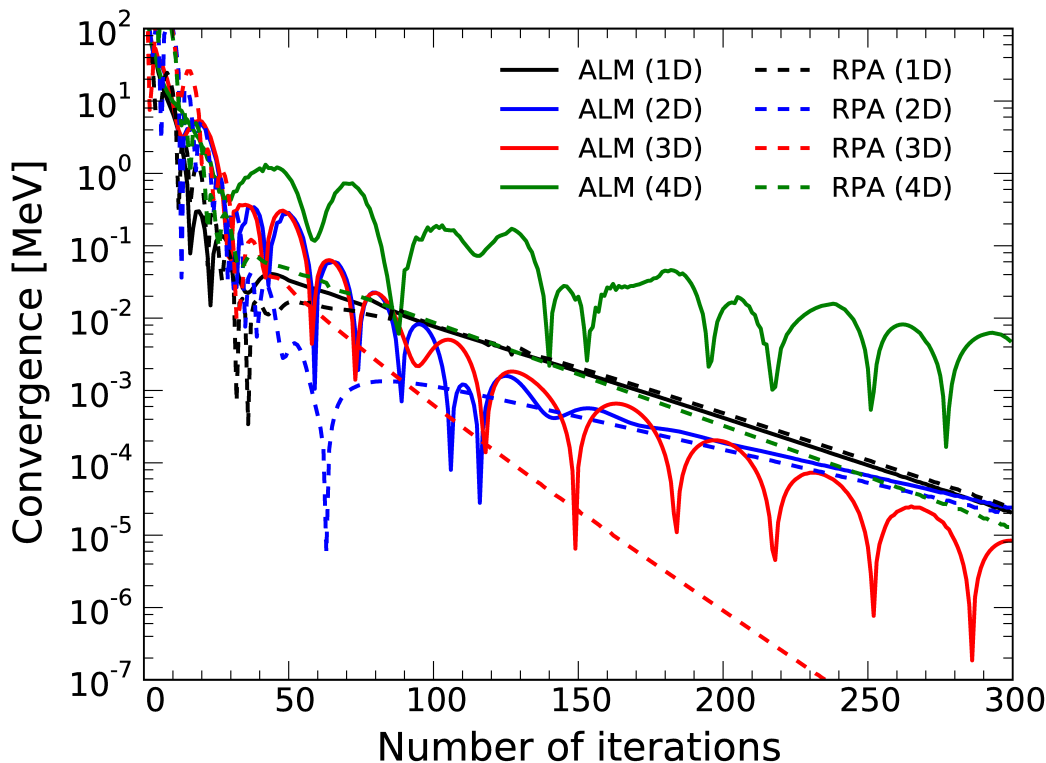}
\caption{(color online) Convergence of the HFB calculation as function of the
number of iterations in $^{240}$Pu for four different sets of constraints, see
text for details.
}
\label{fig:RPA}
\end{figure}

%%%%%%%%%%%%%%%%%%%%%%%%%%%%%%%%%%%%%%%%%%%%%%%%%%%%%%%%%%%%%%%%%%%%%%%%%%%%%%%%

\subsubsection{Comparison with the Augmented Lagrangian Method}

We illustrate in
Fig.~\ref{fig:RPA} the performance of the RPA-based method for multiconstraints
by comparing it with the ALM, see Section~2.2.2 of \cite{[Sch12]}.
Calculations were
all performed in $^{240}$Pu for the SkM* functional using a deformed stretched
HO basis of 20 shells and deformation $\beta_{2} = 0.3$.
Four different configurations were considered: (i) a 1D case with only a
constraint on $\langle \hat{Q}_{20} \rangle = 60$ b, (ii) a 2D case with a
constraint on both $\langle \hat{Q}_{20} \rangle = 60$ b and
$\langle \hat{Q}_{22} \rangle = 0$ b, (iii) a 3D case with the constraints
$\langle \hat{Q}_{20} \rangle = 60$ b, $\langle \hat{Q}_{22} \rangle = 0$ b,
and $\langle \hat{Q}_{40} \rangle = 5$ b$^{2}$, and finally (iv) a 4D case with
the constraints $\langle \hat{Q}_{10} \rangle = 0$ fm,
$\langle \hat{Q}_{20} \rangle = 60$ b, $\langle \hat{Q}_{22} \rangle = 0$ b,
and $\langle \hat{Q}_{30} \rangle = 10$ b$^{3/2}$. The main advantage of the
RPA-based method is that it does take into account correlations between the
constraints: while its performance is comparable to the ALM in the
1D and even 2D case, it becomes more and more efficient when the number of
constraints increases. Note that the constraints need not be limited to
multipole moments: any operator can be included in the list.

%%%%%%%%%%%%%%%%%%%%%%%%%%%%%%%%%%%%%%%%%%%%%%%%%%%%%%%%%%%%%%%%%%%%%%%%%%%%%%%%
%%%%%%%%%%%%%%%%%%%%%%%%%%%%%%%%%%%%%%%%%%%%%%%%%%%%%%%%%%%%%%%%%%%%%%%%%%%%%%%%

\subsection{Fission toolkit}
\label{subsec:fission}

Version {(v\codeversion)} of the code \pr{hfodd} contains a collection of
routines employed primarily in fission studies. Among others, they provide
\begin{itemize}
\item the capability to impose a constraint on the number of particles in the
neck. The Gaussian neck operator is defined as
\begin{equation}
\label{eq:neck}
\hat{Q}_{N} = e^{-(z - z_{N})^{2}/a_{N}^{2}},
\end{equation}
where $z_{N}$ gives the position of the neck (along the $z-$axis of the
intrinsic reference frame) between two nascent fragments. It is defined as the
point near the origin of the intrinsic reference frame where the density is the
lowest. The range $a_{N}$ gives the spatial extent of the neck: in the code, it
is fixed at 1 fm.
\item the charge, mass, total kinetic, nuclear and pairing energy of each
fragment. Expectation values of the multipole moments are also computed,
both in the reference frame of the fissioning nucleus and in those
of each of the individual fragments.
\item the interaction energy between the fragments. Both the nuclear (Skyrme)
interaction energy and the direct Coulomb energy are computed. The direct
Coulomb energy is a measure of the total kinetic energy of fission
fragments.
\item the capability to perform a unitary transformation of q.p.\ operators
so as to maximize the spatial localization of each q.p.\ within a pre-fragment.
\end{itemize}
All these routines are collected in module \tf{hfodd\_fission\_7.f90} and
require (at least) the option \tv{IFRAGM}=1 set in the input file. Below we give
a brief description of each of these features.

%%%%%%%%%%%%%%%%%%%%%%%%%%%%%%%%%%%%%%%%%%%%%%%%%%%%%%%%%%%%%%%%%%%%%%%%%%%%%%%%

\subsubsection{Constraint on the size of the neck}
\label{subsubsec:neck}

The Gaussian neck operator (\ref{eq:neck}) is purely spatial and does not depend on spin or
isospin degrees of freedom. Its expectation value is computed in coordinate
space on the Gauss-Hermite quadrature grid used in \pr{hfodd},
\begin{equation}
\label{eq:neck2}
\langle \hat{Q}_{N} \rangle = \int d^{3}\gras{r} \hat{Q}_{N}\rho(\gras{r}),
\end{equation}
where $\rho(\gras{r})$ is the isoscalar density. The integral is computed by
using the quadrature relation
\begin{equation}
\frac{1}{b_{\mu}} \int d\xi_{\mu} f(\xi_{\mu})\; e^{-2\xi_{\mu}^{2}}
=
\sum_{k_{\mu}} \frac{w_{k_{\mu}}}{b_{\mu}\sqrt{2}}
f \left( \frac{\eta_{k_{\mu}}}{\sqrt{2}}\right)
\end{equation}
valid for any function $f(x_{\mu})$, with $\xi_{\mu} = b_{\mu}x_{\mu}$ the
dimensionless coordinate, $b_{\mu}$ the oscillator length in the direction
$\mu$, $\omega_{k_{\mu}}$ and $\eta_{k_{\mu}}$ the weights and nodes of the
Gauss-Hermite quadrature. In \pr{hfodd}, the isoscalar density
$\rho(\gras{r})$ is represented by the array \tv{DENSIT}, which is defined
at the points $\eta_{k_{\mu}}/\sqrt{2}$. The function $f$ to integrate for
the Gaussian neck operator is, therefore,
\begin{equation}
f(\gras{\xi}) = \rho_{0}(\gras{\xi}) e^{2\gras{\xi}^{2}}
e^{ - (\xi_{z} - \xi_{N})^{2}/\alpha_{N}^{2}},
\end{equation}
with $\alpha_{N} = b_{z}a_{N}$ and $\xi_{N} = b_{z}z_{N}$.

In order to add a constraint on the expectation value of the neck operator,
its matrix elements in the HO basis are also needed. They take the simple
form
\begin{equation}
\langle n_{x} n_{y} n_{z} | \hat{Q}_{N} | m_{x} m_{y} m_{z} \rangle
=
\delta_{n_{x}m_{x}}\delta_{n_{y}m_{y}}I_{n_{z}m_{z}}
\end{equation}
with
\begin{equation}
I_{n_{z}m_{z}} = \sum_{k_{z}=0}^{n_{z}+m_{z}} C^{k_{z}}_{n_{z}m_{z}}(00 )
\int d\xi_{z}\; H_{k_{z}}^{(0)}(\xi_{z}) e^{-\xi_{z}^{2}}
e^{-\left( \frac{\xi_{z} - \xi_{N}}{\alpha_{N}} \right)^{2}}.
\end{equation}
The normalized Hermite polynomials $H_{k_{\mu}}^{(0)}(\xi_{\mu})$ and expansion
coefficients $C^{k}_{nm}(00 )$ are defined in Section~4.3 of Ref.~\cite{[Dob97c]}.
One can show that the integral in $I_{n_{z}m_{z}}$ can be computed exactly. Its
expression is
\begin{equation}
\int d\xi_{z}\;
H_{k_{z}}^{(0)}(\xi_{z}) e^{-\xi_{z}^{2}}
e^{-\left( \frac{\xi_{z} - \xi_{N}}{\alpha_{N}} \right)^{2}}
=
\sqrt{\frac{\pi b_{z} a_{N}^{2}}{G_{z} }}
e^{-\frac{1}{2}\frac{\alpha_{N}^{2}}{G_{z} }}
\frac{1}{G_{z} ^{k/2}}
\psi_{k_{z}} \left( \frac{\alpha_{N}}{\sqrt{G_{z} }} \right),
\end{equation}
with $G_{z} =  1 + b_{z}^{2} a^{2}_{N}$, and $\psi_{k_{z}}$ is the HO
wave-function. From this expression, the matrix elements of $\hat{Q}_{N}$ in
the good-simplex basis are easily found using the relations (84) of
\cite{[Dob97c]}.

%%%%%%%%%%%%%%%%%%%%%%%%%%%%%%%%%%%%%%%%%%%%%%%%%%%%%%%%%%%%%%%%%%%%%%%%%%%%%%%%

\subsubsection{Fission fragment properties}
\label{subsubsec:fragments}

The identification of fission fragments is based on the position of the neck
and the spatial occupation of quasi-particles. We start from the set of q.p.\ in
the compound nucleus defined by the Bogoliubov matrices $U$ and $V$. We may
write the coordinate space representation of the full one-body density matrix
(in coordinate$\otimes$spin space) as
\begin{equation}
\rho(\gras{r}\sigma,\gras{r}'\sigma')
=
\sum_{\mu} \rho_{\mu}(\gras{r}\sigma,\gras{r}'\sigma'),
\label{density}
\end{equation}
with the q.p.\ density operator $\rho_{\mu}(\gras{r}\sigma,\gras{r}'\sigma')$
of q.p.\ $\mu$ defined (at temperature $T$) by
\begin{equation}
\rho_{\mu}(\gras{r}\sigma,\gras{r}'\sigma') = \sum_{ij}
\left[ (1-f_{\mu})V_{i\mu}^{*}V_{j\mu} + f_{\mu} U_{i\mu}U_{j\mu}^{*} \right]
\psi_{i}(\gras{r}\sigma)\psi_{j}^{*}(\gras{r}'\sigma'),
\label{eq:occupation}
\end{equation}
with $\psi_{i}(\gras{r}\sigma)$ the basis functions and $f_{\mu}$ the
Fermi-Dirac statistical occupation at temperature $T$. The
spatial occupation of the q.p.\ $\mu$ is then defined as
\begin{equation}
N_{\mu} = \sum_{\sigma} \int d^{3}\gras{r}\;
\rho_{\mu}(\gras{r}\sigma,\gras{r}\sigma).
\end{equation}
Since the basis $\{ \psi_{i} \}$ is orthonormal, this reduces to the
expression
\begin{equation}
N_{\mu} = \sum_{ij} \left[ (1-f_{\mu})V_{i\mu}^{*}V_{j\mu}
+ f_{\mu} U_{i\mu}U_{j\mu}^{*} \right],
\label{eq:occupation_total}
\end{equation}
with the total number of particles defined as $N = \sum_{\mu} N_{\mu}$.

We can then define the occupation of the q.p.\ $\mu$ in the fragment (1) as
\begin{equation}
N_{1,\mu} = \sum_{ij}
\left[ (1-f_{\mu})V_{i\mu}^{*}V_{j\mu}
+ f_{\mu} U_{i\mu}U_{j\mu}^{*} \right] d_{ij}(z),
\label{eq:occupation_fragment}
\end{equation}
where
\begin{equation}
d_{ij}(z) = \sum_{\sigma}
\int_{-\infty}^{+\infty} dx\int_{-\infty}^{+\infty} dy \int_{-\infty}^{z_{N}}dz\;
\psi_{i}(\gras{r}\sigma)\psi_{j}^{*}(\gras{r}\sigma).
\end{equation}
The occupation of the q.p.\ in the fragment (2) is simply
$N_{2,\mu} = N_{\mu} - N_{1,\mu}$. We then assign the q.p.\ $\mu$ to fragment
(1) if $N_{1,\mu} \geq 0.5 N_{\mu}$, and to fragment (2) if
$N_{1,\mu} <0.5 N_{\mu}$. This gives us two sets of q.p.\ with which we can
define objects analogs to the density matrix and pairing tensor of each of
the fragments.

It was observed that the coordinate representation $\rho_{1}(\gras{r})$ and
$\rho_{2}(\gras{r})$ of the densities in a given fragment has a tail that
extends significantly into the other fragment \cite{[You11]}. This
delocalization of the density can be traced back to the individual
quasi-particles, and can be captured by the following indicator
\begin{equation}
\ell_{\mu} = \frac{| N_{1,\mu} - N_{2,\mu} |}{N_{\mu}^{2}},
\label{eq:localization}
\end{equation}
with $N_{\mu}$ defined by Eq.~(\ref{eq:occupation_total}) and
$N_{1,\mu}, N_{2,\mu}$ by Eq.~(\ref{eq:occupation_fragment}). If
$\ell_{\mu}=0$, the q.p.\ $\mu$ is fully delocalized, if $\ell_{\mu}=1$ it is
fully localized either in the left or in the right fragment. The tails in the
densities are produced by the contributions from the delocalized q.p.\ states
with relatively large occupation and $0 \leq \ell_{\mu} \ll 1$.

%%%%%%%%%%%%%%%%%%%%%%%%%%%%%%%%%%%%%%%%%%%%%%%%%%%%%%%%%%%%%%%%%%%%%%%%%%%%%%%%

\subsubsection{Interaction energies}
\label{subsubsec:interaction_fission}

As mentioned above, the identification of a set of q.p.\ for each fragment
fully defines an analog of the density matrix and pairing tensor for each
fragment. In coordinate$\otimes$spin space, these objects take the form
\begin{eqnarray}
\rho_{\text{f}}(\gras{r}\sigma,\gras{r}'\sigma') =
\sum_{\mu\in(\text{f})}  \sum_{ij}
\left[ (1-f_{\mu})V_{i\mu}^{*}V_{j\mu} + f_{\mu} U_{i\mu}U_{j\mu}^{*} \right]\;
\psi_{i}(\gras{r}\sigma)\psi_{j}^{*}(\gras{r}'\sigma'), \medskip\\
\kappa_{\text{f}}(\gras{r}\sigma,\gras{r}'\sigma') =
\sum_{\mu\in(\text{f})} \sum_{ij}
\left[ (1-f_{\mu})V_{i\mu}^{*}U_{j\mu} + f_{\mu} U_{i\mu}V_{j\mu}^{*} \right]\;
\psi_{i}(\gras{r}\sigma)\psi_{j}^{*}(\gras{r}'\sigma'),
\label{eq:density_fragment}
\end{eqnarray}
where $\text{f}=1,2$ refers to the fragment. Note that these quantities are
not true density operators. In particular, they do not necessarily obey the
usual relations $\rho^{2} + \kappa\kappa^{*}=0$. We refer to them as
pseudodensities. The diagonal components of the pseudodensities (in
coordinate$\otimes$spin space) for each fragment, $\rho_{1}(\gras{r})$,
$\rho_{2}(\gras{r})$, $\kappa_{1}(\gras{r})$ and $\kappa_{2}(\gras{r})$, are
obtained as usual \cite{[Ben03]}.

From these definitions, fragment energies and interaction energies can be
computed in a straightforward manner \cite{[Sch14a]}. The total direct
Coulomb interaction energy is computed as
\begin{equation}
E_{\text{Cou}}^{(\text{dir})} = 2e^{2}
\int d^{3}\gras{r}\int d^{3}\gras{r}'\;
\frac{\rho_{1}(\gras{r})\rho_{2}(\gras{r}')}{|\gras{r} - \gras{r}'|},
\label{eq:direct_energy}
\end{equation}
where $\rho_{f}$ is the proton density in fragment f.
In our calculations, this energy was computed using the Green function
method as in \cite{[Dob97c]}. The total exchange Coulomb energy is defined
as
\begin{equation}
E_{\text{Cou}}^{(\text{dir})} = 2e^{2}
\int d^{3}\gras{r}\int d^{3}\gras{r}'\;
\frac{\rho_{1}(\gras{r},\gras{r}')\rho_{2}(\gras{r}',\gras{r})}{|\gras{r} - \gras{r}'|},
\label{eq:exchange_energy}
\end{equation}
In \pr{hfodd} {(v\codeversion)}, the exchange Coulomb energy of each
fragment is computed at the Slater approximation, while the Coulomb
exchange interaction energy between fragments is computed ``exactly'' by
using an expansion of the Coulomb potential as a sum of Gaussians
\cite{[Dob09d]}. The nuclear Skyrme interaction energy is given by
\begin{align}
E_{\text{int}}^{\text{Skyrme}}
& = E_{\text{int}}^{1\rightarrow 2} + E_{\text{int}}^{2\rightarrow 1} \\
& =
\sum_{t=0,1} \int d^{3}\gras{r} \left\{
C_{t}^{\rho}       \rho_{t}^{(1)}\rho_{t}^{(2)} +
C_{t}^{\Delta\rho} \rho_{t}^{(1)}\Delta\rho_{t}^{(2)} \right.
\nonumber\\
& \qquad +
\left.
C_{t}^{\tau}       \rho_{t}^{(1)}\tau_{t}^{(2)} +
C_{t}^{J}          \sum_{\mu\nu} J_{\mu\nu,t}^{(1)} J_{\mu\nu,t}^{(2)} +
C_{t}^{\nabla J}   \rho_{t}^{(1)}\gras{\nabla}\cdot\gras{J}_{t}^{(2)} \right\}
\nonumber\\
& +
\sum_{t=0,1} \int d^{3}\gras{r} \left\{
C_{t}^{\rho}       \rho_{t}^{(2)}\rho_{t}^{(1)} +
C_{t}^{\Delta\rho} \rho_{t}^{(2)}\Delta\rho_{t}^{(1)} \right.
\nonumber\\
& \qquad +
\left.
C_{t}^{\tau}       \rho_{t}^{(2)}\tau_{t}^{(1)} +
C_{t}^{J}          \sum_{\mu\nu} J_{\mu\nu,t}^{(2)} J_{\mu\nu,t}^{(1)} +
C_{t}^{\nabla J}   \rho_{t}^{(2)}\gras{\nabla}\cdot\gras{J}_{t}^{(1)} \right\},
\end{align}
with the traditional densities $\rho$, $\tau$, $\mathbb{J}$ and $\gras{J}$
built from either set (1) or set (2) of q.p.\ Note that, while some of these
terms are symmetric under an exchange $(1)\leftrightarrow (2)$ (terms
proportional to $C_{t}^{\rho}$, $C_{t}^{\Delta\rho}$, $C_{t}^{J}$), others
are not (those proportional to $C_{t}^{\tau}$ and $C_{t}^{\nabla J}$).

%%%%%%%%%%%%%%%%%%%%%%%%%%%%%%%%%%%%%%%%%%%%%%%%%%%%%%%%%%%%%%%%%%%%%%%%%%%%%%%%

\subsubsection{Rotation of Quasiparticles in Fock space}
\label{subsubsec:localization}

In version {(v\codeversion)}, the code \pr{hfodd} can perform a unitary
transformation of the quasiparticles of the HFB solution. In the context of fission,
this rotation (in the Fock space) was introduced by Younes and Gogny in
Ref.~\cite{[You11]} as a method to ``localize'' the fission fragments. It is the
transposition in nuclei of the localization methods introduced long ago in
electronic structure theory to describe molecular bonding \cite{[You11],[Sch14a]}.
We note here that this localization is a different concept than the one
discussed recently in terms of the cluster localization~\cite{[Zha16]}.
Our unitary transformation is defined by its
action on pairs $(\mu,\nu)$ of q.p. states,
\begin{equation}
\label{eq:unitary}
\begin{array}{rclcrcl}
 U'_{\mu} &=& \phantom{-}\cos\theta_{\mu\nu} U_{\mu} + \sin\theta_{\mu\nu} U_{\nu} &,&  V'_{\mu} &=& \phantom{-}\cos\theta_{\mu\nu} V_{\mu} + \sin\theta_{\mu\nu} V_{\nu} , \\
 U'_{\nu} &=&           -\sin\theta_{\mu\nu} U_{\mu} + \cos\theta_{\mu\nu} U_{\nu} &,&  V'_{\nu} &=&           -\sin\theta_{\mu\nu} V_{\mu} + \cos\theta_{\mu\nu} V_{\nu} .
\end{array}
\end{equation}
Angles of the rotation $\theta_{\mu\nu}$ can be different for
every pair of q.p.\ states. We can, of course, apply a sequence of
unitary transformations (\ref{eq:unitary}) for arbitrarily selected
pairs one after another, and obtain a total unitary transformation
that depends on all angles $\theta_{\mu\nu}$.  In version
{(v\codeversion)}, each unitary transformations (\ref{eq:unitary}) is
chosen so as to maximize the localization (\ref{eq:localization}) in
a given pair of q.p.\ states.

Under transformation (\ref{eq:unitary}), the full density matrix of the
compound nucleus $\rho$ remains unchanged. However, occupation
(\ref{eq:occupation_total}) of a given q.p.\ $\mu$ becomes
\begin{equation}
N'_{\mu}
=
\sum_{ij}
\left[
\cos^{2}\theta_{\mu\nu} V^{*}_{i\mu}V_{j\mu}
+
\sin^{2}\theta_{\mu\nu} V^{*}_{i\nu}V_{j\nu}
+ \sin\theta_{\mu\nu}\cos\theta_{\mu\nu} ( V^{*}_{i\mu} V_{j\nu} + V^{*}_{i\nu}V_{j\mu} ) \right].
\end{equation}
Denoting
\begin{equation}
\omega_{\mu\nu}(z) = \sum_{ij} ( V^{*}_{i\mu} V_{j\nu}
+ V^{*}_{i\nu}V_{j\mu} )d_{ij}(z),
\end{equation}
we find that the occupations of q.p.\ $\mu$ in each of the fragment then reads
\begin{equation}
\begin{array}{l}
\displaystyle N'_{1,\mu} = \cos^{2}\theta_{\mu\nu} N_{1,\mu} + \sin^{2}\theta_{\mu\nu} N_{1,\nu}
+ \sin\theta_{\mu\nu}\cos\theta_{\mu\nu}\; [\omega_{\mu\nu}(-\infty) - \omega_{\mu\nu}(z_{N})],
\medskip\\
\displaystyle N'_{2,\mu} = \cos^{2}\theta_{\mu\nu} N_{2,\mu} + \sin^{2}\theta_{\mu\nu} N_{2,\nu}
+ \sin\theta_{\mu\nu}\cos\theta_{\mu\nu}\; \omega_{\mu\nu}(z_{N}),
\end{array}
\end{equation}
while for q.p.\ $\nu$ they are
\begin{equation}
\begin{array}{l}
\displaystyle N'_{1,\nu} = \cos^{2}\theta_{\mu\nu} N_{1,\nu} + \sin^{2}\theta_{\mu\nu} N_{1,\mu}
 - \sin\theta_{\mu\nu}\cos\theta_{\mu\nu}\; [\omega_{\nu\mu}(-\infty)-\omega_{\nu\mu}(z_{N})],
\medskip\\
\displaystyle N'_{2,\nu} = \cos^{2}\theta_{\mu\nu} N_{2,\nu} + \sin^{2}\theta_{\mu\nu} N_{2,\mu}
- \sin\theta_{\mu\nu}\cos\theta_{\mu\nu}\; \omega_{\nu\mu}(z_{N}).
\end{array}
\end{equation}
In the current implementation of the localization method in the code
\pr{hfodd}, the code first searches for all possible pairs $(\mu, \nu)$ such
that $|\Delta E| = |E_{\mu} - E_{\nu}| \leq \Delta$, the localization of
both q.p.\ is $\ell_{\mu}, \ell_{\nu} \leq \ell_{\mathrm{max}}$ and their
occupation is $N_{\mu}, N_{\nu} \geq N_{\mathrm{min}}$. The quantities
$\Delta$, $\ell_{\mathrm{max}}$, and $N_{\mathrm{min}}$ are user inputs. The
localization of q.p.\ is only performed on those q.p.\ states that match all these
conditions. The code also allows to perform several successive q.p.\ rotations.
Note that, after the first rotation, the HFB matrix is not diagonal anymore:
the definition of $\Delta E$ is, therefore, not based on q.p.\ energies but
on the diagonal elements $E_{\mu\mu}$ and $E_{\nu\nu}$ of the rotated HFB matrix.

%%%%%%%%%%%%%%%%%%%%%%%%%%%%%%%%%%%%%%%%%%%%%%%%%%%%%%%%%%%%%%%%%%%%%%%%%%%%%%%%
%%%%%%%%%%%%%%%%%%%%%%%%%%%%%%%%%%%%%%%%%%%%%%%%%%%%%%%%%%%%%%%%%%%%%%%%%%%%%%%%

\subsection{Interface with \pr{hfbtho}}
\label{subsec:hfbtho}

Since version (v2.49t), the code \pr{hfodd} includes the \pr{hfbtho} DFT solver as a module. In
version  (v2.49t), the \pr{hfbtho} kernel was the one published in \cite{[Sto05]}. In
the current version {(v\codeversion)}, we have upgraded the \pr{hfbtho} kernel to
the version 2.00d of \cite{[Sto12]}. This offers additional capabilities
such as the constraints on (axial) multipole moments, the breaking of
parity to describe asymmetric nuclear shapes, the finite-temperature HFB
theory, etc. In addition to being able to initialize a \pr{hfodd} calculation
with the \pr{hfbtho} solution, version {(v\codeversion)} also offers the reverse
option: the code can read a previous \pr{hfodd} solution, use it to initialize
and run \pr{hfbtho} before to convert back again to \pr{hfodd}. Such an option can
be useful in axial multi-constrained calculations in heavy nuclei (where
\pr{hfodd} calculations can be time- and resource-consuming) which serve as
initial conditions for triaxial calculations. To activate this mode, the
user must set \tv{IF\_THO} to 2 in the input file, and must have the \pr{hfodd}
fields recorded on disk.

%%%%%%%%%%%%%%%%%%%%%%%%%%%%%%%%%%%%%%%%%%%%%%%%%%%%%%%%%%%%%%%%%%%%%%%%%%%%%%%%
%%%%%%%%%%%%%%%%%%%%%%%%%%%%%%%%%%%%%%%%%%%%%%%%%%%%%%%%%%%%%%%%%%%%%%%%%%%%%%%%

\subsection{Parallel capabilities}
\label{subsec:parallel}

Compared to version (v2.49t), several routines have been substantially
accelerated through the use of shared memory parallelism with OpenMP. The
routines affected are \ts{NILASP}, \ts{NILAPN}, \ts{DENMAC}, \ts{SPAVER},
\ts{PNAVER}, \ts{GAUOPT}, \ts{DENSHF}. In addition, several linear algebra
operations such as matrix multiplication or matrix-vector multiplication are
now handled by BLAS routines instead of being hard-coded in the program. Since
several standard implementations of BLAS and LAPACK are multi-threaded (for
instance the Intel MKL or AMD ACML libraries), additional speed-ups can be
thus achieved on multicore architectures.

\begin{figure}[!ht]
\center
\includegraphics[width=0.5\linewidth]{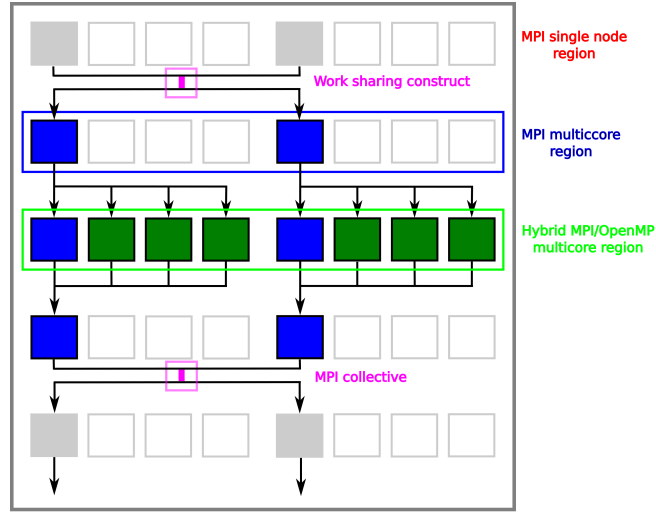}
\caption{(color online) Schematic representation of the hybrid MPI/OpenMP
programming model of \pr{hfodd}. The figure corresponds to the case where a single
HFB calculation is spread across 2 MPI processes, each invoking 4 threads.
Light-gray colored boxes show MPI processes working independently of
each other; dark blue colored boxes represent MPI processes sharing work;
dark green colored boxes show OpenMP threads sharing work of a given MPI
process.
}
\label{fig:mpi}
\end{figure}

In version {(v\codeversion)}, the two time-consuming routines \ts{DENSHF} and \ts{GAUOPT}
have also been explicitly parallelized to take advantage of massively parallel
architectures (very large node count). As a result, the user can choose to
spread one HFB calculation across several MPI processes (typically, between 4
and 8), each MPI process being able to invoke several threads. Note that this
option is only available when running in MPI mode and requires specifying {\it
at compile time} the number of MPI processes handling a given \pr{hfodd}
calculation; see section \ref{subsec:parallel_mode} for details. In such a
case, the program handles three different levels of parallelism for:

\vspace{5mm}
\noindent{\bf Level 1} Number of HFB calculations $N$;

\noindent{\bf Level 2} Number of MPI processes involved in each HFB calculation $N_{\mathrm{p}}$;

\noindent{\bf Level 3} Number of OpenMP threads called in each MPI process $N_{\mathrm{t}}$.

\vspace{5mm}\noindent
The total number of cores required in such a calculation is thus
$N_{\mathrm{cores}} = N \times N_{\mathrm{p}} \times N_{\mathrm{t}}$.
Internally, the program splits {\tt MPI\_COMM\_WORLD} into three sets of
communicators: one corresponding to each group of processes handling a
HFB calculation; one including all the master processors across all groups;
the last one including all the slaves across all groups. MPI collective
operations can then be easily defined within a given group. Note that the
ratio (time of calculation)/(time of communication) remains very large
so that load imbalance issues are negligible at this stage.

Figure~\ref{fig:mpi} schematically illustrates the workflow of a single
HFB calculation in this case. Most of the code execution is serial: all MPI
processes involved perform exactly the same task. When the code enters a
parallelized segment of the code, work is shared explicitly across available
MPI processes. In both routines, the parallel segment involves nested loops:
each MPI process handles different values of the index of the outermost loop.
Some time within the parallel region, each MPI task may enter a
multi-threaded region (OpenMP). At the end of the MPI parallel region, results
from each MPI process are combined and broadcast back to each MPI process.
This step involves blocking collective MPI operations.

In \ts{DENSHF}, the outermost loop involves the node index $k_z$ of the
Gauss-Hermite mesh in the $z$ direction. Each MPI process computes different
$k_z$ values in parallel, hence a sub-array of the full density array.
For example, with 2 MPI tasks, process 0 would compute all densities at all
even values of $k_z$, while process 1 would compute all odd values. At the
end, a call to \ts{mpi\_allgatherv} gathers these different subarrays into
the full density array. In \ts{GAUOPT}, the outermost loop involves a
summation over the quantum number $n_x$. Each MPI process thus performs a
partial summation over a subset of $n_x$ values, and the collective MPI
operation at the end of the routine is thus a call to \ts{mpi\_allreduce},
which sums all contributions from the different MPI processes.

\begin{figure}[!ht]
\center
\includegraphics[width=0.6\linewidth]{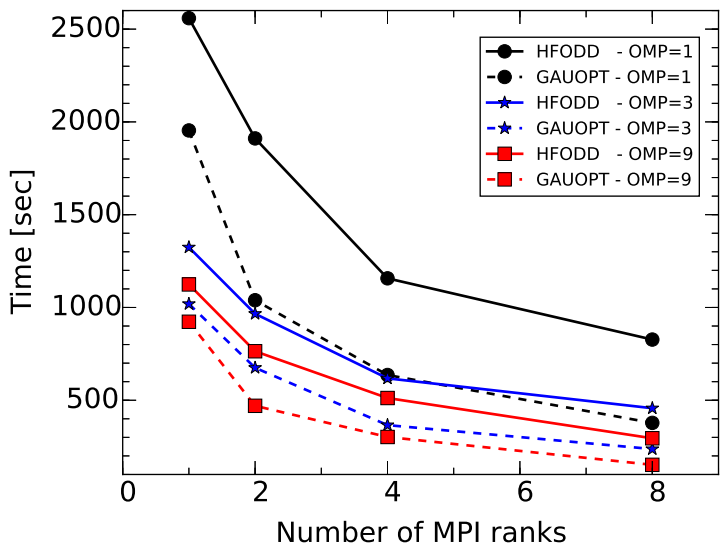}
\caption{(color online) Scaling of execution time for 10 iterations of the
Gogny force as a function of the number of MPI processes per HFB calculation
and the number of OpenMP threads. Curves with plain lines correspond to the
full HFODD execution time, curves with dashed lines for the time spent in
the routine \tf{GAUOPT}; see text for details.
}
\label{fig:scaling}
\end{figure}

Figure~\ref{fig:scaling} shows the speed-up achieved by parallelizing
the kernel of a HFB calculation in the case of the Gogny force. Calculations
were performed in a full spherical basis of 14 major oscillator shells and
the number of Gauss-Hermite points was set to 30 in each Cartesian direction.
The full two-body center of mass correction was included. Results can be
reproduced by using the file \tv{sn120\_gogny.dat} included with the code
and increasing the size of the basis accordingly. With the current
implementation, execution time can be reduced by a factor 8.5 between a pure
serial mode and a hybrid MPI-OpenMP mode with 8 MPI ranks and 9 OpenMP threads;
a speed-up of 3 can be achieved between serial and non-threaded MPI mode.

%%%%%%%%%%%%%%%%%%%%%%%%%%%%%%%%%%%%%%%%%%%%%%%%%%%%%%%%%%%%%%%%%%%%%%%%%%%%%%%%
%%%%%%%%%%%%%%%%%%%%%%%%%%%%%%%%%%%%%%%%%%%%%%%%%%%%%%%%%%%%%%%%%%%%%%%%%%%%%%%%

\subsection{Lipkin translational energy}
\label{subsec:Lipkin_translational}

The Lipkin method was proposed in the early nineteen sixties to approximately
restore broken symmetries at the mean-field level by adding terms in the
functional that cancel out the effects of quantum fluctuations on the total energy \cite{[Lip60]}. The method
requires defining the Lipkin operator to add to the two-body effective Hamiltonian
\cite{[Dob09g],[Sch12]}. In the case of the translational-symmetry restoration, the
Lipkin operator reads
\begin{equation}
\hat{K} = \sum_{i=x,y,z}k_{i}\hat{P}_{i}^{2}
\end{equation}
where $\hat{P}_{i}$ is the total linear momentum in the direction
$i$, and $k_{i}$ is a parameter weighting the dependence of
$\hat{P}_{i}^{2}$ in the energy. In version (2.49t), all parameters
$k_i$ were forced to be equal, and the correction could not be
computed in the presence of pairing correlations (BCS or HFB). In
version {(v\codeversion)}, these restrictions have been removed. Note
that different values of $k_i$ allow us to treat differences of the
linear-momentum fluctuations along the three principal axes of a
deformed nucleus.

The expectation value of $\hat{P}_{i}^{2}$ in the HFB ground-state reads
\begin{equation}
\langle \hat{P}_{i}^{2}\rangle
=
\left[
{\mathrm Tr}\left(P_{i}\rho\right)\right]^{2}
+
{\mathrm Tr}\left[P_{i}^{2}\rho\right]
-
{\mathrm Tr}\left[P_{i}\rho P_{i}\rho\right]
-
{\mathrm Tr}\left[P_{i}\kappa\left(P_{i}\kappa\right)^{*}\right]
\label{eq:2}
\end{equation}
where $P_i$ is the matrix of $\hat{P}_{i}$ in the good-simplex basis
of \pr{hfodd}, $\rho$ is the one-body density matrix, and $\kappa$ is
the pairing tensor. The last term is non-zero only when pairing is
active.

Since operator $\hat{P}_{i}^{2}$ does not break time-reversal, simplex,
parity, or T-simplex symmetries (see \cite{[Dob04]} for a definition of these symmetries),
the code can maintain them during the calculation
except when determining parameters $k_i$ by shifting wave functions. This is
because the shifted wave functions may lose some of the symmetries.
For example, after a shift along the $y$ axis, the wave function no longer conserves
the simplex.

\begin{figure}[!htb]
\centering
\includegraphics[angle=0,width=0.7\columnwidth]{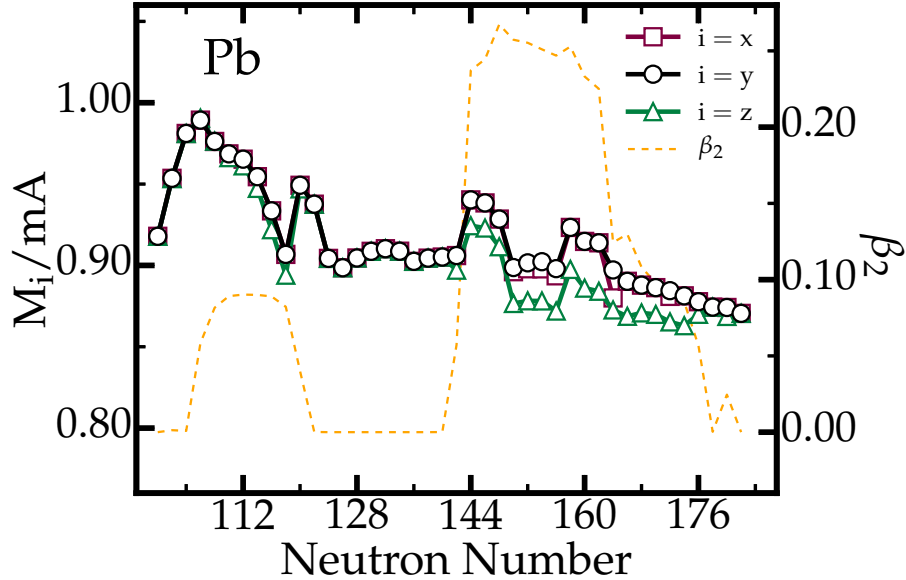}
\caption{Ratios $M_i/mA$ between the renormalized (along $i=x$, $y$ or $z$ axis)
and exact masses for the Pb isotopic chain after Lipkin minimization. For
reference, the $\beta_2$ axial quadrupole deformation is also plotted.}
\label{fig:pb_chain}
\end{figure}

When pairing is active, the determination procedure of parameters $k_i$
is still the same as that outlined in Ref.~\cite{[Dob09g]}. By shifting wave functions in the
$i$th direction by $R_i$, $\left|\Phi\left(R_{i}\right)\right\rangle = \text{\ensuremath{\exp}}
\left(i\hat{P}_{i}R_{i}\right)\left|\Phi\right\rangle $, and calculating overlaps and matrix
elements, respectively,
$\left\langle \Phi\right|\hat{H}\left|\Phi\left(R_{i}\right)\right\rangle $
and
$\left\langle \Phi\right|\hat{P}_{i}^{2}\left|\Phi\left(R_{i}\right)\right\rangle $,
one reaches a new point on the $E=E\left(\langle P_{i}^{2}\rangle\right)$
curve and then one can extract the slope in direction $i$, which is just $k_i$.

Calculation of the parameters $k_i$ can also be performed using the
Gaussian overlap approximation \cite{[Sch12]}. In the case with pairing,
overlaps $\left\langle \Phi|\Phi(R_i)\right\rangle$ are calculated using
the Pfaffian techniques \cite{[Rob09],[Gon11]}.

Fig.~\ref{fig:pb_chain} shows ratios of the renormalized and exact
masses for the Pb isotope chain. Calculations were performed in the space of $N_0=16$ HO
shells and with the SLy4 parametrization of the Skyrme EDF.
A volume zero-range pairing interaction with a cutoff window of
$E_{\mathrm{cut}}=60$\,MeV was used with
the pairing strengths of $V_n=-159$ and
$V_p=-152$\,MeV\,fm$^3$ for neutrons and protons, respectively.
The renormalized mass is defined as
$M_{i}=1/2k_{i}$. As one can see,  if the nucleus is deformed, renormalized masses in different
directions are not equal. Note that even small differences in masses
can have a large impact on the total energy, especially in heavy nuclei.

%%%%%%%%%%%%%%%%%%%%%%%%%%%%%%%%%%%%%%%%%%%%%%%%%%%%%%%%%%%%%%%%%%%%%%%%%%%%%%%%
%%%%%%%%%%%%%%%%%%%%%%%%%%%%%%%%%%%%%%%%%%%%%%%%%%%%%%%%%%%%%%%%%%%%%%%%%%%%%%%%

\subsection{Higher-order Lipkin particle-number corrections}
\label{subsec:higher_order}

Version {(v\codeversion)} of the code \pr{hfodd} allows for the treatment of Lipkin
particle-number corrections to higher orders~\cite{[Wan14]}. The Lipkin method~\cite{[Lip60]}
was proposed as a computationally inexpensive way to obtain an approximate
variation-after-particle energy. For the case of the variation after particle-number projection (VAPNP),
it is realized through an auxiliary Routhian,
\begin{equation}
\label{eq:calK}
\hat{H}'
= \hat{H}-\hat{K}[\hat{N}-N_0] \, ,
\end{equation}
where the Lipkin operator $\hat{K}$ is a function of the shifted
particle-number operator $\hat{N}-N_0$. The role of the Lipkin
operator $\hat{K}$ is to ``flatten" the average Routhian as a
function of the particle number~\cite{[Lip60],[Dob09g],[Wan14]}, that
is, to make it independent of the particle-number fluctuations. As
there exists no such an exact operator, assumptions for the Lipkin
operator $\hat{K}$ have to be made. As proposed by
Lipkin~\cite{[Lip60]}, the simplest and manageable ansatz is in the
form of a polynomial,
\begin{equation}
\label{eq:fN}
\hat{K}[\hat{N}-N_0]=  \sum_{m=1}^M  k_{m} (\hat{N}-N_0)^m \, ,
\end{equation}
where $k_{1}\equiv\lambda$ is the Fermi energy, which is used
as a Lagrange multiplier to fix the average particle number. The
higher-order Lipkin parameters $k_{m}$ for $m>1$, which are used to best
describe the particle-number dependence of the average energies of
projected states, can no longer be regarded as Lagrange multipliers,
but they are determined as follows.

\begin{figure}[!htb]
\centering
\includegraphics[angle=0,width=0.8\columnwidth]{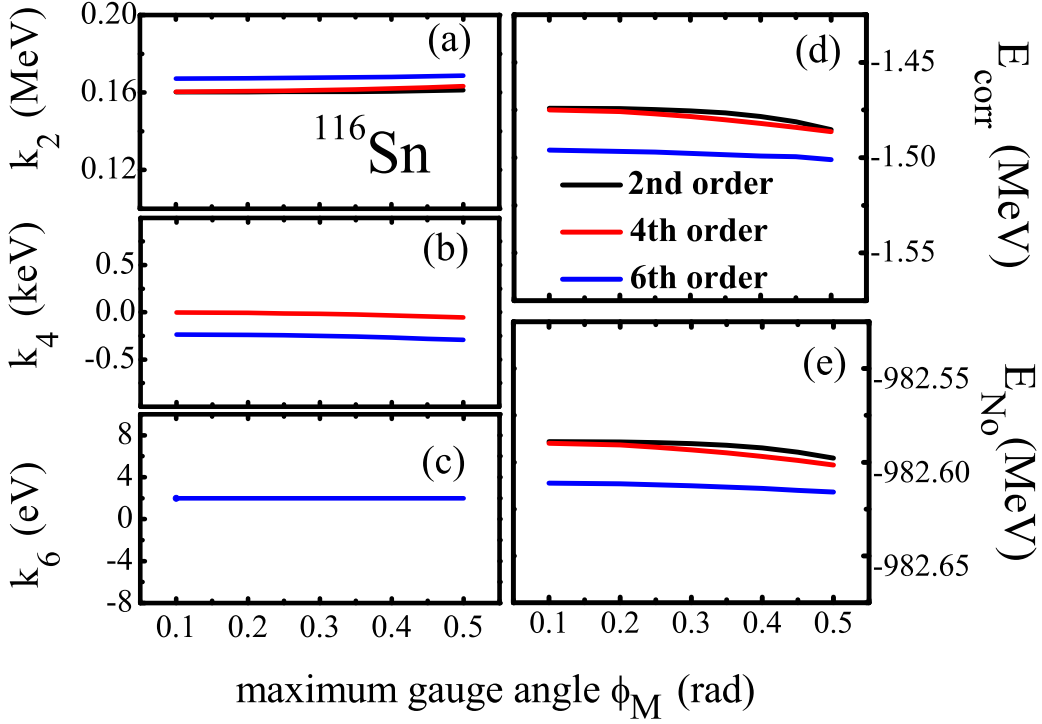}
\vspace*{0.2cm}
\caption{The Lipkin parameters $k_2$ (a), $k_4$ (b), and $k_6$ (c), Lipkin correction
energy $E_{\text{corr}}=-\sum_{m=1}^M  k_{m} n_m(0)$ (d), and Lipkin VAPNP energy
$E_{N_0}=\langle\Phi|\hat{H}-\hat{K}[\hat{N}-N_0]|\Phi\rangle$ (e), determined
for neutrons in $^{116}$Sn at second, fourth, and sixth orders, as functions of
the maximum gauge angle $\phi_{M}$. The Lipkin parameters $k_2$, $k_4$, and $k_6$
are in units of MeV, keV, and eV, respectively, illustrating the rapid convergence
of the power expansions. The particle-hole channel was described with the SIII
parametrization of the Skyrme EDF~\cite{[Bei75]}, and the particle-particle
channel by a volume pairing force with strength $V_0=-155.45$\,MeV. The proton
pairing was switched off.}
\label{fig:sn116}
\end{figure}

We begin by defining the HFB wave functions shifted in the gauge space as
$|\Phi(\phi)\rangle = \exp\left(i\phi(\hat{N}-N_0)\right)|\Phi\rangle$,
which gives us the overlap, energy, and particle-number kernels
$I(\phi) = \langle\Phi|\Phi(\phi)\rangle$,
$H(\phi) = \langle\Phi|\hat{H}|\Phi(\phi)\rangle$, and
$N_m(\phi) =\langle\Phi|(\hat{N}-N_0)^m|\Phi(\phi)\rangle$, respectively.
Then, Eqs.\ (\ref{eq:calK}) and~(\ref{eq:fN}) combine to
\begin{equation}
\label{flat}
h'(\phi) = h(\phi) - \sum_{m=1}^M  k_{m} n_m(\phi) \, ,
\end{equation}
where the reduced kernels are
\begin{equation}
\label{reduced-kernels}
h' (\phi)=\frac{H' (\phi)}{I(\phi)} \, , ~
h  (\phi)=\frac{H  (\phi)}{I(\phi)} \, , ~
n_m(\phi)=\frac{N_m(\phi)}{I(\phi)} \, .
\end{equation}
Assuming that the reduced Routhian kernel $h'(\phi)$ is perfectly flat, that is,
$h'(\phi)\equiv{C}$ for all $\phi$, the Lipkin parameters $k_{m}$ for $m=1,\ldots,M$ can be
determined from
\begin{eqnarray}
C+\sum_m k_m n_m(\phi_i)=h(\phi_i) \, .
\label{eq164}
\end{eqnarray}
This equation is assumed to hold at
gauge angle $\phi=\phi_0=0$ and also at $M$ other nonzero values of
the gauge angle $\phi_i$, which gives a set of linear equations for $k_{m}$.
In practice, equally spaced values of $\phi_i=i\phi_1$ can be
used~\cite{[Wan14]}, which gives the maximum gauge angle $\phi_M=M\phi_1$.
At convergence of the expansion of the Lipkin operator
(\ref{eq:fN}), the resulting Lipkin parameters should not depend on the
maximum gauge angle.

When calculating the projected energies, the largest contributions to the
integrals in the gauge space come from the vicinity of the origin due to
the largest weight~\cite{[Flo97]}. The singularities caused by the
vanishing overlaps have quite small influences on the reduced kernels near
the origin~\cite{[Dob07d]}. Therefore, we evaluate
Lipkin parameters $k_{m}$ using the gauge-shifted HFB states not far
away from the origin, that is, the value of the maximum gauge angle (the largest
gauge angle chosen for the determinations of Lipkin parameters) should be small.

In Fig.~\ref{fig:sn116}, we show an example of the Lipkin parameters
calculated for the neutron states in $^{116}$Sn. The results for the
second, fourth, and sixth orders ar shown as functions of the maximum
gauge angle $\phi_{M}$. For $^{116}$Sn, which is a typical mid-shell
nucleus, the dependence of the Lipkin parameter on $\phi_{M}$ is weak
already at second order, and thus higher-order Lipkin method does not
give much of an improvement, see Ref.~\cite{[Wan14]} for the full discussion.

%%%%%%%%%%%%%%%%%%%%%%%%%%%%%%%%%%%%%%%%%%%%%%%%%%%%%%%%%%%%%%%%%%%%%%%%%%%%%%%%
%%%%%%%%%%%%%%%%%%%%%%%%%%%%%%%%%%%%%%%%%%%%%%%%%%%%%%%%%%%%%%%%%%%%%%%%%%%%%%%%

\subsection{Interface to a plotting program}
\label{subsec:plotting}

The \pr{hfodd} review file \tv{FILREV}, see Section~II-3.9~\cite{[Dob97d]},
is a plain-text file that can be used to
extract information necessary for creating various kinds of
plots. As an example, the present distribution {(v\codeversion)}
contains script \tf{getLevels.py}, which reads the \pr{hfodd} review file
and prepares data files that can be
used to plot single-particle energies or Routhians as functions of
deformation ($Q_{20}$ moment) or rotational
frequency ($\omega$).

In order to use the script, one has to run \pr{hfodd} for a series of
deformations/frequencies, corresponding to different values of
\tv{QASKED} (under the keyword \tk{MULTCONSTR} in the
\pr{hfodd} input file) or \tv{OMEGAY} (under keyword
\tk{OMEGAY} or \tk{OMEGA\_XYZ}).
It should be remembered, however, that multipole moment $q_{20}$
of a converged solution may not be exactly equal to
\tv{QASKED} --- discrepancies can be quite large, but, at least
to some extend, can be controlled by adjusting stiffness parameter
(\tv{STIFFQ} under keyword \tk{MULTCONSTR} in the \pr{hfodd}'s
input data file). Alternatively, one can use the
ALM, see Section~VI-2.2.2~\cite{[Dob09d]}.
Of course, the \pr{hfodd} review file has to contain data describing
single-particle levels, which means the parameter \tv{IREVIE}
(under the keyword \tk{REVIEW} in the \pr{hfodd} input file) must
be set to at least~2.

In the present distribution {(v\codeversion)}, two examples of \pr{hfodd} review files are provided:
\tf{DyDef.rev} corresponds to a~series of deformations
and \tf{DyOme.rev} to a~series of rotational frequencies.
These two review files were obtained by running the code \pr{hfodd}
on input data files \tf{DyDef.dat} and \tf{DyOme.dat}, respectively.

Having a \pr{hfodd} review file prepared, one can run the script
\tf{getLevels.py} with three or four arguments:
\begin{itemize}
    \item The first (required) argument is the name of
        the \pr{hfodd} review file produced by running the code \pr{hfodd} --
        it should correspond to a~series of results obtained from
        runs of the code \pr{hfodd} with input data differing by the value
        of deformation (\tv{QASKED}) or rotational frequency
        (\tv{OMEGAY} or \tv{OMEGAX} or \tv{OMEGAZ}).
        Such a~file can also be merged ``by hand'' from several
        files created in separate runs.
    \item The second (required) argument specifies the quantity
        which plays the role of the independent variable:\\
         \hspace*{2.0em}'d' (or 'D', or 'q', or 'Q') ---
                            deformation (q20);\\
         \hspace*{2.0em}'f' (or 'F', or 'o', or 'O') ---
                             rotational frequency ($\omega$);
                             by default $\omega_y$ is understood,
                             but adding letter 'x' or 'y' or 'z'
                             (e.g., 'fy' or 'oz') another component
                             may be selected
    \item The third (required) argument is:\\
        \hspace*{2.0em}'N' (or 'n') --- neutrons, \\
        \hspace*{2.0em}'Z' (or 'z', or 'P', or 'p') --- protons.
    \item The fourth (optional) argument determines the name of the
        output file; if it is \texttt{name}, then the
        output file will be \tf{name\_XXX\_Y.lev},
        where \texttt{XXX} is \texttt{def} for curves
        vs.~deformation ($q_{20}$) and \texttt{omQ} for curves
        vs.~rotational frequency, with 'Q' equal to 'x', 'y' or 'z',
        while \texttt{Y} is \texttt{N} for neutrons and \texttt{Z}
        for protons. If not specified, \texttt{name} defaults to
        the name of the input file with its extension \texttt{.rev}
        stripped off.
\end{itemize}
For example, with the two review files mentioned above, one can run
\begin{verbatim}
            ./getLevels.py DyDef.rev d n
            ./getLevels.py DyDef.rev d z
            ./getLevels.py DyOme.rev f n
            ./getLevels.py DyOme.rev f z
\end{verbatim}
to get files \tf{DyDef\_def\_N.lev},
\tf{DyDef\_def\_Z.lev}, \tf{DyOme\_omy\_N.lev},
and \tf{DyOme\_omy\_Z.lev}, respectively.

The output file of the script is a~pure text file and has
the following format:
\begin{itemize}
    \item
        The first ten lines constitute a~header containing some
        metadata about the current run; they look like this
        \begin{verbatim}
        # Created on  : 2016-07-21 23:38:22.843367
        # Version     : 1.1
        # Run by      : <user name>
        # Input file  : DyDef.rev
        # Output file : DyDef_def_N.lev
        # N and Z     : 86 66
        # Levels for  : neutrons
        # Curves vs.  : 10 deformations (q20)
        # Phony ene   : 16.0 MeV
        # No of curves: 111
        \end{verbatim}
        For curves in function of frequency, the name of the
        output file will contain \texttt{omy} instead of
        \texttt{def} (line~5) and \texttt{omegas} instead of
        \texttt{deformations} (line~8).
    \item
        After the header, information on all extracted curves
        is written one after another. The number of curves is specified
        in the last line of the header. Data corresponding to each curve are
        preceded by exactly one blank line. A~segment of data
        describing one curve has the following two forms:
        \begin{itemize}
            \item In the case of single-particle energies as functions of deformation:
                \begin{verbatim}
                #  3/2 +1 7
                -11.469401879 -2.0959741301 |4,1,1,3/2>
                1.724126373 -3.8251360412 |6,3,1,3/2>
                7.4711027963 -5.8204478799 |6,5,1,3/2>
                9.6541267607 -6.684000508 |6,5,1,3/2>
                13.883562273 -7.8581971 |6,5,1,3/2>
                23.059344121 -9.0080832634 |4,0,2,3/2>
                39.292223291 -5.9512624571 |4,0,2,3/2>
                41.950864161 -5.3978081111 |4,0,2,3/2>
                44.87486028 -5.7117002799 |6,4,2,3/2>
                48.23253362 -6.5226929621 |6,4,2,3/2>
                \end{verbatim}
                where the three elements in the first, comment line
                denote $j_z$, parity and curve number (starting from~1
                for the lowest state with given $j_z$ and parity).
                Each point of the curve is then represented by three
                space-separated elements: deformation ($q_{20}$),
                energy, and the Nilsson label of the leading component
                of a~given state (the label will not contain any embedded
                spaces.) Note that the label does not have to be the same
                for all points along a~single curve.
            \item In the case of Routhians as functions of the rotational frequency:
                \begin{verbatim}
                # -1 +1 23
                0.001 -7.3419171589 |5,2,1,3/2>
                0.1 -7.3449823643 |5,2,1,3/2>
                0.2 -7.3563104158 |5,2,1,3/2>
                0.3 -7.3826283035 |5,2,1,3/2>
                0.4 -7.4347954666 |5,2,1,3/2>
                0.5 -7.5198781158 |5,2,1,3/2>
                0.6 -7.634943286 |5,2,1,3/2>
                0.7 -7.76903774 |5,2,1,3/2>
                0.8 -7.9093199047 |5,2,1,3/2>
                \end{verbatim}
                the form is similar, but the comment line specifies parity,
                signature and the curve number. The first number on each
                line now denotes the rotational frequency.
        \end{itemize}
\end{itemize}
It may happen that states which belong to a~given orbital are present
at some deformations (frequencies) but are missing at other deformations,
as they were too high to be calculated and/or output by \pr{hfodd}.
In this situation, the script adds ``artificial'' points to the curve
with all quantum numbers in the Nilsson label set to zero and energy
set to \tv{phonyEne}. The value of \tv{phonyEne} can be found
in the ninth line of the header and is guaranteed to be larger by at
least 2\kern0.15em MeV than the largest ``true'' energy present in the
data. Additionally, an asterisk is added at the end of lines
corresponding to these ``artifical'' points. For example, for data
in file \tf{DyDef\_def\_N.lev}, the value of \tv{phonyEne}
is 16\,MeV, and one of the orbital is written as
                \begin{verbatim}
                #  1/2 -1 11
                -29.988366225 16.0 |0,0,0,0/2> *
                -19.994324961 7.0230247553 |5,4,1,1/2>
                -8.59922993 -0.79098160199 |5,4,1,1/2>
                0.54349578927 -2.3664542895 |5,3,0,1/2>
                10.031225775 -4.0948376775 |5,1,0,1/2>
                21.183531098 -2.7050254605 |7,7,0,1/2>
                30.131724613 -6.1232544454 |7,7,0,1/2>
                40.0 -5.758469044 |5,2,1,1/2>
                50.000000013 -7.0317367018 |7,6,1,1/2>
                60.143378687 -8.1834216685 |7,6,1,1/2>
                \end{verbatim}

%%%%%%%%%%%%%%%%%%%%%%%%%%%%%%%%%%%%%%%%%%%%%%%%%%%%%%%%%%%%%%%%%%%%%%%%%%%%%%%%
%%%%%%%%%%%%%%%%%%%%%%%%%%%%%%%%%%%%%%%%%%%%%%%%%%%%%%%%%%%%%%%%%%%%%%%%%%%%%%%%

\subsection{Strong-force isospin-symmetry-breaking terms}%PBaczyk
\label{subsec:ISB}

It is well known that mean-field models involving isospin-invariant
strong force and the Coulomb interaction constituting the only source
of the ISB fail to reproduce both mirror
displacement energies (MDEs)~\cite{[Nol69]} and triplet displacement
energies (TDEs)~\cite{[Sat14]}. These primary indicators of the ISB
effects are defined as
\begin{eqnarray}
\mathrm{MDE} & = & \mathrm{BE}\left(T,  T_3=-T\right) - \mathrm{BE}\left(T,  T_3=+T\right),
\label{eq:MDE}\\
\mathrm{TDE} & = & \mathrm{BE}\left(T=1,T_3=-1\right) + \mathrm{BE}\left(T=1,T_3=+1\right)
-2\mathrm{BE}\left(T=1,T_3=0\right),\label{eq:TDE}
\end{eqnarray}
where $\mathrm{BE}(T,T_3)$ denotes the binding energy of a nucleus
with total isospin $T$ and its projection $T_3$. To account
quantitatively for the MDEs and TDEs, mean-field models must be
extended by including charge dependent components originating from
the strong force.

\begin{figure}[!htb]
\centering
\begin{subfigure}[b]{0.485\textwidth}
\includegraphics[width=\textwidth]{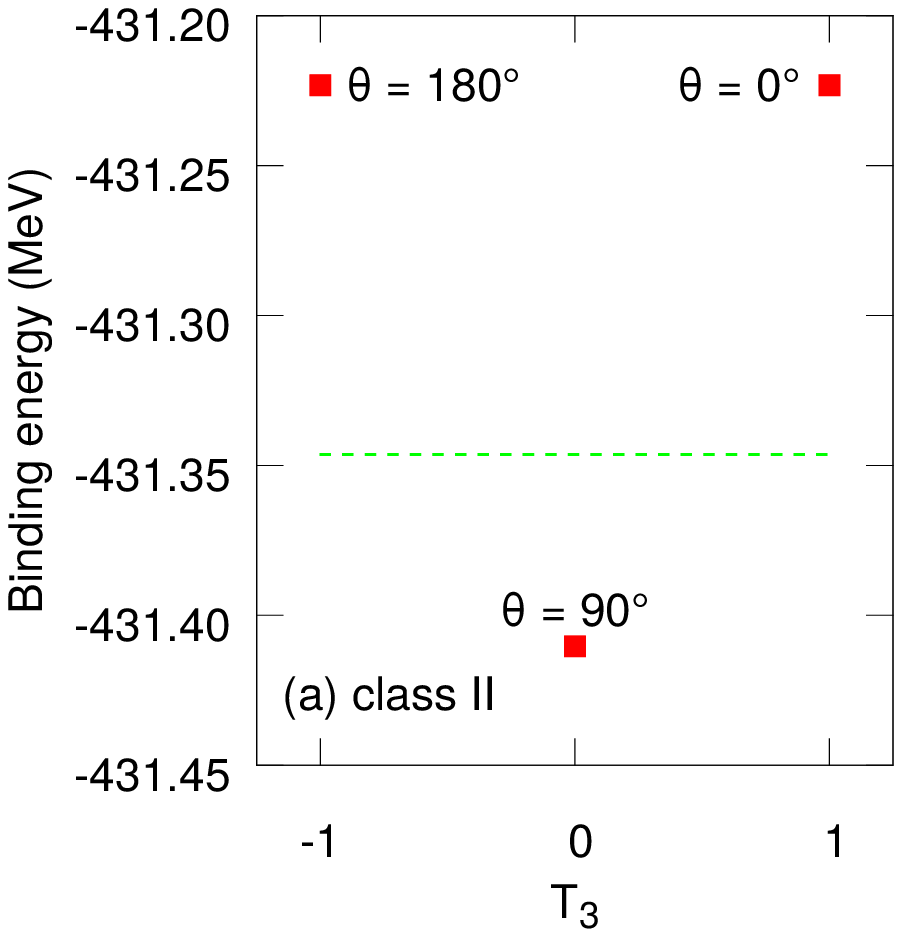}
\end{subfigure}
\hfill
\begin{subfigure}[b]{0.485\textwidth}
\includegraphics[width=\textwidth]{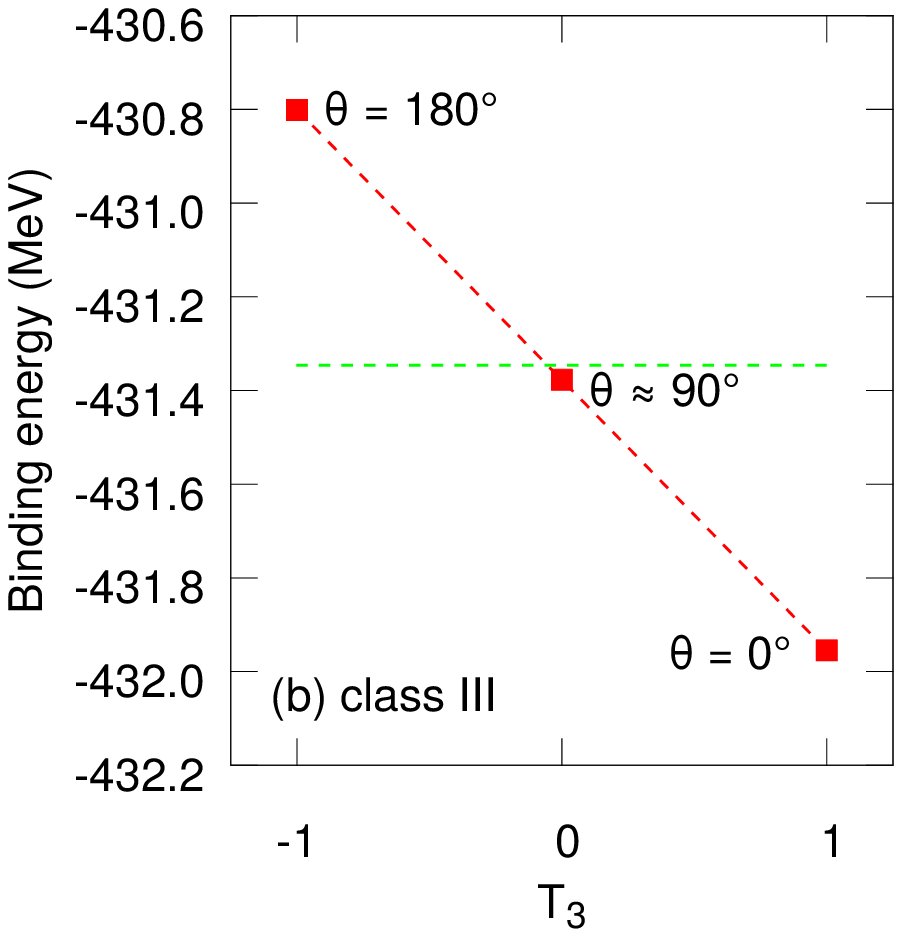}
\end{subfigure}
\caption{Ground-state energies in the $A=42$ isobaric-triplet nuclei.
Calculations were performed without the Coulomb interaction. Panel
(a) shows the result obtained using the class II force only with
$t_{0}^{\mathrm{II}} = +20$\,MeV\,fm$^3$; Panel (b) shows the
result obtained using the class III force only with
$t_{0}^{\mathrm{III}} = -8$\,MeV\,fm$^3$. In both panels, the
horizontal dashed line shows the ground-state energy obtained using
charge-independent model. The slanted dashed line in panel (b)
indicates an almost exact linear trend in the calculated masses.}
\label{fig:classISB}
\end{figure}

\begin{table}[htb!]
\centering
\caption{Ground-state energies in the $A=42$ isobaric-triplet nuclei.
Results obtained without and with the Coulomb interaction are shown
in the second and the third column, respectively. The energies, MDEs
and TDEs are given in MeV. The first column gives values of the
coupling constants $t^\mathrm{II}_0$ and $t^\mathrm{III}_0$ in MeV\,fm$^3$.}
\begin{tabular}{|c c|c|r|c|c|r|c|r|c|c|c|}
\hline
&& \multicolumn{5}{c|}{Without Coulomb} & \multicolumn{5}{c|}{With Coulomb}\\\hline
&& $\theta$~[$^\circ$] & $T_3$ & Energy & MDE & \multicolumn{1}{c|}{TDE} & $\theta$~[$^\circ$] & $T_3$ & Energy & MDE & TDE\\\hline
\parbox[t]{ 4mm}{\multirow{3}{*}{\rotatebox[origin=c]{90}{\everymath{\scriptstyle} $t_0^{\mathrm{II} }=\;0.0$}}}
\parbox[t]{-4mm}{\multirow{3}{*}{\rotatebox[origin=c]{90}{\everymath{\scriptstyle} $t^\mathrm{III}_0=\;0.0$}}}
&&      0.0     &       1       &     $-$431.346        &       0.000   &       0.000   &       0.0     &       1       &     $-$358.274        &       13.789  &       0.159   \\
&&      90.0    &       0       &     $-$431.346        &               &               &       88.9    &       0       &     $-$351.459        &               &               \\
&&      180.0   &    $-$1       &     $-$431.346        &               &               &       180.0   &    $-$1       &     $-$344.485        &               &               \\\hline
\parbox[t]{ 4mm}{\multirow{3}{*}{\rotatebox[origin=c]{90}{\everymath{\scriptstyle} $t^\mathrm{II}_0=\;20.0$}}}
\parbox[t]{-4mm}{\multirow{3}{*}{\rotatebox[origin=c]{90}{\everymath{\scriptstyle} $t^\mathrm{III}_0=\;0.0$}}}
&&      0.0     &       1       &     $-$431.223        &       0.000   &       0.374   &       0.0     &       1       &     $-$358.141        &       13.783  &       0.525   \\
&&      90.0    &       0       &     $-$431.410        &               &               &       88.9    &       0       &     $-$351.512        &               &               \\
&&      180.0   &    $-$1       &     $-$431.223        &               &               &       180.0   &    $-$1       &     $-$344.358        &               &               \\\hline
\parbox[t]{ 4mm}{\multirow{3}{*}{\rotatebox[origin=c]{90}{\everymath{\scriptstyle} $t_0^{\mathrm{II} }=\;0.0 $}}}
\parbox[t]{-4mm}{\multirow{3}{*}{\rotatebox[origin=c]{90}{\everymath{\scriptstyle} $t^\mathrm{III}_0=\,-8.0$}}}
&&      0.0     &       1       &     $-$431.955        &       1.153   &     $-$0.002  &       0.0     &       1       &     $-$359.038        &       14.899  &       0.145   \\
&&      95.5    &       0       &     $-$431.377        &               &               &       94.0    &       0       &     $-$351.661        &               &               \\
&&      180.0   &    $-$1       &     $-$430.801        &               &               &       180.0   &    $-$1       &     $-$344.139        &               &               \\\hline
\parbox[t]{ 4mm}{\multirow{3}{*}{\rotatebox[origin=c]{90}{\everymath{\scriptstyle} $t_0^{\mathrm{II} }=\;20.0$}}}
\parbox[t]{-4mm}{\multirow{3}{*}{\rotatebox[origin=c]{90}{\everymath{\scriptstyle} $t^\mathrm{III}_0=\,-8.0$}}}
&&      0.0     &       1       &     $-$431.827        &       1.147   &       0.372   &       0.0     &       1       &     $-$358.894        &       14.888  &       0.511   \\
&&      95.5    &       0       &     $-$431.440        &               &               &       94.0    &       0       &     $-$351.706        &               &               \\
&&      180.0   &    $-$1       &     $-$430.680        &               &               &       180.0   &    $-$1       &     $-$344.007        &               &               \\
\hline
\end{tabular}

\label{tab:isb}
\end{table}

In version {(v\codeversion)}, the strong-force ISB terms were added
as effective, two-body, zero-range corrections to the conventional
isospin-invariant Skyrme interaction. Contributions of class II and
class III forces (according to the classification of Henley and
Miller~\cite{[Hen79]}) were implemented as follows
\begin{eqnarray}
\hat{V}^{\textit{II}}(i,j) & = &
\frac12 t^\mathrm{II}_0\, \delta\left(\gras{r}_i - \gras{r}_j\right)
\left(1 - x^\mathrm{II}_0\,\hat P^\sigma_{ij}\right)
\left[3\hat{\tau}_3(i)\hat{\tau}_3(j)-\vec{\tau}(i)\circ\vec{\tau}(j)\right],
\label{eq:classII}\\
\hat{V}^{\textit{III}}(i,j) & = &
\frac12 t^\mathrm{III}_0\, \delta\left(\gras{r}_i - \gras{r}_j\right)
\left(1 - x^\mathrm{III}_0\,\hat P^\sigma_{ij}\right)
\left[\hat{\tau}_3(i)+\hat{\tau}_3(j)\right],
\label{eq:classIII}
\end{eqnarray}
where $i$, $j$ label nucleons, $t_{0}^{\mathrm{II}}$,
$x_{0}^{\mathrm{II}}$, $t_{0}^{\mathrm{III}}$, and
$x_{0}^{\mathrm{III}}$ are coupling constants, $\hat P^\sigma_{ij}$
is the usual spin-exchange operator, and $\vec{\tau} \equiv
(\hat{\tau}_1, \hat{\tau}_2, \hat{\tau}_3)$ are the isospin Pauli
matrices. Both forces are charge dependent, but only class III breaks
charge symmetry. The corresponding energy densities read
\begin{eqnarray}
\mathcal{H}^{\textit{II}} & = &
\frac{1}{2}t^\mathrm{II}_0\left(1-x^\mathrm{II}_0\right)
\left(
\rho_n^2+\rho_p^2-2\rho_n\rho_p-2\rho_{np}\rho_{pn}
-
\gras{s}_{n}^2-\gras{s}_{p}^2+2\gras{s}_{n}\cdot\gras{s}_{p}+2\gras{s}_{np}\cdot\gras{s}_{pn}
\right),\\
\mathcal{H}^{\textit{III}} & = &
\frac{1}{2}t^\mathrm{III}_0\left(1-x^\mathrm{III}_0\right)
\left(\rho_n^2-\rho_p^2 - \gras{s}_{n}^2+\gras{s}_{p}^2\right) ,
\end{eqnarray}
and the contributions to the mean-field potentials of Ref.~\cite{[Per04]} are:
\begin{equation}
\begin{array}{rcl}
U^{\textit{II}}_{n} & = & t^\mathrm{II}_0\left(1- x^\mathrm{II}_0\right)
\left(+\rho_n - \rho_p \right),\\[1ex]
U^{\textit{II}}_{p} & = & t^\mathrm{II}_0\left(1- x^\mathrm{II}_0\right)
\left( - \rho_n + \rho_p \right),\\[1ex]
U^{\textit{II}}_{np} & = & t^\mathrm{II}_0\left(1-x^\mathrm{II}_0\right)
\left(-\rho_{np} \right),\\[1ex]
U^{\textit{III}}_{n} & = & t^\mathrm{III}_0\left(1-x^\mathrm{III}_0\right)
\left(+\rho_n \right),\\[1ex]
U^{\textit{III}}_{p} & = & t^\mathrm{III}_0\left(1-x^\mathrm{III}_0\right)
\left(-\rho_p \right),
\end{array}\hspace{2cm}
\begin{array}{rcl}
\gras{\Sigma}^{\textit{II}}_{n} & = & t^\mathrm{II}_0\left(1- x^\mathrm{II}_0\right)
\left(- \gras{s}_{n} + \gras{s}_{p} \right),\\[1ex]
\gras{\Sigma}^{\textit{II}}_{p} & = & t^\mathrm{II}_0\left(1- x^\mathrm{II}_0\right)
\left(+ \gras{s}_{n} - \gras{s}_{p} \right),\\[1ex]
\gras{\Sigma}^{\textit{II}}_{np} & = & t^\mathrm{II}_0\left(1-x^\mathrm{II}_0\right)
\left( + \gras{s}_{np} \right),\\[1ex]
\gras{\Sigma}^{\textit{III}}_{n} & = & t^\mathrm{III}_0\left(1-x^\mathrm{III}_0\right)
\left(-\gras{s}_{n}\right),\\[1ex]
\gras{\Sigma}^{\textit{III}}_{p} & = & t^\mathrm{III}_0\left(1-x^\mathrm{III}_0\right)
\left(+\gras{s}_{p}\right).
\end{array}
\end{equation}
The formulas above indicate that the parameters $x^\mathrm{II}_0$ and
$x^\mathrm{III}_0$ are redundant and can be set to zero what we do
hereafter. However, to maintain compatibility with future
implementations of the finite-range ISB interactions, parameters
$x^\mathrm{II}_0$ and $x^\mathrm{III}_0$ can still be specified
explicitely in the input file, see Sect.~\ref{subsubsec:Interaction}.

Note that the contributions due to class II forces depend explicitly
on the p-n-mixed scalar and vector densities $\rho_{np}$ and
$\gras{s}_{np}$, respectively. Therefore, such forces can only be
used within the mean-field formalism involving p-n mixing, whereas
calculations with class~III forces, which only depend on the standard
isoscalar densities, do not require p-n mixing. For the sake of
consistency, however, all numerical results shown in this Section
have been obtained in the framework involving p-n mixing developed in
Ref.~\cite{[Sat13c]} and described in detail in
Sect.~\ref{subsec:pnmixing}.  Note also, that the spin density
$\gras{s}$ is non-zero only when time-reversal symmetry is internally
broken, which is the case for the odd and odd-odd nuclei.

To verify the influence of the new terms on the HF ground-state
solutions, we have performed test calculations without the Coulomb
interaction. This simplification allows for direct testing of the ISB
effects caused separately by class II and class III terms.
Fig.~\ref{fig:classISB}(a) shows the effect of class II force on the
ground-state energies in the isospin triplet. As anticipated, the
class II interaction is responsible for the curvature of the binding
energies of the triplet. Indeed, in this case the $T_3=\pm 1$ nuclei
are shifted up in the energy by the same value whereas the $T_3=0$
nucleus is shifted down (cf.~Table~\ref{tab:isb}). Results obtained
with class III force are shown in Fig.~\ref{fig:classISB}(b). In this
case the $T_3=0$  nucleus is almost unaffected, whereas the  $T_3=\pm
1$ nuclei are shifted in opposite directions by nearly the same
energy (cf.~Table~\ref{tab:isb}). These results confirm that the
class II (III) forces modify the TDE (MDE) essentially not
influencing the MDE (TDE), respectively.

Table~\ref{tab:isb} shows the calculated MDE and TDE for a
representative example of the  $A=42$ triplet using different
variants of the model. By comparing the calculated values to the
experimental data MDE$_{exp} = 15.007$\,MeV and TDE$_{exp} =
0.590$\,MeV one immediately concludes that the Coulomb interaction
alone is indeed not sufficient to reproduce the data. Taking into
account the ISB strong components clearly improves the agreement
between theory and experiment. More systematic preliminary study
performed in Ref.~\cite{[Bac15]} shows that the class II and III
terms implemented here  allow to reproduce quite well experimental
data on MDE and TDE in a wide range of masses.

\subsection{Augmented Lagrangian Method for calculations with 3D constraints
on angular momentum and isospin}
\label{subsec:alm}

Following the previous implementation in the \pr{hfodd} code of the ALM for
multipole moments, see Section VI-2.2.2, in version {(v\codeversion)}
we implemented the same methodology for constraints on angular
momentum and isospin. To this effect, we added to the total energy
${\cal E}$ the ALM constraints as
\begin{equation}
\label{eq:alm}
{\cal E}' = {\cal E} - \sum_{\mu=x,y,z} \omega_\mu \left( \langle\hat{J}_\mu\rangle - \bar{J}_\mu \right)
                     + \sum_{\mu=x,y,z}      C_\mu \left( \langle\hat{J}_\mu\rangle - \bar{J}_\mu \right)^2
                     - \sum_{  k=1,2,3}\lambda_k   \left( \langle\hat{t}_k  \rangle - \bar{t}_k   \right)
                     + \sum_{  k=1,2,3}      C_k   \left( \langle\hat{t}_k  \rangle - \bar{t}_k   \right)^2.
\end{equation}
In fact, the constraints on the angular momentum were already
implemented in version (2.08i), see Section~VI-2.3~\cite{[Dob09d]}, so the ALM only
required introducing corrective terms $\delta\omega_\mu$, and
updating the angular frequencies as
$\omega_\mu=\omega^0_\mu+\delta\omega_\mu$, with $\omega^0_\mu$
denoting the previous fixed values. A similar technology was used for the
constraints on the isospin, whereby the previous fixed values of the
isocranking frequencies $\lambda^0_k$, see
Section~\ref{subsec:pnmixing}, were updated as $\lambda_k =
\lambda^0_k+\delta\lambda_k$.

%%%%%%%%%%%%%%%%%%%%%%%%%%%%%%%%%%%%%%%%%%%%%%%%%%%%%%%%%%%%%%%%%%%%%%%%%%%%%%%%
%%%%%%%%%%%%%%%%%%%%%%%%%%%%%%%%%%%%%%%%%%%%%%%%%%%%%%%%%%%%%%%%%%%%%%%%%%%%%%%%

\subsection{Corrected errors}
\label{subsec:bugs}

In the present version {(v\codeversion)}, we have corrected the following errors
of the previous published version (v2.49t)~\cite{[Sch12]}.

%%%%%%%%%%%%%%%%%%%%%%%%%%%%%%%%%%%%%%%%%%%%%%%%%%%%%%%%%%%%%%%%%%%%%%%%%%%%%%%%

\subsubsection{Entropy}
\label{subsubsec:entropy}

The entropy calculated in version (v2.49t) was too small by a factor two. As a
consequence, the Maxwell relations of thermodynamics could not be satisfied.

%%%%%%%%%%%%%%%%%%%%%%%%%%%%%%%%%%%%%%%%%%%%%%%%%%%%%%%%%%%%%%%%%%%%%%%%%%%%%%%%

\subsubsection{Finite Temperature BCS Calculations}
\label{subsubsec:bcs}

In the extension of the Hartree-Fock with BCS pairing correlations at finite
temperature, the spectral gap is defined by
\begin{equation}
\Delta = \frac{\sum_{n} u_{n}v_{n} \Delta_{n}}{\sum_{n}{u_{n}v_{n}}},
\end{equation}
where $u_{n}$ and $v_{n}$ are the usual BCS occupations of single-particle
states. When temperature increases, pairing correlations vanish and the
denominator of this expression can become zero. In version (v2.49t), the value
of the denominator was not tested, which could lead to undefined values.

%%%%%%%%%%%%%%%%%%%%%%%%%%%%%%%%%%%%%%%%%%%%%%%%%%%%%%%%%%%%%%%%%%%%%%%%%%%%%%%%

\subsubsection{Symmetries}
\label{subsubsec:symmetries}

When calculating contributions to the mean field from the
finite-range Coulomb (exchange), Yukawa, or Gogny interactions, the
densities are computed directly in the configuration space
$\rho_{ij}$ (that is, on the HO basis), which involves the whole
$\mathbb{R}^{3}$ domain, irrespective of symmetries of the problem.
Conversely, the Skyrme-type mean fields, including the
density-dependent terms, are expressed as functions of densities.
Therefore, they are computed in the coordinate-space representation
$\rho(\gras{r})$, and thus are explicitly symmetrized, so as to
benefit from symmetries of the problem. When both types of mean
fields are simultaneously present, this may create an inconsistency
between the two contributions. In version (v2.49t), in the case of
the Gogny force, the resulting small numerical inconsistency was
building up along the self-consistent iterations and led to
divergences. Enforcing the calculation of unsymmetrized
coordinate-space densities resolved the problem.

%%%%%%%%%%%%%%%%%%%%%%%%%%%%%%%%%%%%%%%%%%%%%%%%%%%%%%%%%%%%%%%%%%%%%%%%%%%%%%%%

\subsubsection{Skyrme parameters}
\label{subsubsec:parameters}

The Skyrme-force parameter sets predefined for
acronyms SLY4, SLY5, SLY7 and sly4, sly5, sly7 were coded in an
opposite way with respect to what was presented in Section~VI-2.4.2~\cite{[Dob09d]}.
Moreover, parameter sets predefined for acronyms
SLY6 and SLY7 were incorrectly accompanied  by  the switch
\tv{IETACM}=1  (two-body center-of-mass  correction  after
variation), whereas these forces have been fitted with the
center-of-mass correction before variation, and should  have been
accompanied  the switch \tv{IETACM}=2. In addition, for
acronym UDF0, the values predefined for the Skyrme force UNEDF0 corresponded
to preliminary results and not to the final values given in Ref.~\cite{[Kor10b]}.

%%%%%%%%%%%%%%%%%%%%%%%%%%%%%%%%%%%%%%%%%%%%%%%%%%%%%%%%%%%%%%%%%%%%%%%%%%%%%%%%

\subsubsection{HO basis}
\label{subsubsec:basis}

In  case  of the HO basis definition with \tv{NLIMIT}<0, see
Section~II-3.6~\cite{[Dob97d]}, that is, when the basis  was  supposed to  be  cut
based on the energies of the HO states, an incorrect safety check
performed for the number of HO states was stopping the code.

%%%%%%%%%%%%%%%%%%%%%%%%%%%%%%%%%%%%%%%%%%%%%%%%%%%%%%%%%%%%%%%%%%%%%%%%%%%%%%%%

\subsubsection{Shell correction}
\label{subsubsec:shell}

In the parallel mode, the proton smoothing factor for the  shell
correction  was equal to that for neutrons.

%%%%%%%%%%%%%%%%%%%%%%%%%%%%%%%%%%%%%%%%%%%%%%%%%%%%%%%%%%%%%%%%%%%%%%%%%%%%%%%%

\subsubsection{Initialized Lagrange parameters}
\label{subsubsec:Lagrange}

For \tv{IACONT}=0, see Section~VI-3.2~\cite{[Dob09d]}, the
initialized Lagrange parameters were not printed.

%%%%%%%%%%%%%%%%%%%%%%%%%%%%%%%%%%%%%%%%%%%%%%%%%%%%%%%%%%%%%%%%%%%%%%%%%%%%%%%%

\subsubsection{Iterations}
\label{subsubsec:Iterations}

In subroutines \ts{SKFILD}, \ts{SKPAIR}, and \ts{LINMIX}, the
slow-down parameters were incorrectly implemented, and as a result
the code was sometimes iterating with an incorrect slow-down or
without any slow-down, and could then crash.

%%%%%%%%%%%%%%%%%%%%%%%%%%%%%%%%%%%%%%%%%%%%%%%%%%%%%%%%%%%%%%%%%%%%%%%%%%%%%%%%

\subsubsection{Occupation numbers}
\label{subsubsec:Occupation}

In subroutines \ts{CANQUA} and \ts{CANQUZ}, canonical occupation
numbers were treated differently. As a result, during the iterations,
results obtained for conserved and broken simplex symmetry could be
different. Fortunately, the differences disappeared for converged
results.

%%%%%%%%%%%%%%%%%%%%%%%%%%%%%%%%%%%%%%%%%%%%%%%%%%%%%%%%%%%%%%%%%%%%%%%%%%%%%%%%

\subsubsection{Reduced transition probabilities}
\label{subsubsec:Reduced}

Reduced transition probabilities, see Section~V-2.2~\cite{[Dob05]}, have been
calculated for the scaled multipole operators, see Table~III-5~\cite{[Dob00c]}, and
not for the standard electric multipole operators.

%%%%%%%%%%%%%%%%%%%%%%%%%%%%%%%%%%%%%%%%%%%%%%%%%%%%%%%%%%%%%%%%%%%%%%%%%%%%%%%%

\subsubsection{Lipkin-Nogami method}
\label{subsubsec:LN}

In versions (v2.40h) and (v.2.49t), for the Lipkin-Nogami
calculations, the slowing-down of convergence was performed in a
different way than described in Section IV-3.2. First, the
Lipkin-Nogami parameters $\lambda_2$ were slowed-down twice, which
amounted to the true slowing-down parameter of $\epsilon(2-\epsilon)$
instead of the value of $\epsilon$ provided by the user
(\keywo{SLOWLIPKIN}). For example, the input value of 0.5 resulted in
a slower convergence corresponding to the value of 0.75. Second,
density matrices defining the Lipkin-Nogami corrections were also
slowed-down by the same factor of $\epsilon$. In the present version
{(v\codeversion)}, this latter feature is maintained, but another
slowing-down parameter is used to this effect, see parameter \tv{SLOWLD}
under \keywo{SLOWLIPMTD} in Section~\ref{subsubsec:Symmetries}.

%%%%%%%%%%%%%%%%%%%%%%%%%%%%%%%%%%%%%%%%%%%%%%%%%%%%%%%%%%%%%%%%%%%%%%%%%%%%%%%%
%%%%%%%%%%%%%%%%%%%%%%%%%%%%%%%%%%%%%%%%%%%%%%%%%%%%%%%%%%%%%%%%%%%%%%%%%%%%%%%%
%%%%%%%%%%%%%%%%%%%%%%%%%%%%%%%%%%%%%%%%%%%%%%%%%%%%%%%%%%%%%%%%%%%%%%%%%%%%%%%%
%%%%%%%%%%%%%%%%%%%%%%%%%%%%%%%%%%%%%%%%%%%%%%%%%%%%%%%%%%%%%%%%%%%%%%%%%%%%%%%%

\section{Input Data File}
\label{sec:input_file}

%%%%%%%%%%%%%%%%%%%%%%%%%%%%%%%%%%%%%%%%%%%%%%%%%%%%%%%%%%%%%%%%%%%%%%%%%%%%%%%%
%%%%%%%%%%%%%%%%%%%%%%%%%%%%%%%%%%%%%%%%%%%%%%%%%%%%%%%%%%%%%%%%%%%%%%%%%%%%%%%%

\subsection{Input data for serial mode}
\label{subsec:serial}

The structure of the input data file has been described in
Section~II-3~\cite{[Dob97d]}; in version {(v\codeversion)} of the code \pr{hfodd} this
structure is exactly the same. All previous items of the input data
file remain valid, and several new items were added, as described in
Sections~\ref{subsubsec:Interaction}--\ref{subsubsec:Starting}. For
some pervious items, new features or new values of variables were
added (Section~\ref{subsubsec:Previous}), whereas some other items,
although still active and allowed, have become obsolete and their
further use is not recommended (Section~\ref{subsubsec:Obsolete}).

%%%%%%%%%%%%%%%%%%%%%%%%%%%%%%%%%%%%%%%%%%%%%%%%%%%%%%%%%%%%%%%%%%%%%%%%%%%%%%%%

\subsubsection{Interaction}
\label{subsubsec:Interaction}

\key{GOGNY\_SET} D1S =
                \tv{GOGNAM}
\keyspace

\noindent The keyword \tv{GOGNAM} specifies the name of the
          parametrization of the Gogny interaction. In version
          {(v\codeversion)}, the D1S and D1N parametrizations are
          supported. Additional parametrizations can be predefined in
          subroutine \ts{PARGOG}. Code \pr{hfodd} treats the
          density-dependent term of the Gogny interaction as a term
          of the Skyrme interaction. Therefore, to avoid
          inconsistencies, for a given choice of \tv{GOGNAM},
          variable \tv{SKYRME} under keyword \tk{SKYRME\_SET} must be
          set to the same value.

\key{GOGNY} 0 =
                \tv{I\_GOGA}
\keyspace

\noindent For \tv{I\_GOGA>0}, the average value of the finite-range
          Gogny interaction in the particle-hole channel is
          calculated In addition, for \tv{I\_GOGA=2} or 3, direct
          contributions to the mean field are included in the
          calculation, and for \tv{I\_GOGA=2} or 4, exchange
          contributions to the mean-field are included in the
          calculation. Therefore, to perform typical self-consistent
          calculations for the Gogny interaction one sets
          \tv{I\_GOGA=2}.

\key{GOGNY\_PAIR} 0 =
                \tv{IGOGPA}
\keyspace

\noindent For \tv{IGOGPA>0}, the average value of the finite-range
          Gogny interaction in the particle-particle channel is
          calculated. In addition, for \tv{IGOGPA=2}, the
          contributions to the pairing mean field are also included.
          Therefore, to perform typical self-consistent HF or HFB
          calculations for the Gogny interaction one sets
          \tv{IGOGPA=0} or \tv{IGOGPA=2}, respectively.
          \tv{IGOGPA>0} requires \tv{I\_GOGA>0}.

\key{CHARBREAK2} 0., 0., 0 =
                \tv{T02CBR}, \tv{X02CBR}, \tv{I02CBR}

\keyspace

\noindent
For \tv{I02CBR}=1, class II ISB terms are included in the calculation
with parameters $t^\mathrm{II}_0$=\tv{T02CBR} and
$x^\mathrm{II}_0$=\tv{X02CBR}. Note, that the interaction of class II
requires p-n mixing (\tv{IPNMIX}=1).

\key{CHARBREAK3} 0., 0., 0 =
                \tv{T03CBR}, \tv{X03CBR}, \tv{I03CBR}
\keyspace

\noindent
For \tv{I03CBR}=1, class III ISB terms are included in the calculation with
parameters $t^\mathrm{III}_0$=\tv{T03CBR} and $x^\mathrm{III}_0$=\tv{X03CBR}.

\key{POWERDENSI} 1., 1 =
                \tv{POWERD}, \tv{KETA\_P}

For \tv{KETA\_P}=1, the code uses the density-dependent term with the
power of the density dependence predefined for a given Skyrme
interaction, whereas for \tv{KETA\_P}=2, the predefined value is
overwritten by the value of \tv{POWERD}.

\keyspace

\key{TIMEREPAIR} 0 =
                \tv{ITIREP}

For \tv{ITIREP}=1, the code neglects time-odd (imaginary) parts of the pairing
densities.

\keyspace

%%%%%%%%%%%%%%%%%%%%%%%%%%%%%%%%%%%%%%%%%%%%%%%%%%%%%%%%%%%%%%%%%%%%%%%%%%%%%%%%

\subsubsection{Symmetries}
\label{subsubsec:Symmetries}

\key{PROTNEUMIX} 0 =
                        \tv{IPNMIX}
\keyspace

\noindent For \tv{IPNMIX}=1, the p-n mixing calculation is performed,
          in which single-particle states are expressed as
          superpositions of the proton and neutron components. In
          version {(v\codeversion)}, the p-n mixing is implemented
          for the HF calculations (no pairing correlations) and at
          zero temperature only, that is, \tv{IPNMIX}=1 requires
          \tv{IPAIRI}=0 and \tv{IFTEMP}=0. Moreover, \tv{IPNMIX}=1
          requires \tv{IBROYD}=0, \tv{I\_YUKA}=0, \tv{I\_GOGA}=0,
          \tv{IF\_RPA}=0, \tv{IFSHEL}=0, \tv{IFRAGM}=0, and
          \tv{MIN\_QP}=0.

\key{SLOWLIPMTD} 0.5, 0.5, 0.5, 0.5 =
                 \tv{SLOWLD}, \tv{SLOWTP}, \tv{SLOWRP}, \tv{SLOWLM}
\keyspace

\noindent Variable \tv{SLOWLD} is a slow-down mixing fraction
          $\epsilon$ used for slowing-down density matrices
          determining the Lipkin-Nogami corrections, see
          Section~\ref{subsubsec:LN}. Variable \tv{SLOWTP} is the
          analogous mixing fraction used for the Lipkin parameter in
          the Lipkin translational-energy correction. Variable
          \tv{SLOWRP} is reserved for a similar role in future
          implementations of the Lipkin rotational energy correction.
          Variable \tv{SLOWLM} is a mixing fraction used for
          slowing-down density matrices determining the Lipkin
          center-of-mass or rotational corrections.

\key{LIPORDER} 0, 0 = \tv{ILIPON}, \tv{ILIPOP}

\keyspace

\noindent For \tv{ILIPON}$>$0 or \tv{ILIPOP}$>$0, the code performs
          calculations with the Lipkin VAPNP method for neutrons or
          protons, respectively. Values of \tv{ILIPON} and
          \tv{ILIPOP} give orders of the Lipkin operators. In version
          {(v\codeversion)}, only even orders 2, 4, and 6 (second,
          fourth, and sixth) are allowed The present implementation
          of the Lipkin VAPNP method requires conservation of the
          simplex (\tv{ISIMPY}=1) and time-reversal (\tv{IROTAT}=0)
          symmetries. \tv{ILIPON}$>$0 or \tv{ILIPOP}$>$0 requires
          \tv{LIPKIN}$=$0 and \tv{LIPKIP}$=$0, that is, the Lipkin-Nogami
          (Section VI-2.9) and Lipkin VAPNP methods cannot be used
          simultaneously.

\key{GAUGESHIFT} 0.123 = \tv{GAUSHI}

\keyspace

\noindent \tv{GAUSHI} gives the value of the maximum gauge angle
          $\phi_M$ of the Lipkin VAPNP method,
          Section~\ref{subsec:higher_order}. The Lipkin VAPNP methods
          for neutrons and protons share the same value of the
          maximum gauge angle. \tv{GAUSHI} must be larger than 0 and
          smaller than $2\pi$.

\key{GAUGEFRACT} -1 = \tv{MAXGAU}

\noindent For \tv{MAXGAU}>0 or \tv{MAXGAU}=0, the codes sets
          \tv{GAUSHI}=$2\pi$/\tv{MAXGAU} or \tv{GAUSHI}=$2\pi$/51,
          respectively, whereas values of \tv{MAXGAU}<0 are ignored.

\keyspace

\key{TRANSLMASS} 1., 1., 1. =
                          \tv{HBMRIN}(1), \tv{HBMRIN}(2), \tv{HBMRIN}(3)
\keyspace

\noindent For \tv{KETACM}=2, see Section~\ref{subsubsec:Previous},
the two-body center-of-mass correction is included before variation
for values of translational masses in three Cartesian directions $x$,
$y$, and $z$ that are scaled by factors \tv{HBMRIN}(1), \tv{HBMRIN}(2), and
\tv{HBMRIN}(3), respectively.

\key{TWOBODYLIN} 0 =
                        \tv{ITWOLI}
\keyspace

\noindent For \tv{KETACM}=2, see Section~\ref{subsubsec:Previous},
          the mean-field terms generated by the variation of the
          two-body center-of-mass correction break time-reversal,
          signature, and simplex symmetries. Therefore, for
          \tv{KETACM}=2 and \tv{ITWOLI}=1 the code stops unless these
          symmetries are broken, that is, unless \tv{IROTAT}=1,
          \tv{ISIMPY}=0, and \tv{ISIQTY}=0. However, usually these
          symmetry-breaking terms do not induce symmetry breaking on
          their own, and thus for self-consistent solutions their
          contributions vanish. Hence, value of \tv{ITWOLI}=0 (which
          is the default) allows for simply neglecting these terms
          and for performing calculations with the symmetries
          conserved, which requires much less CPU time. In addition,
          value of \tv{ITWOLI}=-1 allows for taking into account only
          those symmetry-breaking terms, which are compatible with
          selected conserved symmetries.

%%%%%%%%%%%%%%%%%%%%%%%%%%%%%%%%%%%%%%%%%%%%%%%%%%%%%%%%%%%%%%%%%%%%%%%%%%%%%%%%

\subsubsection{Configurations}
\label{subsubsec:Configurations}

\key{VACSIG\_NUC} 38, 38, 38, 38 =
                \begin{tabular}[t]{l}
                \tv{KVAMIG(0,0)}, \tv{KVAMIG(0,1)}, \\
                \tv{KVAMIG(1,0)}, \tv{KVAMIG(1,1)}
                \end{tabular}
\keyspace

\noindent Numbers of lowest p-n mixed nucleon states occupied in the four
          parity-signature blocks, denoted by $(+,+), (+,-), (-,+),$ and
          $(-,-)$, of given (parity, signature) combinations, i.e.,
          $(\pi, r)=(+1,+i), (+1,-i), (-1,+i)$, and $(-1,-i)$, respectively.
          These numbers define the parity-signature reference configuration
          from which the particle-hole excitations are counted. The
          definitions of parity-signature reference configuration and
          excitations are ignored unless \tv{IPNMIX}=1, \tv{ISIMPY}=1,
          \tv{ISIGNY}=1, and \tv{IPAIRI}=0.

\key{VACSIM\_NUC} 76, 76 =
                \tv{KVAMIM(0)}, \tv{KVAMIM(1)}
\keyspace

\noindent Numbers of lowest p-n mixed nucleon states occupied in the two
          simplex blocks, denoted by $(+)$ and $(-)$, of given simplexes,
          $s=+i$ and $s=-i$, respectively. These numbers define the simplex
          reference configuration from which the particle-hole excitations
          are counted. The definitions of simplex reference configuration and
          excitations are ignored unless \tv{IPNMIX}=1, \tv{ISIMPY}=1,
          \tv{ISIGNY}=0, and \tv{IPAIRI}=0.

\key{VACPAR\_NUC} 76, 76 =
                \tv{KVAMPA(0)}, \tv{KVAMPA(1)}
\keyspace

\noindent Numbers of lowest p-n mixed nucleon states occupied in the two
          parity blocks, denoted by $(+)$ and $(-)$, of given parities,
          $\pi=+1$ and $\pi=-1$, respectively. These numbers define the
          parity reference configuration from which the particle-hole
          excitations are counted. The definitions of parity reference
          configuration and excitations are ignored unless \tv{IPNMIX}=1,
          \tv{ISIMPY}=0, \tv{IPARTY}=1, and \tv{IPAIRI}=0.

\key{PHSIGN\_NUC} 1, 0, 0, 0, 0, 0, 0, 0, 0 =
                \tv{NUPAHO},
                \begin{tabular}[t]{l}
                \tv{LPPPSP}, \tv{LPPPSM}, \\
                \tv{LPPMSP}, \tv{LPPMSM}, \\
                \tv{LHPPSP}, \tv{LHPPSM}, \\
                \tv{LHPMSP}, \tv{LHPMSM},
                \end{tabular}
\keyspace

\noindent Nucleon particle-hole excitations in the parity-signature blocks for the
          p-n mixing calculation. Basic principles are the same as those for the
          excitations in the parity-signature blocks for no p-n mixing calculation,
          defined by the keywords \tv{PHSIGN\_NEU} and \tv{PHSIGN\_PRO}. \tv{NUPAHO}
          is the consecutive number from 1 to 5 (up to five sets of excitations can
          be specified in separate items). Particles are removed from the \tv{LHPPSP}-th
          state in the $(+,+)$ block, from the \tv{LHPPSM}-th state in the $(+,-)$
          block, from the \tv{LHPMSP}-th state in the $(-,+)$ block, and from the
          \tv{LHPMSM}-th state in the $(-,-)$ block, and put in the \tv{LPPPSP}-th
          state in the $(+,+)$ block, in the \tv{LPPPSM}-th state in the $(+,-)$ block,
          in the \tv{LPPMSP}-th state in the $(-,+)$ block, and in the \tv{LPPMSM}-th
          state in the $(-,-)$ block. These particle-hole excitations are ignored
          unless \tv{IPNMIX}=1, \tv{ISIMPY}=1, \tv{ISIGNY}=1, and \tv{IPAIRI}=0.

\key{PHSIMP\_NUC} 1, 0, 0, 0, 0 =
                \tv{NUPAHO},
                \tv{LPSIMP}, \tv{LPSIMM},
                \tv{LHSIMP}, \tv{LHSIMM},
\keyspace

\noindent Nucleon particle-hole excitations in the simplex blocks for the p-n mixing
          calculation. Basic principles are the same as those for the excitations in
          the simplex blocks for no p-n mixing calculation, defined by the keywords
          \tv{PHSIMP\_NEU} and \tv{PHSIMP\_PRO}. \tv{NUPAHO} is the consecutive number
          from 1 to 5 (up to five sets of excitations can be specified in separate
          items). Particles are removed from the \tv{LHSIMP}-th state in the $(+)$
          block, and from the \tv{LHSIMM}-th state in the $(-)$ block, and put in the
          \tv{LPSIMP}-th state in the $(+)$ block, and in the \tv{LPSIMM}-th state in
          the $(-)$ block. These particle-hole excitations are ignored unless
          \tv{IPNMIX}=1, \tv{ISIMPY}=1, \tv{ISIGNY}=0, and \tv{IPAIRI}=0.

\key{PHPARI\_NUC} 1, 0, 0, 0, 0 =
                \tv{NUPAHO},
                \tv{LPSIQP}, \tv{LPSIQM},
                \tv{LHSIQP}, \tv{LHSIQM},
\keyspace

\noindent Nucleon particle-hole excitations in the parity blocks for the p-n mixing
          calculation. Basic principles are the same as those for the excitations in
          the parity blocks for no p-n mixing calculation, defined by the keywords
          \tv{PHSIQP\_NEU} and \tv{PHSIQP\_PRO}. \tv{NUPAHO} is the consecutive number
          from 1 to 5 (up to five sets of excitations can be specified in separate
          items). Particles are removed from the \tv{LHSIQP}-th state in the $(+)$
          block, and from the \tv{LHSIQM}-th state in the $(-)$ block, and put in the
          \tv{LPSIQP}-th state in the $(+)$ block, and in the \tv{LPSIQM}-th state in
          the $(-)$ block. These particle-hole excitations are ignored unless
          \tv{IPNMIX}=1, \tv{ISIMPY}=0, \tv{IPARTY}=1, and \tv{IPAIRI}=0.

\key{PHNONE\_NUC} 1, 0, 0 =
                \tv{NUPAHO}, \tv{LPNONE}, \tv{LHNONE}
\keyspace

\noindent Nucleon particle-hole excitations for the p-n mixing calculation with no
          conserved simplex, parity, or parity symmetry. Basic principles are the
          same as those for the excitations for no p-n mixing calculation, defined
          by the keywords \tv{PHNONE\_NEU} and \tv{PHNONE\_PRO}. \tv{NUPAHO} is the
          consecutive number from 1 to 5 (up to five sets of excitations can be
          specified in separate items). Particles are removed from the \tv{LHNONE}-th
          state and put in the \tv{LPNONE}-th state. These particle-hole excitations
          are ignored unless \tv{IPNMIX}=1, \tv{ISIMPY}=0, \tv{IPARTY}=0, and
          \tv{IPAIRI}=0.

\key{DIASIG\_NUC} 2, 2, 2, 2, 1, 1, 1, 1, 0, 0, 0, 0 =
                \begin{tabular}[t]{l}
                \tv{KPMLIG(0,0)}, \tv{KPMLIG(0,1)}, \\
                \tv{KPMLIG(1,0)}, \tv{KPMLIG(1,1)}, \\
                \tv{KHMLIG(0,0)}, \tv{KHMLIG(0,1)}, \\
                \tv{KHMLIG(1,0)}, \tv{KHMLIG(1,1)}, \\
                \tv{KOMLIG(0,0)}, \tv{KOMLIG(0,1)}, \\
                \tv{KOMLIG(1,0)}, \tv{KOMLIG(1,1)},
                \end{tabular}
\keyspace

\noindent The diabatic blocking of p-n mixed single-particle parity-signature
          configurations. Matrices \tv{KPMLIG} contain the indices of particle
          states in the four parity-signature blocks denoted by $(+,+)$, $(+,-)$,
          $(-,+)$, and $(-,-)$, of given (parity, signature) combinations, i.e.,
          $(\pi, r)=(+1,+i)$, $(+1,-i)$, $(-1,+i)$, and $(-1,-i)$, respectively.
          Matrices \tv{KHMLIG} contain analogous indices of hole states. The
          type of blocking is defined by matrices \tv{KOMLIG} according to Table~III-6~\cite{[Dob00c]}.
          In addition, the following option is also
          available for the p-n mixing calculation:
\begin{align}
\tv{KOMLIG}=+11 &\Longleftrightarrow \mbox{The state which has the } \mbox{\it larger} \mbox{ isospin z-alignment is occupied.} \notag \\
\tv{KOMLIG}=-11 &\Longleftrightarrow \mbox{The state which has the } \mbox{\it smaller} \mbox{ isospin z-alignment is occupied.} \notag
\end{align}
\noindent

\key{DIASIM\_NUC} 2, 2, 1, 1, 0, 0 =
                \begin{tabular}[t]{l}
                \tv{KPMLIM(0)}, \tv{KPMLIM(1)}, \\
                \tv{KHMLIM(0)}, \tv{KHMLIM(1)}, \\
                \tv{KOMLIM(0)}, \tv{KOMLIM(1)},
                \end{tabular}
\keyspace

\noindent The diabatic blocking of p-n mixed single-particle simplex
          configurations. Matrices \tv{KPMLIM} contain the indices of
          particle states in the two simplex blocks denoted by $(+)$ and
          $(-)$, of given simplex values, i.e., $s=+i$, and $-i$,
          respectively. Matrices \tv{KHMLIM} contain analogous indices of
          hole states, and matrices \tv{KOMLIM} define the type of blocking
          in analogy to \tv{KOMLIG}.

\key{DIAPAR\_NUC} 2, 2, 1, 1, 0, 0 =
                \begin{tabular}[t]{l}
                \tv{KPMLIQ(0)}, \tv{KPMLIQ(1)}, \\
                \tv{KHMLIQ(0)}, \tv{KHMLIQ(1)}, \\
                \tv{KOMLIQ(0)}, \tv{KOMLIQ(1)},
                \end{tabular}
\keyspace

\noindent The diabatic blocking of p-n mixed single-particle parity configurations.
          Matrices \tv{KPMLIQ} contain the indices of particle states in the two
          parity blocks denoted by $(+)$ and $(-)$, of given parities, i.e., $\pi=+1$,
          and $-1$, respectively. Matrices \tv{KHMLIQ} contain analogous indices of
          hole states, and matrices \tv{KOMLIQ} define the type of blocking in analogy
          to \tv{KOMLIG}.

\key{DIANON\_NUC} 2, 1, 0 =
                \tv{KPMLIZ}, \tv{KHMLIZ}, \tv{KOMLIZ}
\keyspace

\noindent The diabatic blocking of p-n mixed single-particle configurations in
          the situation when all nucleons are in one common block. \tv{KPMLIZ}
          and \tv{KHMLIQ} contain the indices of a particle state and a hole state,
          respectively. \tv{KOMLIZ} defines the type of blocking in analogy to
          \tv{KOMLIG}.

\key{VACUUMCONF} 0 =
                         \tv{IVACUM}
\keyspace

\noindent For \tv{IVACUM}=1, the HF calculations are performed by occupying in
each iteration the states having the lowest single-particle energies
(the vacuum configuration), and the configuration data are ignored.
The user should be aware that code may then diverge if during the
iterations levels cross at the Fermi energy. \tv{IVACUM}=1 is ignored
unless \tv{IPAIRI}=0 and is not yet implemented for \tv{IPNMIX}=1.

%%%%%%%%%%%%%%%%%%%%%%%%%%%%%%%%%%%%%%%%%%%%%%%%%%%%%%%%%%%%%%%%%%%%%%%%%%%%%%%%

\subsubsection{Numerical parameters}
\label{subsubsec:Numerical}

\key{FREQBASIS} 1.0, 1.0, 1.0, 0 =
                \tv{HBARIX}, \tv{HBARIY}, \tv{HBARIZ}, \tv{INPOME}
\keyspace

\noindent For \tv{INPOME}=1, the frequencies of the HO basis are set as
$\hbar\omega_x$=\tv{HBARIX},
$\hbar\omega_y$=\tv{HBARIY}, and
$\hbar\omega_z$=\tv{HBARIZ}, and the standard rules of defining
these frequencies (see Section I-3.6) are ignored.

\key{BASISAUTOM} 0 =
                         \tv{IBASIS}
\keyspace

\noindent For \tv{IBASIS}=1, the code determines the spherical frequency
$\hbar\omega_0$ and axial deformation $\alpha_{20}$ of the HO basis according
to the following empirical formulas,
\begin{equation}
\hbar\omega_0 =
\left\{ \begin{array}{l}
8.1464\quad\mbox{for}\quad |Q_{20}|\leq30\,\mathrm{b}, \\
6.5 + 0.1 Q_{20} \exp(-0.02 Q_{20})\quad\mbox{for}\quad  |Q_{20}|>30\,\mathrm{b} ,
\end{array} \right.
\end{equation}
and $\alpha_{20} = 0.05\sqrt{Q_{20}}$, where $\hbar\omega_0$ is in
MeV and the axial component of quadrupole moment $Q_{20}$ is in b.
Frequencies $\hbar\omega_x$, $\hbar\omega_y$, and $\hbar\omega_z$
are extracted from $\hbar\omega_0$ and $\alpha_{20}$ following the
prescription given in Section II-2.

%%%%%%%%%%%%%%%%%%%%%%%%%%%%%%%%%%%%%%%%%%%%%%%%%%%%%%%%%%%%%%%%%%%%%%%%%%%%%%%%

\subsubsection{Constraints}
\label{subsubsec:Constraints}

\key{NECK\_CONST} 0, 1.0, 0.0 =
                          \tv{IFNECK}, \tv{Q0NECK}, \tv{G\_NECK}
\keyspace

\noindent For \tv{IFNECK}=1, the code uses the constraint on the
          number of particles in the neck. The expectation value
          $\langle \hat{Q}_{N}\rangle $ (\ref{eq:neck2}) of the Gaussian neck operator
          (\ref{eq:neck}) is constrained to the value of \tv{Q0NECK}.
          When \tv{IF\_RPA}=1 (keyword \tk{RPA\_CONSTR}), which is
          strongly recommended, the constraint on the neck is handled
          simultaneously with other possible constraints on the
          multipole moments by using the linear RPA method.
          \tv{G\_NECK} is then the value of the Lagrange multiplier.
          When \tv{IACONT}=1 (keyword \tk{CONTAUGMEN}), the value of
          the Lagrange multiplier is read from the record file.

\key{ISO\_FERMI} 0.0, 0.0, 0.0 =
                \tv{FERISO}(1), \tv{FERISO}(2), \tv{FERISO}(3)
\keyspace

\noindent
Values of isocranking frequencies $\lambda_1$=\tv{FERISO}(1),
$\lambda_2$=\tv{FERISO}(2), and $\lambda_3$=\tv{FERISO}(3) in
Eq.~(\ref{eq:isofrequency}). They are ignored unless \tv{IPNMIX}=1.
\tv{FERISO}(2) $\neq 0.0$ requires broken time-reversal symmetry, that is, \tv{ITIREV}=0
or \tv{IROTAT}=1.

\key{FERMI\_RTP} 0.0, 0.0, 0.0, 0.0 =
                \tv{FE\_RAD}, \tv{FE\_THE}, \tv{FE\_PHI}, \tv{FE\_OFF}
\keyspace

\noindent Values of parameters $\lambda^\prime$=\tv{FE\_RAD},
          $\theta^\prime$=\tv{FE\_THE}, $\phi$=\tv{FE\_PHI}, and
          $\lambda_{\rm off}$=\tv{FE\_OFF} in
          Eq.~(\ref{eq:isofrequency}). They are ignored unless
          \tv{IPNMIX}=1. Values of \tv{FE\_THE} and \tv{FE\_PHI} are
          in degrees. \tv{FE\_PHI} $\neq 0.0$ requires \tv{ITIREV}=0.
          Values of \tv{FE\_RAD}, \tv{FE\_THE}, \tv{FE\_PHI},
          and \tv{FE\_OFF} simply redefine values of \tv{FERISO(1)},
          \tv{FERISO(2)}, and \tv{FERISO(3)}, so keywords
          \tk{FERMI\_RTP} and \tk{ISO\_FERMI} can be used
          interchangeably.

\key{FERMI\_SCA} 0.0 =
                \tv{FERISO(0)}

\noindent Value of the isoscalar Fermi energy $\lambda_0$=\tv{FERISO}(0),
          which is introduced here for compatibility with future
          implementation of the p-n mixing with pairing.

\keyspace

\key{SPINLAGRAN}  0.0  0 =
                 \tv{DALSPI}(2),\tv{IFLALI}(2)

\keyspace

\noindent For \tv{IFLALI}(2)=1, the ALM corrections are included for the
simplex-conserving $y$ component of the angular momentum, whereby
\tv{DALSPI}(2) is the initial value of the corresponding ALM
correction $\delta\omega_y$ to angular frequency $\omega_y$,
see Section~\ref{subsec:alm}. \tv{IFLALI}(2)=1 requires \tv{IFLAGI}(2)=1,
see keyword \tk{SPINCONSTR} or \tk{SPICON\_XYZ}, where the target
value of the average angular momentum $\langle\hat{J}_y\rangle$
must be specified in \tv{ASKEDI}(2). \tv{IFLALI}(2)=1 cannot be used
together with the adjustable direction of the angular frequency \tv{IMOVAX}=1,
see keyword \tk{OMEGA\_TURN} and Section~IV-2.3~\cite{[Dob04]}.

 \key{SPINLA\_XYZ}  0.0  0 = \tv{DALSPI}(1),\tv{IFLALI}(1)
\keyp{SPINLA\_XYZ}  0.0  0 = \tv{DALSPI}(2),\tv{IFLALI}(2)
\keyp{SPINLA\_XYZ}  0.0  0 = \tv{DALSPI}(3),\tv{IFLALI}(3)

\keyspace

\noindent The three consecutive lines are the analogues of the line
described for \keywo{SPINLAGRAN}, and correspond to the components
$\mu=x$, $y$, and $z$ of the average angular momentum
$\langle\hat{J}_\mu\rangle$, respectively. To read values of
\tv{DALSPI}(2) and \tv{IFLALI}(2), keywords \tk{SPINLA\_XYZ} and
\tk{SPINLAGRAN} can be used interchangeably.

 \key{ISO\_CONSTR}  0.0  0.0  0 =  \tv{STIFFT}(1),\tv{ASKEDT}(1),\tv{IFLAGT}(1)
\keyp{ISO\_CONSTR}  0.0  0.0  0 =  \tv{STIFFT}(2),\tv{ASKEDT}(2),\tv{IFLAGT}(2)
\keyp{ISO\_CONSTR}  0.0  0.0  0 =  \tv{STIFFT}(3),\tv{ASKEDT}(3),\tv{IFLAGT}(3)

\keyspace

\noindent The three consecutive lines are the analogues of the lines
defining the 3D angular-momentum constraints, described for
\keywo{SPICON\_XYZ} in Section~IV-3.5~\cite{[Dob04]}, and correspond to the 3D
isospin constraints on the average isospin $\langle\hat{t}_k\rangle$,
see Section~\ref{subsec:alm}. For \tv{IFLAGT}=1, the quadratic
constraint on one of the Cartesian components of isospin is used
together with the linear constraint. Values of \tv{STIFFT} and
\tv{ASKEDT} correspond, respectively, to $C_k$ and $\bar{t}_k$ in
Eq.~(\ref{eq:alm}), where $k=1$, $2$, or $3$. For \tv{IFLAGT}=0,
there is no quadratic constraint on a given component.
\tv{IFLAGT}(1)=1, \tv{IFLAGT}(2)=1, or \tv{IFLAGT}(3)=1 requires p-n
mixing, that is, \tv{IPNMIX}=1. \tv{IFLAGT}(2)=1 requires broken
time-reversal symmetry, that is, \tv{ITIREV}=0 or \tv{IROTAT}=1.

 \key{ISO\_LAGRAN}  0.0  0 = \tv{DALISO}(1),\tv{IFLALT}(1)
\keyp{ISO\_LAGRAN}  0.0  0 = \tv{DALISO}(2),\tv{IFLALT}(2)
\keyp{ISO\_LAGRAN}  0.0  0 = \tv{DALISO}(3),\tv{IFLALT}(3)

\keyspace

\noindent The three consecutive lines are the analogues of the lines
defining the ALM for the 3D angular-momentum constraints, described
for \keywo{SPINLA\_XYZ} above, and correspond to the ALM for the 3D
isospin constraints on the average isospin $\langle\hat{t}_k\rangle$,
see Section~\ref{subsec:alm}. For \tv{IFLALT}(k)=1, the ALM
corrections are included for the $k$th component of the isospin,
whereby \tv{DALISO}(k) is the initial value of the corresponding ALM
correction $\delta\lambda_k$ to isocranking frequency $\lambda_k$, see
Section~\ref{subsec:alm}. \tv{IFLALT}(k)=1 requires \tv{IFLAGT}(k)=1,
see keyword \tk{ISO\_CONSTR}, where the target value of the average
isospin $\langle\hat{t}_k\rangle$ must be specified in
\tv{ASKEDT}(2).

\keyspace

%%%%%%%%%%%%%%%%%%%%%%%%%%%%%%%%%%%%%%%%%%%%%%%%%%%%%%%%%%%%%%%%%%%%%%%%%%%%%%%%

\subsubsection{Output parameters}
\label{subsubsec:Output}

\key{MASSFRAGME} 0 =
                         \tv{IFRAGM}
\keyspace

\noindent For \tv{IFRAGM}=1, \ti{fission fragment} properties are computed at
          the last iteration. The code determines the position of the
          \ti{neck}, that is, the point near the center of the nucleus where the
          density is the lowest. The neck divides the compound nucleus in two
          parts: the occupation of each quasi-particle $\mu$ is computed for
          both fragments, $N_{1,\mu}$ and $N_{2,\mu}$ (with the total
          occupation $N_{\mu} = N_{1,\mu} + N_{2,\mu}$). If
          $N_{1,\mu} \geq 0.5N_{\mu}$, then the q.p.\ $\mu$ is assigned to
          fragment (1), else to fragment (2). The set of all q.p.\ is thus
          divided in two subsets corresponding to the two fragments. These
          subsets define, in turn, the \ti{local densities}
          $\rho_{1}(\boldsymbol{r})$ and $\rho_{1}(\boldsymbol{r})$ of each
          fragment, which allow to compute the total charge, mass, Skyrme
          energy, \tii{Coulomb energy}{(fission fragments)},
          \tii{pairing energy}{(fission fragments)} and
          \tii{multipole moment}{(fission fragments)} expectation values within
          each fragment. The \ti{nuclear interaction energy} and \ti{Coulomb
          energy}{(fission fragments)} between the fragments is also computed.
          A table that gives \ti{localization indicator} $\ell_{\mu}$ for
          each q.p.\ is also printed in the output.

\key{QPROTATION} 0 =
                         \tv{MIN\_QP}, \tv{DELTAE}, \tv{XLOCMX}, \tv{V2\_MIN}, \tv{ITRMAX}, \tv{NTHETA}
\keyspace

\noindent For \tv{MIN\_QP}=1, pairs of quasiparticles are rotated in order
          to minimize the tail of the density in each of the two fission
          fragments. All pairs of quasiparticles with
          $|\Delta E| = |E_{\mu} - E_{\nu}| \leq \Delta$, localization
          $\ell_{\mu}, \ell_{\nu} \leq \ell_{\mathrm{max}}$ and occupation
          $N_{\mu}, N_{\nu} \geq N_{\mathrm{min}}$ are considered for rotation.
          The quantities $\Delta$, $\ell_{\mathrm{max}}$, and $N_{\mathrm{min}}$
          are \tv{DELTAE}, \tv{XLOCMX}, \tv{V2\_MIN}, respectively. The localization
          can be performed several times successively, with \tv{ITERMAX} the number
          of iterations. At each iteration of the rotation method, the code searches
          over \tv{NTHETA} value of the rotation angle $\theta_{\mu\nu}$.

\key{PRINT\_SYME} 0 = \tv{ISYMDE}

\keyspace

\noindent For \tv{ISYMDE}=1, the code determines and prints tables
of spatial symmetries of particle-hole and particle-particle densities.
\tv{ISYMDE}=1 cannot be used together with projection, that is,
with \tv{IPRGCM}>0, see \keywo{PROJECTGCM} in Section~VI-3.2~\cite{[Dob09d]}.

%%%%%%%%%%%%%%%%%%%%%%%%%%%%%%%%%%%%%%%%%%%%%%%%%%%%%%%%%%%%%%%%%%%%%%%%%%%%%%%%

\subsubsection{Starting and restarting the iteration}
\label{subsubsec:Starting}

\key{CONTCMCORR} 0 =
                 \tv{IMCONT}
\keyspace

\noindent For \tv{IMCONT}=1, average values of linear momenta, which are stored in the \pr{hfodd}
replay file, see Section~II-3.9~\cite{[Dob97d]}, will be used in the first iteration of calculations that
include two-body center-of-mass correction before variation, see
Section~\ref{subsubsec:Previous}; otherwise these average values are
put to zero. Using the stored values allows for a smooth continuation
of iterations. For \tv{IMCONT}=1 and \tv{IRENMA}>0, also the stored
values of the renormalized translational masses, see
\keywo{RENORMASS} in Section~VII-3.1~\cite{[Sch12]}, will be used. \tv{IMCONT}=1
requires switching on the center-of-mass correction before variation,
that is, \tv{KETACM}=2. \tv{IMCONT}=1 requires continuation of iterations,
that is, \tv{ICONTI}=1, see Section~II-3.10~\cite{[Dob97d]}.

\key{CONTAUGSPI} 0 =
                 \tv{ISCONT}
\keyspace

\noindent For \tv{ISCONT}=1, the ALM corrections $\delta\omega_\mu$
to angular frequencies $\omega_\mu$, see Section~\ref{subsec:alm},
which are stored in the \pr{hfodd} replay file, see Section~II-3.9~\cite{[Dob97d]},
will be used in the first iteration; otherwise the values read with
keywords \tk{SPINLA\_XYZ} or \tk{SPINLAGRAN} are used. Using the stored values allows for
a smooth continuation of iterations.
\tv{ISCONT}=1 requires switching on the ALM for angular momentum,
that is, \tv{IFLALI}(1)=1, \tv{IFLALI}(2)=1, or \tv{IFLALI}(3)=1.
\tv{ISCONT}=1 requires continuation of iterations,
that is, \tv{ICONTI}=1, see Section~II-3.10~\cite{[Dob97d]}.

\key{CONTAUGISO} 0 =
                 \tv{ITCONT}
\keyspace

\noindent For \tv{ITCONT}=1, the ALM corrections $\delta\lambda_k$ to
isocranking frequencies $\lambda_k$, see Section~\ref{subsec:alm},
which are stored in the \pr{hfodd} replay file, see Section~II-3.9~\cite{[Dob97d]},
will be used in the first iteration; otherwise the values read with
keyword \tk{ISO\_LAGRAN} are used. Using the
stored values allows for a smooth continuation of iterations.
\tv{ITCONT}=1 requires switching on the ALM for isospin,
that is, \tv{IFLALT}(1)=1, \tv{IFLALT}(2)=1, or \tv{IFLALT}(3)=1.
\tv{ITCONT}=1 requires continuation of iterations, that is,
\tv{ICONTI}=1, see Section~II-3.10~\cite{[Dob97d]}.

%%%%%%%%%%%%%%%%%%%%%%%%%%%%%%%%%%%%%%%%%%%%%%%%%%%%%%%%%%%%%%%%%%%%%%%%%%%%%%%%

\subsubsection{New features of previous items}
\label{subsubsec:Previous}

\key{SKYRME-STD} 0, 1, 0, 0, 0 = \tv{ISTAND},\tv{KETA\_J},\tv{KETA\_W},\tv{KETACM},\tv{KETA\_M}

\keyspace

\noindent In version {(v\codeversion)}, the two-body center-of-mass
          correction before variation was fully implemented both in
          particle-hole and particle-particle channel. Value of
          \tv{KETACM}=2 is now thus allowed, see Section~IV-3.1~\cite{[Dob04]}, and
          it has also be encoded for the standard Skyrme forces that
          use this option.

\key{TWOBASIS} 0 = \tv{ITWOBA}

\keyspace

\noindent In version {(v\codeversion)}, the two-basis method, see
          Sections~VII-2.2.1 and VII-3.2~\cite{[Sch12]}, has been implemented for
          all symmetries, and thus \tv{ITWOBA}=1 does not any more
          requires \tv{ISIMPY}=0 and \tv{IPARTY}=0.

\key{SKYRME-SET} SKM* = \tv{SKYRME}

\keyspace

\noindent In addition to acronyms of Skyrme forces listed in Section~IV-3.1~\cite{[Dob04]},
in version {(v\codeversion)}, the following ones were added:
SII, SVI, SKI2, SKI3, SKI4, SKI5, SLY6, SLY7, UDF0, UDF1, UDF2, SAMi, and SD1S.

\key{ONE\_LINE} 1 = \tv{I1LINE}

\keyspace

\noindent In addition to values of \tv{I1LINE}=1 or 2, described
in Section~VI-3.6~\cite{[Dob09d]}, for \tv{I1LINE}=3 the code prints values of
entropy, neutron and proton Fermi energies, and neutron and proton pairing energies;
and for \tv{I1LINE}=4, 5, or 6 it prints values of components $k=1$, 2, or 3,
respectively, of the total isospin and isocranking frequency.

\key{RENORMASS} 0, 0.0, 0.0, 0.0 =
                    \tv{IRENMA}, \tv{DISTAX}, \tv{DISTAY}, \tv{DISTAZ}

\keyspace

\noindent As discussed in Section~\ref{subsec:Lipkin_translational},
for \tv{IRENMA}>0, the Lipkin translational corrections introduced in
Section~VII-3.1~\cite{[Sch12]} can now be performed for paired
states. However, for the q.p.\ blocking in odd nuclei, see
Section~VI-2.7~\cite{[Dob09d]}, these corrections have not yet been
implemented.

%%%%%%%%%%%%%%%%%%%%%%%%%%%%%%%%%%%%%%%%%%%%%%%%%%%%%%%%%%%%%%%%%%%%%%%%%%%%%%%%

\subsubsection{Obsolete items}
\label{subsubsec:Obsolete}

In version {(v\codeversion)}, for options involving
calculations with the Yukawa forces or for the Lipkin-Nogami method,
the code can be restarted using matrix elements of the mean field
saved in the field file, see \keywo{FIELD\_SAVE} in Section~VI-3.7~\cite{[Dob09d]}.
Therefore, keywords pertaining to saving files specific to these
options have become obsolete, although they are still active and
allowed. This concerns keywords \tk{YUKAWASAVE}, \tk{LIPKINSAVE},
\tk{REPYUKFILE}, \tk{RECYUKFILE}, \tk{REPLIPFILE}, \tk{RECLIPFILE},
and \tk{CONTYUKAWA}. For \tv{IFCONT}=1 (\keywo{CONTFIELDS}), see
Section~VI-3.8~\cite{[Dob09d]}, when the code is restarted, the Yukawa or
Lipkin-Nogami calculations can be automatically smoothly continued,
and the contents of the Yukawa and Lipkin files is ignored. Then, for
\tv{ILCONT}=0 or 1 (\keywo{CONTLIPKIN}), the Lipkin parameters are
read from the input or record file, respectively. For \tv{IFCONT}=0,
by using the Yukawa or Lipkin files, a smooth continuation of the
Yukawa or Lipkin-Nogami calculations, respectively, is still
possible, but this option is not any more recommended.

%%%%%%%%%%%%%%%%%%%%%%%%%%%%%%%%%%%%%%%%%%%%%%%%%%%%%%%%%%%%%%%%%%%%%%%%%%%%%%%%
%%%%%%%%%%%%%%%%%%%%%%%%%%%%%%%%%%%%%%%%%%%%%%%%%%%%%%%%%%%%%%%%%%%%%%%%%%%%%%%%

\subsection{Input data for parallel mode}
\label{subsec:parallel_data}

%%%%%%%%%%%%%%%%%%%%%%%%%%%%%%%%%%%%%%%%%%%%%%%%%%%%%%%%%%%%%%%%%%%%%%%%%%%%%%%%

\setcounter{mysubsubsection}{0}

\subsubsection{Updated list of active keywords in hfodd.d}
\label{subsubsec:hfodd.d}

In parallel mode, the code \pr{hfodd} in version {(v\codeversion)} reads all
user-defined sequential data from the input file named \tv{hfodd.d}. Since
version (v2.49t), a few additional \pr{hfodd} options have been activated in
parallel mode. The updated subset of \pr{hfodd} keywords that can be activated
is given below:
\begin{itemize}
\item {\bf General data and iterations - }
\tv{ITERATIONS}, \tv{BROYDEN}, \tv{SLOW\_DOWN},
\tv{SLOW\_PAIR}, \newline\tv{SLOWLIPKIN}, \tv{ITERAT\_EPS},
\tv{MAXANTIOSC}, \tv{PING\_PONG}, \tv{CHAOTIC},
\item {\bf Specific features - }
\tv{FINITETEMP}, \tv{SHELLCORCT}, \tv{SHELLPARAM},
\tv{HFBTHOISON}, \tv{COULOMBPAR}, \tv{MASSFRAGME},
\item {\bf Interaction - }
\tv{UNEDF\_PROJ},\tv{SKYRME-SET}, \tv{SKYRME\_STD},
\item {\bf Pairing - }
\tv{PAIRING}, \tv{HFB}, \tv{CUTOFF}, \tv{BCS},
\tv{HFBMEANFLD}, \tv{LIPKIN}, \tv{PAIR\_INTER},
\tv{PAIRNINTER}, \tv{PAIRPINTER},
\item {\bf Symmetries - }
\tv{SIMPLEXY}, \tv{SIGNATUREY}, \tv{PARITY},
\tv{ROTATION}, \tv{TSIMPLEX3D},
\item {\bf Parameters of the HO basis - }
\tv{BASISAUTOM}, \tv{BASIS\_SIZE}, \tv{HOMEGAZERO},
\tv{OPTI\_GAUSS}, \tv{GAUSHERMIT}, \tv{SURFAC\_DEF},
\item {\bf Multipole moments - }
\tv{RPA\_CONSTR},\tv{MAX\_MULTIP}
\item {\bf Angular momentum - }
\tv{OMEGAY},
\item {\bf Output-file  parameters - }
\tv{ONE\_LINE}, \tv{NILSSONLAB}, \tv{BOHR\_BETAS},
\item {\bf I/O Flags - }
\tv{FIELD\_SAVE}, \tv{REVIEW},
\item {\bf Starting the iteration - }
\tv{RESTART}, \tv{CONT\_PAIRI}, \tv{CONTLIPKIN},
\tv{CONTFIELDS}, \tv{CONTAUGMEN}, \tv{EXECUTE}.
\end{itemize}

In principle, these options provide enough flexibility to cover the majority of
\pr{hfodd} applications in parallel mode. The user interested in some specific
option which could not be activated by one of the keywords above can still
manually modify the routine \tv{PREDEF} prior to compilation. This routine
pre-defines all \pr{hfodd} input data.

%%%%%%%%%%%%%%%%%%%%%%%%%%%%%%%%%%%%%%%%%%%%%%%%%%%%%%%%%%%%%%%%%%%%%%%%%%%%%%%%

\subsubsection{Structure of hfodd\_mpiio.d}
\label{subsubsec:hfodd_mpiio.d}

In version 2.49t, the keyword \tv{CONSTR\_LIN} was used to activate the
readjustment of the constraints using the RPA method. This keyword is not
active anymore: the user should use the keyword \tv{RPA\_CONSTR} in
\tv{hfodd.d} instead.

\key{CALCULMODE} 1, 0 =
                                \tv{mpidef}, \tv{mpibas}
\keyspace

\noindent Compared to version (v2.49t), there are now 3 more possible values for
          the variable \tv{mpidef}, which provide additional flexibility when
          running large-scale parallel calculations of potential energy surfaces.
          In the cases of \tv{mpidef}=2 and \tv{mpidef}=4, the program assumes that
          \begin{itemize}
          \item In the
          sub-directory {\tt ./restart} of the directory where the program is
          executed, there exists a set of valid \pr{hfodd} record files. \newline Names of
          these files must be of the type \tf{HFODD\_XXXXXXXX.REC},
          where {\tt XXXXXXXX} is a 8-digit integer number (padded with 0 if needed).
          \item There also exists sub-directory {\tt ./rec}, where the new
          record files would are written.
          \item If the Lipkin-Nogami prescription is requested, e.g., when UNEDF
          functionals are used, there should also exist sub-directory
          {\tt ./lic}, which would contain Lipkin files.
          \end{itemize}
          These various sub-directories must be created by the user, or the
          code would fail to execute properly.\\

\noindent Providing the proper directory structure has been set up as described
          above, the different options to use the \pr{hfodd} record files for
          restart in a new parallel run are the following:
          \begin{itemize}
          \item For \tv{mpidef=2}, the list of record files in {\tt ./restart}
          must correspond {\it exactly} to the deformation grid defined under
          keyword \tk{MULTICONST}. The actual deformation grid for the run is
          then defined under keyword \tk{MULTIRESTA}. The new grid cannot be
          smaller than the old one.
          \item For \tv{mpidef=3}, the program computes an arbitrary path in
          a user-defined collective space. Calculations are initialized from
          scratch. This mode requires a file named \tf{hfodd\_path.d} containing
          a list of points that define the path. The first line contains two
          integers, the number of different constraints $k$, and the total number
          of points in the path $n$. Each line $i$ then has the structure
          $$
          \lambda^{(i)}_{1} \ \ \mu^{(n)}_{1}\ \ Q^{(i)}_{\lambda\mu,1}\ \
          \lambda^{(i)}_{2} \ \ \mu^{(n)}_{2}\ \ Q^{(i)}_{\lambda\mu,2}\ \ \dots
          \lambda^{(i)}_{k} \ \ \mu^{(n)}_{k}\ \ Q^{(i)}_{\lambda\mu,k}
          $$
          For example, the following content of \tf{hfodd\_path.d}
          \begin{verbatim}
          3 2
          1 0   0.000 2 0  300.000   3 0   0.000
          1 0   0.000 2 0  315.000   3 0   0.000
          \end{verbatim}
          defines a list of 2 points in the collective space, with
          $\langle \hat{Q}_{10} \rangle = \langle \hat{Q}_{30} \rangle = 0$,
          and $\langle \hat{Q}_{20} \rangle$ taking the values from 300.0 b and
          315.0 b. By convention, the multipolarity $\lambda = 0$, $\mu=0$
          corresponds to constraints on the size of the neck.
          \item For \tv{mpidef=4}, the code also computes an arbitrary path
          in the collective space. However, this path is now defined in the file
          \tf{hfodd\_path\_new.d}, which has the same structure as described above.
          In addition, calculations in this mode are initialized from existing,
          valid, \pr{hfodd} record files that correspond to another, pre-calculated path.
          This initial path is defined (still using the same conventions as before)
          in the file \tf{hfodd\_path.d}. If the initial path has $n$ points, the
          program assumes that the directory {\tt restart/} contains $n$ valid \pr{hfodd}
          record files, numbered from 1 to $n$ in the format \tf{HFODD\_XXXXXXXX.REC}
          which correspond exactly to the $n$ points defined in the file \tf{hfodd\_path.d}.
          For example, suppose the file \tf{hfodd\_path.d} has the structure
          \begin{verbatim}
          3 2
          1 0   0.000 2 0  300.000   3 0   5.000
          1 0   0.000 2 0  315.000   3 0   0.000
          \end{verbatim}
          and the file \tf{hfodd\_path\_new.d} is
          \begin{verbatim}
          3 1
          1 0   0.000 2 0  300.000   3 0   0.000
          \end{verbatim}
          In this case, the program assumes that the directory {\tt restart/} contains two
          files named \tf{HFODD\_00000001.REC} and \tf{HFODD\_00000002.REC}, the
          first one associated to the point $\langle \hat{Q}_{20} \rangle$ = 300\,b,
          $\langle \hat{Q}_{30} \rangle$ = 5\,b$^{3/2}$, the second one to the point
          $\langle \hat{Q}_{20} \rangle$ = 315\,b, $\langle \hat{Q}_{30} \rangle$ = 0.
          In this particular example, the point with constraints
          $\langle \hat{Q}_{20} \rangle$ = 300\,b and $\langle \hat{Q}_{30} \rangle$ = 0
          will be computed by restarting from the file \tf{HFODD\_00000001.REC}.
          [For each point in the new collective path, the program will automatically
          determines which point in the old path is the closest, and will use the
          corresponding \pr{hfodd} record file as restart file.]
          \end{itemize}

\noindent To summarize: \tv{mpidef}=2 uses a regular grid as restart points
          to compute a new regular grid; \tv{mpidef}=3 computes an irregular grid
          from scratch; \tv{mpidef}=4 uses an irregular grid to compute an irregular
          grid. Refer to the examples included with the program for more information.

\key{ALL\_FORCES} 1, SKM* =
                                \tv{numero}, \tv{skyrme}
\keyspace

\noindent This keyword (i) overwrites the definition of the Skyrme force in the
          \tf{hfodd.d} file containing process-independent input data and (ii)
          allows to define a parallel job containing calculations with several
          different Skyrme functionals. To run a calculation with a single
          Skyrme functional, set \tv{numero=1} and \tv{skyrme} to one of the
          keys for Skyrme functionals; to run a calculation with more than one
          Skyrme functional, add one line per functional, all but the last line
          having a negative value for \tv{numero} (same convention as for
          multipole moments)

\key{BATCH\_MODE} 0, 1 =
                                \tv{ibatch}, \tv{nbatch}
\keyspace

\key{BATCH\_SPEC} 3, 0 =
                                \tv{lambda}, \tv{miu}
\keyspace

\noindent In batch mode, the code will attempt to converge to a given solution
          by successive steps. Let us assume a target value
          $\bar{Q}_{\lambda\mu} \equiv \langle \hat{Q}_{\lambda\mu} \rangle$
          for the multipolarity $\lambda, \mu$, and a requested number of
          iterations equal to $N$. The code will first divide the interval
          $[0, \bar{Q}_{\lambda\mu}]$ in $p$ segments. The first $N/p$
          iterations will have the target value $\bar{Q}_{\lambda\mu}/N$, and
          will start from scratch; the iterations $kN/p$ to $(k+1)N/p$ will
          have the target value $(k+1)\bar{Q}_{\lambda\mu}/N$ and will restart
          from the solution obtained for the previous target value
          $k\bar{Q}_{\lambda\mu}/N$. This multi-step process allows to converge
          to a solution that is quite distant from the initial point without
          crashing the iteration process. In practice, \tv{ibatch=1} activates
          the batch mode, and \tv{nbatch}$=p$. The batch mode is compatible
          with multiple constraints, but only one of them, defined by
          \tv{lambda} and \tv{miu}, will be segmented as described.

\key{MULTIRESTA} 2, 0, 10.0, 10.0, 4 =
                                \tv{lambda\_res}, \tv{miu\_res}, \tv{qBegin\_res}, \tv{qFin\_res}, \tv{numberQ\_res}
\keyspace

\noindent This keyword is only active when \tv{mpidef=2}. It
          defines the new regular deformation grid that will be computed in
          this run.

\key{BASIS-NSTA} 165, 2, 1 =
                                \tv{nsmini}, \tv{nsstep}, \tv{nofsta}
\keyspace

\noindent For \tv{mpibas}=1, the total number of states in the HO basis,
          $N_{\text{states}}$, can take different values of the form
          \begin{equation}
          N_{\text{states}}(k) = N_{\text{states}}(0)
          + (k-1)\delta N_{\text{states}}, \ \ \  k = 1, \dots, N_{S}.
          \end{equation}
          Then, $N_{\text{states}}(0) :=$ \tv{nsmini},
          $\delta N_{\text{states}} :=$ \tv{nsstep}, $N_{S} :=$ \tv{nofsta}.

%%%%%%%%%%%%%%%%%%%%%%%%%%%%%%%%%%%%%%%%%%%%%%%%%%%%%%%%%%%%%%%%%%%%%%%%%%%%%%%%
%%%%%%%%%%%%%%%%%%%%%%%%%%%%%%%%%%%%%%%%%%%%%%%%%%%%%%%%%%%%%%%%%%%%%%%%%%%%%%%%
%%%%%%%%%%%%%%%%%%%%%%%%%%%%%%%%%%%%%%%%%%%%%%%%%%%%%%%%%%%%%%%%%%%%%%%%%%%%%%%%
%%%%%%%%%%%%%%%%%%%%%%%%%%%%%%%%%%%%%%%%%%%%%%%%%%%%%%%%%%%%%%%%%%%%%%%%%%%%%%%%

\section{Fortran Source Files}
\label{sec:source_files}

The \ti{FORTRAN source} of version {(v\codeversion)} of the code \pr{hfodd}
is provided in the file \tf{hfodd.f} and its accompanying modules
\begin{itemize}
\item \tf{hfodd\_fission\_{\fissionversion}.f90}: Toolkit for fission
calculations. Contains several routines to compute fission fragments properties
such as charge, mass, total energy, interaction energy; the routines needed to
use a constraint on the number of particles in the neck; the routines used for
the quantum localization method.
\item \tf{hfodd\_fits\_{\fitsversion}.f90}: Fit module. Allows the code
\pr{hfodd} to work as a routine in an external program.
\item \tf{hfodd\_functional\_{\functionalversion}.f90}: Interface to UNEDF
functionals.
\item \tf{hfodd\_hfbtho\_{\hfbthoversion}.f90}: \pr{hfbtho} DFT solver based
on version 200d published in Ref. \cite{[Sto12]}.
\item \tf{hfodd\_interface\_{\interfaceversion}.f90}: Interface between the
\pr{hfbtho} and \pr{hfodd} solvers. Contains the routine to transform the HFB
matrix from the HO basis used in \pr{hfbtho} (\pr{hfodd}) to the basis used
in \pr{hfodd} (\pr{hfbtho}).
\item \tf{hfodd\_lipcorr\_{\lipkinversion}.f90}: Toolkit for the Lipkin method.
\item \tf{hfodd\_modules\_{\modulesversion}.f}: Definition of memory-consuming
modules. Defines, among others, the matrices of the Bogoliubov transformation,
the eigenvectors of the HF equations, etc.
\item \tf{hfodd\_mpiio\_{\mpiioversion}.f90}: IO interface in MPI calculations.
Contains the routine to read input data for parallel \pr{hfodd} calculations.
\item \tf{hfodd\_mpimanager\_{\mpimanagerversion}.f90}: MPI toolkit. Defines
the list of MPI tasks based on the data read in the parallel input file
\tf{hfodd\_mpiio.d}.
\item \tf{hfodd\_pnp\_{\pnpversion}.f90}: Toolkit for particle number projection.
\item \tf{hfodd\_shell\_{\shellversion}.f}: Toolkit for the shell correction.
\item \tf{hfodd\_sizes\_{\sizeversion}.f90}: Static array size declaration.
Contains all {\tt PARAMETER} statements controlling the size of all the
statically (and some of the dynamically allocated) arrays used in the code.
\item \tf{hfodd\_SLsiz\_{\scalapackversion}.f90}: ScaLAPACK interface.
Experimental module testing the ScaLAPACK diagonalization routine for
simplex-breaking HFB calculations.
\end{itemize}

The \ti{FORTRAN source} of version {(v\codeversion)} of the code
\pr{hfodd} contains numerous undocumented and untested features that
are under development. The user should not attempt to activate or
inverse-engineer these features, because this can certainly lead
to an unpredictable behaviour of the code, and even to a damage to
computer hard drive.

%%%%%%%%%%%%%%%%%%%%%%%%%%%%%%%%%%%%%%%%%%%%%%%%%%%%%%%%%%%%%%%%%%%%%%%%%%%%%%%
%%%%%%%%%%%%%%%%%%%%%%%%%%%%%%%%%%%%%%%%%%%%%%%%%%%%%%%%%%%%%%%%%%%%%%%%%%%%%%%%

\subsection{Standard Libraries}
\label{subsec:libraries}

The code \pr{hfodd} requires an implementation of the BLAS and LAPACK libraries
to function correctly, see Section~V-5.2~\cite{[Dob05]} for details. In this
version, the interface to older NAGLIB routines has been discontinued.

%%%%%%%%%%%%%%%%%%%%%%%%%%%%%%%%%%%%%%%%%%%%%%%%%%%%%%%%%%%%%%%%%%%%%%%%%%%%%%%%
%%%%%%%%%%%%%%%%%%%%%%%%%%%%%%%%%%%%%%%%%%%%%%%%%%%%%%%%%%%%%%%%%%%%%%%%%%%%%%%%

\subsection{Parallel Mode}
\label{subsec:parallel_mode}

We recall that a parallel machine is made of a certain number of sockets, each
containing one processor. Every processor contains a number of CPU units, or
cores, sharing the same memory.

%%%%%%%%%%%%%%%%%%%%%%%%%%%%%%%%%%%%%%%%%%%%%%%%%%%%%%%%%%%%%%%%%%%%%%%%%%%%%%%%

\subsubsection{Basic MPI}

To activate multi-core calculations, \pr{hfodd} requires an implementation of
the Message Passing Interface (MPI). The current version was tested on two
different implementations:
\begin{itemize}
\item MPICH-1 and MPICH-2, available at:

http://www.mcs.anl.gov/research/projects/mpich2/
\item Open MPI available at: http://www.open-mpi.org/
\end{itemize}
In parallel mode, the code \pr{hfodd} is compiled by setting \tv{USE\_MPI} to
1 in the project Makefile. Typically, the executable is run as follows
(\tv{bash} syntax):\\

\tv{mpiexec -np [number of processes] hf{\execversion} < /dev/null >\& hf{\execversion}.out}\\

\noindent where \tv{hf{\execversion}.out} is a redirection file for the output and
files \tv{hfodd.d} and \tv{hfodd\_mpiio.d} must be in the directory where this
command is run. The code will automatically generate output files with the
names \tf{hfodd\_XXXXXX.out}, where {\tt XXXXXX} is between 1 and the number
$N$ of MPI processes requested. Note that the naming convention is slightly
different from version (2.49t) where the filenames were numbered from 0 to $N-1$.

%%%%%%%%%%%%%%%%%%%%%%%%%%%%%%%%%%%%%%%%%%%%%%%%%%%%%%%%%%%%%%%%%%%%%%%%%%%%%%%%

\subsubsection{Basic Hybrid OpenMP/MPI Mode}

Multi-threading is activated by switching the \tv{USE\_OPENMP} to 1 in the
project Makefile. This option can be used on its own, or in combination with
\tv{USE\_MPI}=1, in which case the programming model is hybrid MPI/OpenMP. We
recall that to activate multi-threading, the environment variable
\tv{OMP\_NUM\_THREADS} must be set to the required number of threads prior to
execution. If every processor has 6 cores, then to run 12 MPI processes with 3
threads each, the following command line (in the OPENMPI implementation) should
be executed:\\

\tv{export OMP\_NUM\_THREADS = 3}\\

\tv{mpiexec -np 12 -npersocket 2 hf{\execversion} < /dev/null >\& hf{\execversion}.out}\\

\noindent Therefore, instead of the 12 MPI processes being executed by all the
12 cores of 2 full processors, the \tv{-npersocket 2} option imposes that only
2 cores within a given socket are actually used, leaving the remaining 4
available when multi-threading kicks in. Such an instruction requires 6
processors instead of 2 in the pure MPI mode, and up to (12 processes)$\times$
(3 threads) = 36 cores may be active at a given time .

%%%%%%%%%%%%%%%%%%%%%%%%%%%%%%%%%%%%%%%%%%%%%%%%%%%%%%%%%%%%%%%%%%%%%%%%%%%%%%%%

\subsubsection{Advanced Hybrid OpenMP/MPI Mode}

In advanced MPI mode, each HFB calculation is spread across several MPI
processes, see section \ref{subsec:parallel} above. This option is activated
by setting the {\tt Makefile}| variable {\tt USE\_MANYCORES} to 1. In addition,
the user must specify {\it at compile time} the number of MPI per HFB
calculation. This is done by setting the variables {\tt M\_GRID} and
{\tt N\_GRID}. The total number of MPI processes per HFB calculation
$N_{\mathrm{p}}$ is $N_{\mathrm{p}} = \mathtt{N\_GRID}\times\mathtt{M\_GRID}$.
The calling sequence of the program is the same as in the basic MPI mode.
However, file handling will be different: the output and record files will
embed in their name both the job number (corresponding to one of the $N$
requested HFB calculations) and the MPI process number within a group. For
example, the output file will take the form \tf{procXXX\_hfoddXXXXXX.out},
where {\tt XXX} is between 1 and $N_{\mathrm{p}}$ and {\tt XXXXXX} between
1 and $N$. Similar file naming convention holds for the record file, the
Lipkin files and the \pr{hfbtho} file.

%%%%%%%%%%%%%%%%%%%%%%%%%%%%%%%%%%%%%%%%%%%%%%%%%%%%%%%%%%%%%%%%%%%%%%%%%%%%%%%%
%%%%%%%%%%%%%%%%%%%%%%%%%%%%%%%%%%%%%%%%%%%%%%%%%%%%%%%%%%%%%%%%%%%%%%%%%%%%%%%%
%%%%%%%%%%%%%%%%%%%%%%%%%%%%%%%%%%%%%%%%%%%%%%%%%%%%%%%%%%%%%%%%%%%%%%%%%%%%%%%%
%%%%%%%%%%%%%%%%%%%%%%%%%%%%%%%%%%%%%%%%%%%%%%%%%%%%%%%%%%%%%%%%%%%%%%%%%%%%%%%%

\section{Acknowledgments}
\label{sec:acknowledgments}

\bigskip
Discussions with Hai Ah Nam and Robert Parrish are warmly
acknowledged. X.B.\ Wang wishes to thank the National Natural Science
Foundation of China for support under Grant Nos.\ 11505056 and
11605054. This work was supported in part by the THEXO JRA within the
EU-FP7-IA project ENSAR (No.\ 262010), by the ERANET-NuPNET grant
SARFEN of the Polish National Centre for Research and Development
(NCBiR), by the Polish National Science Centre (NCN) under Contracts
Nos.\ 2014/15/N/ST2/03454 and 2015/17/N/ST2/04025, by the Academy of
Finland and University of Jyv\"askyl\"a within the FIDIPRO program,
and by the U.S. Department of Energy under Contract Nos.\
DE-AC52-07NA27344 (Lawrence Livermore National Laboratory) and
DE-SC0008499 (NUCLEI SciDAC Collaboration).

Computing support for this work came from the Lawrence Livermore National
Laboratory (LLNL) Institutional Computing Grand Challenge program.
Computational resources were also provided by a computational grant from
the Interdisciplinary Centre for Mathematical and Computational
Modeling (ICM) of the Warsaw University and by the Swierk Computing
Centre (CIS) at the National Centre for Nuclear Research (NCBJ). We acknowledge
an award of computer time by the Innovative and Novel Computational
Impact on Theory and Experiment (INCITE) program. This research used
resources of the Oak Ridge Leadership Computing Facility located in the
Oak Ridge National Laboratory, which is supported by the Office of Science
of the U.S. Department of Energy under Contract No.\ DE-AC05-00OR22725,
and of the National Energy Research Scientific Computing Center, which is
supported by the Office of Science of the U.S.\ Department of Energy under
Contract No.\ DE-AC02-05CH11231. We acknowledge the CSC-IT Center for
Science Ltd., Finland, for the allocation of computational resources.

%%%%%%%%%%%%%%%%%%%%%%%%%%%%%%%%%%%%%%%%%%%%%%%%%%%%%%%%%%%%%%%%%%%%%%%%%%%%%%%%
%%%%%%%%%%%%%%%%%%%%%%%%%%%%%%%%%%%%%%%%%%%%%%%%%%%%%%%%%%%%%%%%%%%%%%%%%%%%%%%%
%%%%%%%%%%%%%%%%%%%%%%%%%%%%%%%%%%%%%%%%%%%%%%%%%%%%%%%%%%%%%%%%%%%%%%%%%%%%%%%%
%%%%%%%%%%%%%%%%%%%%%%%%%%%%%%%%%%%%%%%%%%%%%%%%%%%%%%%%%%%%%%%%%%%%%%%%%%%%%%%%

%\bibliography{hfodd}
%\bibliography{jacwit31}
\bibliography{hfodd,/Actual/LaTeX/Latex.all/jacwit33}
%\bibliography{/Actual/LaTeX/Latex.all/jacwit32}
\bibliographystyle{cpc}

\end{document}